\documentclass[11pt,a4paper]{article}
\usepackage[english]{babel}
\usepackage[T1]{fontenc}
\usepackage[utf8]{inputenc}
\usepackage{float}   
\usepackage{url}     
\usepackage{amsfonts,amsthm,amsmath,amssymb,graphicx}
\usepackage{geometry}
\usepackage{tikz}
\usepackage{csquotes}
\usepackage{enumitem}  
\usepackage{color}
\usepackage[flushleft]{threeparttable}
\usepackage{tabularx} 
\usepackage{lscape} 
\usepackage{rotating}
\usepackage{array}
\usepackage{booktabs}
\usepackage{hyperref}
\usepackage{xurl}
\usepackage{bm}
\usepackage{caption}
\usepackage{subcaption}
\usepackage{setspace}
\usepackage[export]{adjustbox}
\usepackage{xcolor}
\usepackage{soul}
\usepackage{bbm}

\newcommand{\indep}{\perp \!\!\! \perp}
\theoremstyle{plain}
\newtheorem{thm}{Theorem}
\newtheorem{prop}{Proposition}

\newtheorem{cor}{Corollary}
\newtheorem{lem}{Lemma}
\newtheorem{ass}{Assumption}
\newtheorem{definition}{Definition}

\usepackage[
authordate,bibencoding=auto,strict,backend=biber,natbib,maxbibnames=99,maxcitenames=2,doi=false,isbn=false,url=false,eprint=false]{biblatex-chicago}
\title{Production function estimation using subjective expectations data}
\author{Agnes Norris Keiller\thanks{London School of Economics}
	\and \'{A}ureo de Paula\thanks{University College London, Institute for Fiscal Studies and CeMMAP}
	\and John Van Reenen\thanks{London School of Economics and MIT}}
\date{\today}

\begin{document}

	\maketitle

	\begin{abstract}
	\noindent	Standard proxy methods for estimating production functions in the \citet{olley_dynamics_1996} tradition require assumptions on input choices. We introduce a new method that exploits (increasingly available) data on firms' expectations of their future output and inputs that allows us to obtain consistent production function parameter estimates while relaxing these input demand assumptions. In contrast to both proxy and dynamic panel methods like \citet{blundell_gmm_2000}, our proposed estimator can be implemented on a single cross-section of data and Monte Carlo simulations show it outperforms alternative estimators when firms' material input choices are subject to optimization error. Implementing a range of production function estimators on UK panel data, we find our proposed estimator yields results that are either similar to or more credible than commonly-used alternatives. These differences are larger in industries where material inputs appear harder to optimize. We show that the share of cross-firm TFP dispersion accounted for by persistent productivity differences is substantially larger when calculated using parameter estimates from our proposed estimator.

\bigskip

\bigskip
\newline
\noindent \footnotesize \textbf{Acknowledgements: } We would like to thank  Orazio Attanasio, Nick Bloom, Richard Blundell, Moshe Buchinsky, Michele Fioretti, Bruce Hansen, Bo Honor\'{e}, Joel Horowitz, Isabela Manelici, Chuck Manski, Ariel Pakes, Joris Pinkse, Maarten de Ridder, Jean-Marc Robin and participants at several institutional seminars and conferences for comments. Generous funding has come from the ESRC through the Programme for Innovation and Diffusion (Grant Ref: ES/V009478/1) and through the ESRC Institute for the Microeconomic Analysis of Public Policy (Grant Ref: ES/T014334/1) and from UK Research and Innovation (UKRI) under the UK government’s Horizon Europe funding guarantee (Grant Ref: EP/X02931X/1). This work was undertaken in the Office for National Statistics Secure Research Service using data from ONS and other owners and does not imply the endorsement of the ONS or other data owners.
  
\end{abstract}

\maketitle

\newpage

\onehalfspacing

\section{Introduction}
	
The ``production function'' -- a representation of the process by which inputs are turned into outputs -- has long been an object of great economic interest (\citealp{cobb_theory_1928,strom_production_1999}). Production functions are critical to examining a wide range of topics including technological change, productivity dispersion, firm markups and the impact of policy. Research on these topics has gained greater salience in recent years, in part due to the productivity growth slowdown, particularly since the Global Financial Crisis. Before one can analyze such topics, however, it is necessary to consistently estimate a production function, which has proven no easy task.

Econometric research on production function estimation has had a renaissance in recent years.\footnote{For surveys see \citet{ackerberg_chapter_2007} or \citet{de_loecker_industrial_2021}. Recent contributions include  \citet{ackerberg_identification_2015}, \citet{collard-wexler_productivity_2021}, \citet{de_loecker_product_2011}, \citet{de_loecker_prices_2016}, \citet{doraszelski_rd_2013, doraszelski_measuring_2018}, \citet{gandhi_identification_2020},  \citet{orr_within-firm_2022}, \citet{de_roux_estimating_2021} and \citet{valmari2023}.} Estimation is complicated by a number of long-known issues, most notably the endogeneity of inputs: because a firm's productivity is unobservable and likely correlated with input choices, straightforward OLS estimation methods will be inconsistent (\citealp{marschak_random_1944,zellner_specification_1966}). 
 
Standard methods to deal with these problems included controlling for firm fixed effects  by differencing \citep{mundlak_empirical_1961} and instrumenting with lagged input values \citep{anderson_estimation_1981}, for example. Such approaches generally find implausibly low estimates for relevant parameters, especially on the output elasticity with respect to capital.\footnote{This has generally been thought to be because of the high persistence of the capital stock. Differencing removes all cross sectional information on capital, and much of the remaining time series variation may be measurement error. Moreover, lags will be poor predictors of the change in the capital stock, if the level of capital is close to a random walk.} \citet{blundell_gmm_2000} consider an alternative approach by including lagged differences as instruments for the levels of factor inputs. This nonetheless requires conditioning on at least three consecutive time series observations on a firm, which in many empirical settings loses a considerable subset of data. Moreover, it relies on exact parametric specification of the productivity process and requires a strong stationarity assumption (or a restriction on initial conditions, itself a ``stationarity restriction'', see \cite{blundellbond2023}), making the method potentially inappropriate for younger and fast-growing firms.

The drawbacks of these dynamic panel data methods have contributed to the popularity of an alternative suite of ‘proxy variable’ estimators that use a non-parametric function of various observables to control for unobserved productivity. The pioneers of this method were \citet{olley_dynamics_1996} (OP), who control for productivity using a flexible function of investment and capital that represents the inverse of a firm's investment policy. The reasoning behind their approach is that if firms' investment policy function can be written as an invertible function of pre-determined capital and the persistent component of unobserved productivity, then the latter can be proxied with a flexible function of capital and investment. \citet{levinsohn_estimating_2003} (LP) instead propose using the inverse of firms’ material input demand as a proxy for productivity to address the concern that there are often many observations of zero investment, which is an issue for the OP approach. 

Noting that a production function’s labor input parameter is unidentified using the LP methodology under plausible assumptions, \citet{ackerberg_identification_2015} (ACF) outline a refinement on timing assumptions and proxy variable arguments to address this. Unlike LP, who rely on a material input demand function conditioned solely on capital, ACF propose controlling for unobserved productivity by inverting a material input demand function conditioned on labor as well as capital and then recovering both input elasticities in a second estimation stage. \citet{gandhi_identification_2020} (GNR) propose an alternative estimation strategy based on the implications of price-taking firms’ optimality conditions. \citet{bond_adjustment_2005} provide another estimator, showing that in the presence of adjustment costs on all inputs the parameters of a Cobb-Douglas production function can be recovered by using lagged levels of inputs as instruments. This approach incorporates aspects of both the dynamic panel literature -- in using lags as instruments and specifying the productivity process -- and the proxy variable approach -- by relying on the implications of optimal firm input decisions to yield identification. However, simulation results show that their proposed method is sensitive to the form and magnitude of adjustment costs, which may explain why it has not been widely deployed. More in-depth reviews of alternative production function estimation strategies are provided in \citet{ackerberg_chapter_2007} and \citet{de_loecker_industrial_2021}.

Despite their differences, proxy methods such as OP, LP and ACF all rely on the existence of a strictly
monotonic relationship between a firm’s (conditional) input demand and productivity -- an assumption usually justified with recourse to models of firms’ decisions that yield optimal policies satisfying monotonicity. The performance of these estimation methods is therefore threatened by any unobserved factor that violates the required relationship between productivity and the input used to generate its proxy. This includes input adjustment costs, prices that vary across firms and optimization errors.\footnote{\citet{gandhi_identification_2020}, by contrast, explicitly requires firms’ flexible input demands to be optimal and is therefore compromised by any factor causing deviations from optimality.}

Our paper contributes to the literature on production function estimation by showing how data on a firm's perception of its future output and inputs can be used to recover consistent production function parameter estimates while relaxing assumptions on firms’ input demand policies.  We leverage recent surveys that collect detailed information on firms' perceived \emph{probabilistic distribution} of output (e.g.,revenues) and inputs (e.g., employment) in the future. The intuition underpinning our approach is that a firm's expectations regarding future output and inputs contain information about its expected future productivity which, in turn, contains information about current productivity. Unlike dynamic panel estimators, which require parametric specification of the productivity process, we require the relatively common assumption that persistent productivity follows a first-order Markov process -- an assumption imposed by OP, LP, ACF and GNR.

Combined with assumptions that persistent productivity is uni-dimensional (which is also imposed by standard proxy variable methods) and that there is a monotonic relationship between current and expected future productivity, firms’ expectations can be used to control for unobserved productivity and thereby recover consistent parameter estimates.\footnote{The requirement that persistent productivity be uni-dimensional confines us to a setting where productivity shocks are Hicks-neutral. While this is conventional in the literature, two notable exceptions that accommodate factor-augmenting technology shocks are \citet{doraszelski_measuring_2018} and \citet{demirer_production_2022}. \citet{doraszelski_measuring_2018} relaxes the assumption of unidimensional productivity by trying to leverage data on firm-level input prices while \citet{demirer_production_2022} does so by imposing assumptions on firms’ input demands. While both these papers are valuable contributions and argue the importance of factor-augmenting productivity shocks, identifying two sources of technical change is more challenging (e.g., average wages reflect labor composition changes as well as pure labor cost shocks).}  Our proposed method, which we label ``NPR'', is therefore similar to OP/LP/ACF as it requires a monotonic relationship between productivity and observables, but is distinct as it leverages data on firm expectations rather than input quantities. It is therefore robust to a range of factors -- such as unobserved firm-specific input prices and optimization error -- that would undermine alternative estimators by breaking the one-to-one link between input demands and productivity. Moreover, unlike the dynamic-panel and the control function estimators, our proposed method can be implemented in a single cross-section of data. This enables production function estimation in contexts where panel data would be infeasible to collect for either cost or logistical reasons.  We also note that the moments we exploit here can be seen as complementary to those already in use in the traditional methods whenever data is available to implement both, thus lending more precision or over-identifying restrictions to existing methods.

Monte Carlo simulations show that our proposed estimation method recovers precise estimates of production function parameters under a range of data generating processes. Notably, it retains consistency when firms' input decisions are subject to optimization error whereas other approaches generally do not. We also note that our estimation method can accommodate biased input expectations as long as such bias is also reflected in firms' expected output.  While our proposed estimator is undermined if firms' reported expectations are measured with error, we leverage insights from \citet{evdokimov_zeleneev_2025} to show our basic estimation algorithm can be extended to retain consistency. We call this extension ``EZ-NPR'' and note that it (and others) can also accommodate certain forms of bias in firms' expected next-period productivity which, similarly to measurement error in expectations, undermines our ``baseline'' NPR. If the bias in productivity is uncorrelated with all observables used in the estimator, it operates like a measurement error and can be dealt with by EZ-NPR. If it is instead a function of a time-invariant attribute, and if one possesses panel data on realizations and expectations of revenue and employment, we propose an alternative extension that retains consistency. Finally, we also highlight that a similar ``wedge'' between expected output, the true production function and expected productivity may arise in some circumstances when firms have imperfect knowledge of the production technology and that this issue can be dealt with by similar extensions to our basic approach.

To test the empirical performance of our method, we leverage the UK's Management and Expectations Survey \citep{office_for_national_statistics_mesmanagement_2022}. The MES records information on firms' inputs, output and their one-year-ahead expectations of these objects in 2017 and 2020. We focus on three industries -- electronics manufacturing, retail/wholesale and restaurants -- and estimate industry-specific production functions using a range of methods. The estimates recovered using our proposed method are broadly similar for the electronics and retail production functions but differ in non-negligible yet plausible ways for the restaurant sector. To rationalize these results, we show material inputs are subject to particularly large within-year revisions among the restaurant sector, which suggests optimization is particularly hard for these firms and hence the LP and ACF monotonicity assumption is less likely to hold. 

Finally, we use the alternative production function estimates to recover total factor productivity (TFP), which the control function and proposed estimators allow us to decompose into persistent and transitory components. Comparing variance shares across estimators, we find that our proposed estimator consistently attributes a larger share of total TFP variance to the persistent component. In electronics, for example, our estimates imply that 64\% of total TFP variance is persistent, compared to around 43\% under ACF, despite both estimators yielding similar overall TFP variance. The corresponding figures are 77\% versus 49\% in retail and 52\% versus 33\% in restaurants.

Combined with the Monte Carlo evidence, our empirical application demonstrates the utility of expectations data in the context of production function estimation and thereby contributes to a more general literature
documenting the value of expectations data. Starting in the 1990s much of this literature's initial focus was on income dynamics: the seminal work by \citet{dominitz_using_1997}, for example, demonstrates how surveys can be used to elicit subjective income expectations, while \citet{pistaferri_superior_2001}, shows the econometric benefits of such additional information as a means to separately identify permanent and transitory shocks to income. \citet{manski_measuring_2004} summarised these early advances and argued that data on expectations could be useful both as a means to relax and to validate assumptions within various economic models. Of the subsequent work that has examined the value of subjective expectations data in a wide range of contexts, our work is related to \citet{gennaioli_expectations_2016}, in that it demonstrates insights that can be gained from firms' expectations rather than those of individuals (in particular, \cite{bloom_rationalizing}). Our work is also related to recent work by \citet{attanasio_modelling_2022}, who return to the literature's early focus on subjective income expectations and show how such data can be used to estimate income processes in a flexible manner that relaxes commonly-imposed parametric assumptions. Similar to their work, we document that the additional information contained within data on subjective expectations allows one to relax particular assumptions that underpin conventional production function estimators and thus implicitly allows for more flexible models of firm behavior.  

The remainder of the paper is as follows. In Section \ref{sect:methodology} we show how expectations data identify production function parameters, describe our proposed estimation methodology and compare it to other standard methods. Section \ref{sect:montecarlo} outlines the Monte Carlo setup we use to compare alternative estimators and discusses the results across various data generating processes. Section \ref{sect:data} describes the data we use in our empirical application, the results of which are described in Section \ref{sect:results}. Section \ref{sect:conclusion} concludes. In online Appendices we provide a discussion of expectation formation processes (\ref{adix:expectations}), a convergence proof of our estimator (\ref{adix:convergence}), NPR implementation details (\ref{adix:implementation}), extensions to the estimator, in particular to measurement error in expectations and to biased expectations (\ref{adix:bias}), Monte Carlo details (\ref{adix:mc_setup}) and Data and Supplementary results (\ref{adix:supp_results}).

\section{Production Function Estimation Using Firms' Expectations}\label{sect:methodology}

Consider a general production function of the following form
\begin{equation}\label{eq:prodfunct_general}
y_{it}=f(k_{it},l_{it}; \beta)+\omega_{it}+\epsilon_{it},
\end{equation}
where subscript $i$ denotes firm and subscript $t$ denotes time. Lower case letters denote logs, so $y$ is the log of output, $k$ is the log of capital, $l$ is the log of labor and $f(\cdot;\beta)$ is some general function of the two with parameters $\beta$, which captures the process by which they are combined during production.\footnote{For the general exposition of this subsection, output may either be value added (i.e., net output) or sales revenue (i.e., gross output). In the latter case, the omission of materials as an input can be justified under a Leontief model of production in which labor and capital are combined in a fixed proportion with materials \citep{ackerberg_identification_2015}. In practice, the distinction will influence how data on firms' expectations are treated, which we return to in Section \ref{sect:data}.} The scalar variables $\omega$ 
and $\epsilon$ are unobserved by the econometrician. The variable $\omega$ represents idiosyncratic productivity that is known by the firm at the time period-$t$ input and investment decisions are made. In contrast, $\epsilon$ are unanticipated mean-zero disturbances representing productivity shocks, such as extreme weather events or machine failures, which only become observable to the firm \emph{after} its period-$t$ decisions have been made. In addition, $\epsilon$ can include mean-zero measurement error, which does not affect the firm but may pose problems to the econometrician \citep{strom_production_1999}.    
Capital evolves according to
\begin{equation*}\label{eq:capital_lom}
K_{it}=(1-\delta)K_{it-1}+I_{it-1},
\end{equation*}
where $\delta$ is the depreciation rate and $I_{it-1}$ is investment.
Unobserved productivity $\omega$ follows a Markov process

\begin{equation*}
\omega_{it}=\mathbb{E}[\omega_{it}|\Omega_{it-1}]+\xi_{it}=\mathbb{E}[\omega_{it}|\omega_{it-1}]+\xi_{it}=g(\omega_{it-1})+\xi_{it},
\end{equation*}
where $\mathbb{E}[\xi_{it}|\Omega_{it-1}]=0$ and $\Omega_{it-1}$ represents the firm's information set at $t-1$. The information includes $k_{it-1},l_{it-1},\omega_{it-1},I_{it-1}$ (and thus $k_{it}$) but also additional variables such as input and output prices and demand factors. 

\subsection{Identification}\label{subsect:identification}

The method we propose relies on a simple intuition, and there are two elements to this. First, data on a firm's expectations about its own output and inputs one period ahead ($t+1$) are informative about expected productivity one period ahead ($t+1$). If productivity follows a Markov process, those data thus also carry information about the firm's own productivity in the current period ($t$). Second, once period $t$ productivity is controlled for using expectations for $t+1$, current inputs are orthogonal to the remaining error, which corresponds to random shocks to productivity not observed by the firm before making input choices. The expectations data are therefore useful to proxy for unobserved productivity in the spirit of other production function proxy methods. We regard the first element of this intuition as the more fundamental contribution, as it highlights how a new type of data can aid the estimation of production functions in contrast to traditional approaches that rely on panel data of inputs and output. 

To formalize this intuition, suppose firms form expectations about their period $t+1$ production and inputs at the end of period $t$ conditional on their information set $\Omega_{it}\supset\{k_{it},l_{it},I_{it},\omega_{it},k_{it+1}\}$. As is usually the case, we assume that firms know their production function $f(\cdot; \beta)$ and its parameter values, so they both belong to the information set. We discuss imperfect knowledge in subsection \ref{subsect:method_bias}. 

Following the notation in \citet{pesaranweale2006}, we append the subscripts $i$ and $t$ to highlight that the relevant variables are obtained with respect to the subjective probabilities reported by decision makers in firm $i$ at period $t$ (i.e., $\mathbb{E}_{it}[ \mathbf{x}_{it+1} | \Omega_{it} ] = \int \mathbf{x}_{t+1} h_{it}(\mathbf{x}_{t+1}| \Omega_{it})d\mathbf{x}_{t+1}$, where $\mathbf{x}_{t+1}$ is a vector and $h_{it}(\cdot)$ represents firm $i$'s distribution for it in period $t$ given their information $\Omega_{it}$). As long as they satisfy the identification conditions in Theorem \ref{thm:ident} and Corollary \ref{cor:ident} (see below), we do not take a stand on the expectation formation process for output or inputs and treat those as data. One may nonetheless envision different scenarios on the expectation process for productivity and we discuss different expectation formation protocols in Appendix \ref{adix:expectations}, such as rational expectations, static expectations, ``return to normality'' and adaptive expectations. 

For simplicity, in what follows we assume firms' expectations to be rational (i.e., $\mathbb{E}_{it}[ \omega_{it} | \Omega_{it-1} ] = \mathbb{E}[\omega_{it}|\Omega_{it-1}]$), which implies
\begin{equation}\label{eq:eprodfunct_general}
\begin{split}
\mathbb{E}_{it}[y_{it+1}|\Omega_{it}] & =\int f(k_{it+1},l_{it+1}; \beta)dF_{it}(l_{it+1})+\mathbb{E}_{it}[\omega_{it+1}|\Omega_{it}]+\mathbb{E}_{it}[\epsilon_{it+1}|\Omega_{it}]\\
& =\int f(k_{it+1},l_{it+1}; \beta)dF_{it}(l_{it+1})+g(\omega_{it}), \\
\end{split}
\end{equation}
where $F_{it}(l_{it+1})$ represents firm $i$'s subjective probability distribution over their next-period labor input given its information $\Omega_{it}$ (i.e., following the notation above, $F_{it}'(l_{it+1}) \equiv h_{it}(l_{t+1}|\Omega_{it})$) and the second equality follows from the assumptions that $\mathbb{E}_{it}[\epsilon_{it}|\Omega_{it-1}]$ and $\mathbb{E}_{it}[\xi_{it}|\Omega_{it-1}]$ are equal to zero. The former ($\mathbb{E}_{it}[\epsilon_{it}|\Omega_{it-1}]=0$) relies on $\epsilon_{it}$ being unanticipated by definition and thus orthogonal to the information set $\Omega_{it}$ as discussed earlier. This also rules out measurement error in elicited expectations and distributions, which we return to in subsection \ref{subsect:measurement_error}.  The latter ($\mathbb{E}_{it}[\xi_{it}|\Omega_{it-1}]=0$) follows from rational expectations about $\omega_{it}$ and thus $\mathbb{E}_{it}[\xi_{it}|\Omega_{it-1}]=\mathbb{E}[\xi_{it}|\Omega_{it-1}]=0$.

Rearranging equation (\ref{eq:eprodfunct_general}) for $g(\omega_{it})$ obtains

\begin{equation}\label{eq:gomega_general}
g(\omega_{it}) = \mathbb{E}_{it}[y_{it+1}|\Omega_{it}] - \int f(k_{it+1},l_{it+1}; \beta)dF_{it}(l_{it+1}). 
\end{equation}

Like other proxy variable approaches, we now require a monotonicity assumption. In our case this assumption is that the right hand side of equation (\ref{eq:gomega_general}) is strictly increasing in $\omega_{it}$. Or, in words: given a firm's current (persistent) productivity there is a single level of productivity they expect next period and this single level can be uniquely inferred from their expectations about next-period output, labor and the deterministic level of next-period capital. A related assumption -- i.e., that the conditional distribution of $\omega_{it+1}$ given $\omega_{it}$ be increasing in the latter -- appears in ACF (see their Assumption 2).  It is used there and in OP to guarantee their invertibility condition in an optimal firm decision model (see \citet{pakes1994}), but it is not needed in their estimation protocol and typically it is not imposed.  This assumption is implied by the linearity assumptions on the productivity process in other methodologies (e.g., \cite{blundell_gmm_2000}).  Whereas we think the assumption is generally reasonable, this remains of course a matter to be examined according to the context at hand.  Elsewhere, for instance, stochastic processes for interest rates or commodity prices have been documented to be non-monotonic due to regime switches or equilibrium conditions (see, e.g., \cite{angbekaert2000} or \cite{deatonlaroque1992}, respectively).  

Under the strict monotonicity assumption, the scalar $\omega_{it}$ can be recovered as

\begin{equation}\label{eq:gomega_inverse}
\omega_{it} = g^{-1}\left(\mathbb{E}_{it}[y_{it+1}|\Omega_{it}] - \int f(k_{it+1},l_{it+1}; \beta)dF_{it}(l_{it+1})\right).
\end{equation}

If output ($y_{it}$), inputs ($k_{it}$, $k_{it+1}$ and $l_{it}$), and beliefs ($\mathbb{E}_{it}[y_{it+1}|\Omega_{it}]$ and $F_{it}(\cdot)$) are observable, we can combine equations (\ref{eq:prodfunct_general}) and (\ref{eq:gomega_inverse}) to obtain a moment condition that can be used to recover the parameters of interest:

\begin{equation}\label{eq:moment1}
\begin{split}
& \mathbb{E}[\epsilon_{it}|\Omega_{it}] = \mathbb{E}[y_{it}-f(k_{it},l_{it}; \beta)-\omega_{it}|\Omega_{it}] \\
&= \mathbb{E}\left[y_{it}-f(k_{it},l_{it}; \beta) -\Psi\left(\mathbb{E}_{it}[y_{it+1}|\Omega_{it}] - \int f(k_{it+1},l_{it+1}; \beta)dF_{it}(l_{it+1})\right)\Big|\Omega_{it} \right]  \\
&=0,
\end{split}
\end{equation}    

\noindent where $\Psi$ is some non-parametric function representing $g^{-1}(\cdot)$. 

Consider for example the case of a Cobb-Douglas production function where $\beta = (\alpha,\beta_k,\beta_l)$ and an AR(1) process for $\omega$ with auto-regressive parameter $\rho$ (i.e., $g(\omega)=\rho \omega$), then equation (\ref{eq:moment1}) becomes
\begin{equation}
\mathbb{E}\left[y_{it}|\Omega_{it} \right]- \frac{\alpha (\rho-1)}{\rho}-\beta_kk_{it}-\beta_ll_{it}-\frac{1}{\rho} \mathbb{E}_{it}[y_{it+1}|\Omega_{it}] + \frac{\beta_k}{\rho} k_{it+1} + \frac{\beta_l}{\rho} \mathbb{E}_{it}[l_{it+1}|\Omega_{it}] =0,
\end{equation}
as long as $\rho \ne 0$.  The model would then identify $\theta=(\beta,g)$ (where $g(\omega)=\rho \omega$) if, for example, $\mathbb{E}[x_{it}x_{it}^\top]$, where $x_{it}=(1,k_{it},l_{it},$ $\mathbb{E}_{it}[y_{it+1}|\Omega_{it}],k_{it+1},\mathbb{E}_{it}[l_{it+1}|\Omega_{it}])$, has full rank. This example highlights that identification of $\theta$ by equation (\ref{eq:moment1}) will depend on the specifications of the production function $f(\cdot; \beta)$, the Markov process encoded in $g(\cdot)$ and on the degree of variation observed in the data. 

For more general specifications, one can establish that:

\begin{thm} \label{thm:ident} Assume that $g:\mathbb{R} \rightarrow \mathbb{R}$ is strictly monotonic and $\mathbb{E}[\epsilon_{it}|\Omega_{it}]=0$.  Let $x_{it} = (k_{it},l_{it})$ and $z_{it} = (\mathbb{E}_{it}[y_{it+1}|\Omega_{it}],k_{it+1},F_{it}(\cdot))$ and denote by $\theta_0 = (\beta_0,g_0)$ the data generating parameters. Then, if
\begin{equation}\label{eq:nlls} 
f(x_{it}; \beta_0)-\mathbb{E}\big[f(x_{it}; \beta_0)|z_{it}\big] \neq f(x_{it}; \beta)-\mathbb{E}\big[f(x_{it}; \beta)|z_{it}\big]
\end{equation}
with positive probability for any $\beta \neq \beta_0$, the parameter vector $\theta_0 = (\beta_0,g_0)$ is identified.
\end{thm}

\noindent \emph{Proof.} Notice that
$y_{it} = f(x_{it}; \beta_0) + h_0(z_{it}) + \epsilon_{it}$, where 
$$h_0(z_{it}) = \Psi_0\left(\mathbb{E}_{it}[y_{it+1}|\Omega_{it}] - \int f(k_{it+1},l_{it+1}; \beta_0)dF_{it}(l_{it+1})\right).$$ Taking expectations conditional on $z_{it}$ on both sides of $y_{it} = f(x_{it}; \beta_0) + h_0(z_{it}) + \epsilon_{it}$ and subtracting, one obtains that
$$\underbrace{y_{it} - \mathbb{E}(y_{it}|z_{it})}_{\equiv w_{it}} = \underbrace{f(x_{it}; \beta_0) - \mathbb{E}(f(x_{it}; \beta_0)|z_{it})}_{\equiv m(x_{it},z_{it};\beta_0)} + \epsilon_{it},$$
where we use the fact that $h_0(z_{it}) = \mathbb{E}[h_0(z_{it}) | z_{it}]$.  Since $\mathbb{E}[\epsilon_{it}|\Omega_{it}]=0$ and $\{ x_{it},z_{it} \} \subset \Omega_{it}$, we have that $\mathbb{E}[\epsilon_{it}|x_{it},z_{it}]=0$ and $m(x_{it},z_{it};\beta_0)=\mathbb{E}(w_{it}|x_{it},z_{it})$.  It thus uniquely solves $\min_{\tilde m(\cdot)} \mathbb{E}[(w_{it}-\tilde m(x_{it},z_{it}))^2]$ as long as condition (\ref{eq:nlls}) is satisfied with positive probability, which implies that $\beta_0$ is identified.
\newline \newline
The function $\Psi_0(\cdot) \equiv g_0^{-1}(\cdot)$ is then identified since
$$\underbrace{y_{it}-f(x_{it}; \beta_0)}_{\equiv \tilde y_{it}}=\Psi_0\left(\underbrace{\mathbb{E}_{it}[y_{it+1}|\Omega_{it}] - \int f(k_{it+1},l_{it+1}; \beta_0)dF_{it}(l_{it+1})}_{\equiv \tilde z_{it}}\right)+\epsilon_{it}.$$
Since $\tilde z_{it} \subset \Omega_{it}$, we thus have that $\mathbb{E}[\epsilon_{it}|\tilde z_{it}]=0$ and $\Psi_0(\tilde z_{it})=\mathbb{E}(\tilde y_{it}|\tilde z_{it})$ and $g_0(\cdot)=\Psi_0^{-1}(\cdot)$. \qquad $\blacksquare$
\newline \newline
The identification result generalizes ideas in \citet{robinson_root-n-consistent_1988}, who deals with partially linear models where $f(\cdot; \beta)$ is linear. 
 Condition (\ref{eq:nlls}) is a conventional identification assumption used in the context of (nonlinear) least squares applied to the parametric function $m(\cdot;\beta)$, which can be obtained from $f(\cdot; \beta)$ and the observable distribution of $x_{it}$ given $z_{it}$. It would fail, for instance, if expectations (in $z_{it}$) are static and injective functions \emph{solely} of inputs (in $x_{it}$), but is likely to hold in more general settings where expectations depend on other elements in the information set $\Omega_{it}$.  In fact, if $f(\cdot; \beta)$ is linear in parameters (e.g., Cobb-Douglas and translog), the result boils down to that in \citet{robinson_root-n-consistent_1988}:

\begin{cor} \label{cor:ident} Under the assumptions of Theorem \ref{thm:ident}, if $f(\cdot; \beta)$ is linear in parameters and, in place of Condition (\ref{eq:nlls}), suppose that
\begin{equation} \label{eq:pl}
\mathbb{E}\left\{ [ x_{it}-\mathbb{E}(x_{it}|z_{it}) ][ x_{it}-\mathbb{E}(x_{it}|z_{it}) ]^\top \right\}
\end{equation}
is non-singular. Then the parameter vector $\theta_0 = (\beta_0,g_0)$ is identified.
\end{cor}

\noindent \emph{Proof.} Since $f(\cdot; \beta)$ is linear in parameters we can represent it as $f(x_{it};\beta)=x_{it}^\top \beta$. The result obtains as Condition (\ref{eq:pl}) implies Condition (\ref{eq:nlls}). Suppose that there exists $\beta \neq \beta_0$ such that
\begin{equation*}
\begin{split}
 & \mathbb{P}\left( f(x_{it}; \beta_0)-\mathbb{E}\big[f(x_{it}; \beta_0)|z_{it}\big] = f(x_{it}; \beta)-\mathbb{E}\big[f(x_{it}; \beta)|z_{it}\big] \right) \\
 & = \mathbb{P}\left( [ x_{it}-\mathbb{E}(x_{it}|z_{it})]^\top(\beta_0 - \beta) = 0 \right) = 1.
\end{split}
\end{equation*}
This then implies that
$$\mathbb{E}\left\{ [ x_{it}-\mathbb{E}(x_{it}|z_{it})][ x_{it}-\mathbb{E}(x_{it}|z_{it})]^\top\right\}(\beta_0 - \beta)=0.$$
This means that $\beta_0 - \beta \neq 0$ is in the nullspace of $\mathbb{E}\left\{ [ x_{it}-\mathbb{E}(x_{it}|z_{it})][ x_{it}-\mathbb{E}(x_{it}|z_{it})]^\top \right\}$ thus implying that this matrix is singular. Hence, Condition (\ref{eq:pl}) implies Condition (\ref{eq:nlls}) and the result follows from Theorem \ref{thm:ident}. \qquad $\blacksquare$ \newline

\noindent This can also be obtained by directly applying the results in \citet{robinson_root-n-consistent_1988}. As discussed there (see p.940), Condition (\ref{eq:pl}) prevents any element of $x_{it}$ (inputs, in our case) from being almost surely perfectly predictable by $z_{it}$ (expectations, in our case) in the least squares sense, although it does not preclude (nonlinear) functional relations among $x_{it}$ elements and identification is possible even if $x_{it}$ uniquely defines $z_{it}$, when the converse is not true. If $x_{it}$ (inputs, in our case) and $z_{it}$ (expectations, in our case) are observable, perfect predictability of the former by the latter is in principle testable.  This condition would fail, as already mentioned above, if expectations about future output and inputs (which appear in $z_{it}$) are static and injective functions \emph{solely} of inputs (which appear in $x_{it}$), but is likely to hold in more general settings where expectations depend on other elements in the information set $\Omega_{it}$ for example.

\subsection{Estimation}\label{subsect:estimation}

As highlighted in the previous section, the protocol we put forward builds on the idea that expectations about future output and inputs contain information about future productivity and, under Markovianity, about current productivity.  This can be leveraged to proxy for current productivity using observed expectations and thus address econometric endogeneity concerns.  Equation (\ref{eq:moment1}) can then be used to recover estimates of $(\beta,g)$ that are either fully parametric or semi-parametric depending on whether one specifies the functional form of $g(\cdot)$. In the remainder of this paper, we follow a semi-parametric approach, which allows us to avoid imposing structure on $g(\cdot)$ and yields a novel estimation methodology. This section focuses on the case of Cobb-Douglas production technology to outline our proposed methodology, although it generalizes to other specifications (e.g.,translog), which we examine in our empirical application.

Cobb-Douglas production implies:
\begin{equation}\label{eq:prodfunct_cd}
\begin{split}
y_{it}&=\beta_0+\beta_kk_{it}+\beta_ll_{it}+\omega_{it}+\epsilon_{it} \\
&=\beta_0+\beta_kk_{it}+\beta_ll_{it}+\Psi\left(\mathbb{E}_{it}[y_{it+1}|\Omega_{it}] - \beta_0-\beta_kk_{it+1}-\beta_l\mathbb{E}[l_{it+1}|\Omega_{it}]\right)+\epsilon_{it}. 
\end{split}
\end{equation}

Assuming $\Psi$ is a smooth function, equation (\ref{eq:prodfunct_cd}) is an example of a generalized additive model, early explorations of which were provided for instance by \citet{hastie_generalized_1986}, and the partially linear model studied by \citet{robinson_root-n-consistent_1988} among others. Hastie and Tibshirani characterize the non-linear part of the model -- in our case, the $\Psi$ function -- as a weighted sum of unspecified smooth functions, the parameters and weighting of which can be recovered using (quasi-)maximum-likelihood-based estimation. As in a linear regression, maximum likelihood based on normal errors amounts to least squares minimization here. 

There are two specific features of the model of equation (\ref{eq:prodfunct_cd}) that depart from the standard generalized additive model. First, we know that $\Psi$ is monotonic, which amounts to imposing constraints on the derivative of $\Psi$. The exact form of these constraints and the consequent optimization problem are derived by \citet{pya_shape_2015}, who also present an algorithm to estimate such ``shape constrained'' generalized additive models that we deploy in our empirical application.

Second, the argument of the smooth $\Psi$ function is itself a function of the model's parameters. To address this issue, we take inspiration from \citet{friedman_projection_1981}, who develop an iterative ``backfitting'' algorithm that recovers parameter estimates in additive models where the arguments of the smooth functions are linear functions of parameters (see also \citet{ichimura_chapter_2007}). Adapting the algorithm to our setting yields the following iterative estimation procedure:
\begin{enumerate}
\item Pick initial parameter values $(\hat{\beta}_{k0},\hat{\beta}_{l0})$.
\item For iteration $j$, calculate $Z_{ij}=\mathbb{E}_{it}[y_{it+1}|\Omega_{it}] - \hat \beta_{kj-1}k_{it+1}-\hat \beta_{lj-1}\mathbb{E}_{it}[l_{it+1}|\Omega_{it}]$.
\item Fit the model $y_{it}=\beta_kk_{it}+\beta_ll_{it}+\Psi\left(Z_{ij}\right)+\epsilon_{it}$ using the shape constrained estimation protocol of \citet{pya_shape_2015} to obtain $(\hat{\beta}_{j},\hat{\Psi}_j)$.\footnote{The constant $\hat{\alpha}_{j}$ is not included in the calculation of $Z_{ij}$ as any constant term in the smooth function cannot be separately identified from the constant term of the linear part of the model.}
\item Calculate the Euclidean distance between $(\hat{\beta}_{kj},\hat{\beta}_{lj})$ and $(\hat{\beta}_{kj-1},\hat{\beta}_{lj-1})$. If the distance is below some tolerance level, stop and treat $\hat{\beta}_{j}$ as the model's parameter estimates. If not then update the iteration number $j\leftarrow j+1$ and repeat from step \ref{algo:repeat}.
\end{enumerate}
For more general production functions, one instead should use $Z_{ij} = \mathbb{E}_{it}[y_{it+1}|\Omega_{it}] - \int f(k_{it+1},l_{it+1}; \hat \beta)dF_{it}(l_{it+1})$ in step 2 and  $y_{it}=f(k_{it},l_{it}; \beta)+\Psi\left(Z_{ij}\right)+\epsilon_{it}$ in step 3. General treatments for the convergence of related procedures are examined, for example, in \citet{pastorelloetal2003} and \citet{dominitzsherman2005} and we explore theoretical conditions for convergence in Appendix \ref{adix:convergence}.  We henceforth refer to this iterative algorithm as ``NPR'' and provide further implementation details in Appendix \ref{adix:implementation}. 

\subsection{Comparison with Other Methods} \label{subsect:methodcomp}

Given the range of existing production function estimation methods, we emphasize three aspects that distinguish our NPR estimation algorithm.

First, unlike the widely-used proxy variable approaches of OP, LP and ACF, NPR does not require that firm input decisions be invertible.\footnote{In standard models, assumptions over the information set and optimality of input choices generate the key econometric assumptions that the input demand equation is strictly monotonic in productivity and thus invertible when there is one scalar persistent unobservable.  That said, even if the decision rules are not the solution to an optimization problem, as long as invertibility holds one can generate a proxy for $\omega$.  This invertibility fails if not all determinants of the firm's decision are observed by the econometrician. We see it as an advantage of our approach that we impose assumptions on a primitive object rather than an equilibrium outcome.} To see this note that the $\Psi$ function in equation (\ref{eq:moment1}) plays a role analogous to the proxy function used by OP and obtained from firms' investment policy, or by LP and ACF where it is obtained from firms' material input choice. In OP, LP and ACF, this proxy function $\Phi$ (using ACF's notation) is a function of current inputs used to control for current $\omega$. The success of this approach therefore hinges on the existence of a monotonic relationship between contemporaneous productivity and inputs, which is typically assumed with recourse to models of optimal firm behavior that imply such relationships are monotonic (though it may still hold even if decision rules are not the solution to an optimization problem). NPR, by contrast, requires the assumption that firms' expectations align with the true production technology but allows one to remain agnostic about how firms make their input decisions.\footnote{While the NPR algorithm requires firms' expectations align with the true production technology, the moment condition of equation (\ref{eq:moment1}) may still yield a consistent estimator for $\beta$ in contexts where this does not hold owing to bias in firms' expectations. This is discussed in subsection \ref{subsect:method_bias}.} As confirmed by the Monte Carlo simulations discussed in Section \ref{sect:montecarlo}, this distinction means NPR remains consistent when firms' decisions are subject to optimization error or when there are additional unobservable variables influencing input decision, whereas other proxy variable methods do not.\footnote{This feature also favors NPR over `index number' methods discussed by \citet{van_biesebroeck_robustness_2007}, such as those proposed by \citet{solow_technical_1957} and \citet{hall_relation_1988}, which derive equations expressing production function parameters as functions of observables under the assumption of optimal firm behavior.} Some papers have extended proxy methods to include additional variables to the law of motion for $\omega$, such as \citet{de_loecker_markups_2012} and \citet{eslava_2024}.

A second point of distinction is that NPR can accommodate non-linear productivity dynamics, whereas the dynamic panel methods of \citet{blundell_gmm_2000} typically require linearity (see \cite{bondetal2026}). Such non-linearity is enabled by the flexible form of the $\Psi$ function at the core of the NPR method, although it is worth noting the monotonicity constraint required by NPR demands that $\omega$ follow a first order Markov process. In theory, a relative strength of dynamic panel methods is that they can be used in situations where $\omega$ follows a Markov process of higher order, but in practice this requires the researcher to correctly specify both the AR and MA components of the linear productivity process and requires a longer, and hence more selected, data panel. In practice, the use of longer lags as instruments typically generates estimation problems, especially on the capital coefficient as the series is usually highly persistent.

The final point is that NPR can identify the production function parameters $\beta$ and the transition law ($g$) from a single cross section of data. While repeated observations of current-period and next-period inputs and outputs would accommodate more general models than that presented in subsection \ref{subsect:identification}, such as a production function with firm fixed effects,\footnote{See, for instance, \citet{attanasio_modelling_2022} for an elaboration on this point in the context of earnings dynamics where expectations data helps resolve important issues in dynamic panel data models, such as ``Nickell bias''.} for this baseline -- which is standard in the literature -- a single observation per firm is adequate. Both the proxy variable and dynamic panel approaches, by contrast, require multiple observations per firm.  These are used, for instance, in the second estimation stage in OP (as in LP or ACF) where, in addition, a selection correction to handle firm exit is incorporated.  Interestingly, as noted in \citet{ackerberg_chapter_2007} though, ``the first stage of the OP procedure is not affected by selection. The reason is that by construction \dots the residual in the first stage equation \dots represents unobservables that are not observed
(or predictable) by the firm before input and exit decisions. Thus there is no selection
problem \dots Intuitively, the fact that in the first stage we are able to
completely proxy $\omega_{jt}$ means that we can control for both endogenous input choice and
endogenous exit'' (p.4217).  In our case, just as in the first stage of OP, we completely proxy $\omega_{jt}$ and simultaneously control for both endogenous input choice and endogenous exit.  The use of expectations data allows us to circumvent the selection problems that appear in OP's second stage estimation insofar as the expectations correspond to distributions over $\omega_{jt}$ that are not conditional on survival.  Methods do exist for correcting for the selection problem in those other approaches, but we regard the absence of any such requirement for NPR as highly attractive.\footnote {\citet{blundell_initial_1998} discuss how selection may be controlled for by a firm fixed effects and   \citet{olley_dynamics_1996} focus on a proxy variable approach. But the absence of an external instrument in the selection equation may pose identification issues for these approaches.}

A disadvantage of the NPR approach is that it requires data on firms' subjective expectations. Since the vast majority of production functions are estimated in logs, we require information of firms' subjective expectation \textit{distributions} because a single value of firms' expected output, for example, would be insufficient to recover firms' expected \textit{log} output. However, questions that provide such information are increasingly being included in firm surveys such as the UK Management and Expectations Survey (MES) \citep{office_for_national_statistics_mesmanagement_2022} that we use in our empirical application below. We offer further considerations on the emerging data landscape in section \ref{sect:data}. As such data become increasingly available, we believe the three features of NPR discussed above bring notable advantages that warrant its addition to the established suite of production function estimators.

\subsection{Accommodating Measurement Error in Expectations Data}\label{subsect:measurement_error}

The expectations data that enable the proposed NPR estimator are, as with all data, susceptible to measurement error. Similarly to the case of measurement error in materials discussed in \citet{ackerberg_identification_2015}, measurement error in firms' subjective expectations will violate the invertibility condition required by NPR and thereby undermine the estimator's consistency. While this is true of the baseline NPR estimator outlined above, we can draw on the insights of \citet{evdokimov_zeleneev_2025} (EZ) to develop an alternative estimator, ``EZ-NPR'', which we show below is robust to moderate levels of measurement error in expected output and employment. 

The key insight of EZ is that non-linear moments can be ``corrected'' for measurement error by taking Taylor expansions of the moments with respect to the potentially mis-measured variables around the point of zero measurement error. To demonstrate the application of this in our context, consider the moment conditions that follow from equation \eqref{eq:moment1} in the case of Cobb-Douglas production (note that the general EZ approach can be extended to other functional forms).  Consider
\begin{equation}\label{eq:generalmoment}
    M^x_{it}(Z_{it})=\left(y_{it}-\beta_k k_{it}-\beta_l l_{it} -\Psi\left(Z_{it}\right)\right)x_{it},
\end{equation}
where $x_{it}$ is any variable in firm-$i$'s information set $\Omega_{it}$ and
\begin{equation*}
    Z_{it} = \mathbb{E}_{it}[y_{it+1}|\Omega_{it}] -\beta_k k_{it+1}-\beta_l \mathbb{E}_{it}[l_{it+1}|\Omega_{it}].
\end{equation*}
Under the assumptions outlined in Section \ref{subsect:identification}, such quantities are zero in expectation and can be used to recover consistent estimates of $\beta$. If expectations data are subject to measurement error, however, then 
\begin{equation*}
    \begin{split}
        Z_{it} &= \mathbb{E}^\ast_{it}[y_{it+1}|\Omega_{it}] +\upsilon_{yit} -\beta_k k_{it+1}-\beta_l(\mathbb{E}^\ast_{it}[l_{it+1}|\Omega_{it}]+\upsilon_{lit}) \\
        &= Z^\ast_{it}+\upsilon_{yit}-\beta_l\upsilon_{lit},
    \end{split}
\end{equation*}
where asterisks denote true values and $\upsilon_y$ and $\upsilon_l$ represent measurement error in expected output and labor, respectively. In general, this means that equation \eqref{eq:generalmoment} will not equal zero in expectation when evaluated at the true parameters using observed data, and we lose identification. 

To progress, we follow EZ and take a second-order Taylor expansion of $ M^x_{it}(Z_{it})=M^x_{it}(Z^\ast_{it}+\upsilon_{yit}-\beta_l\upsilon_{lit})$ around the point $(\upsilon_{yit},\upsilon_{lit})=0$ to obtain
\begin{equation} \label{eq:measus_error}
    \begin{split}
         \mathbb{E}&[M^x_{it}(Z_{it})]\approx \mathbb{E}[M^x_{it}(Z^\ast_{it})]+\mathbb{E}\left[\frac{\partial M^x_{it}(Z^\ast_{it})}{\partial \mathbb{E}_{it}[y_{it+1}]}\right]\mathbb{E}[\upsilon_y]+\mathbb{E}\left[\frac{\partial^2 M^x_{it}(Z^\ast_{it})}{\partial \mathbb{E}_{it}[y_{it+1}]^2}\right]\frac{\mathbb{E}[\upsilon_y^2]}{2}\\
         & +\mathbb{E}\left[\frac{\partial M^x_{it}(Z^\ast_{it})}{\partial \mathbb{E}_{it}[l_{it+1}]}\right]\mathbb{E}[\upsilon_l]+\mathbb{E}\left[\frac{\partial^2 M^x_{it}(Z^\ast_{it})}{\partial \mathbb{E}_{it}[l_{it+1}]^2}\right]\frac{\mathbb{E}[\upsilon_l^2]}{2} +\mathbb{E}\left[\frac{\partial^2 M^x_{it}(Z^\ast_{it})}{\partial \mathbb{E}_{it}[y_{it+1}] \partial \mathbb{E}_{it}[l_{it+1}]}\right]\mathbb{E}[\upsilon_y\upsilon_l] \\
         &=\mathbb{E}\left[M^x_{it}(Z^\ast_{it})+\frac{\partial^2 M^x_{it}(Z^\ast_{it})}{\partial \mathbb{E}_{it}[y_{it+1}]^2}\frac{\mathbb{E}[\upsilon_y^2]}{2}+\frac{\partial^2 M^x_{it}(Z^\ast_{it})}{\partial \mathbb{E}_{it}[l_{it+1}]^2}\frac{\mathbb{E}[\upsilon_l^2]}{2}\right],
     \end{split}
\end{equation}
where we assume independence between measurement errors and true variables and the second equality assumes that both measurement errors are mean-zero and uncorrelated with one another. (\citet{evdokimov_zeleneev_2025} also consider more general settings with non-classical and correlated measurement error.)  EZ show that, up to higher-order terms, the average derivatives evaluated at observed values are consistent estimators of the expected derivatives evaluated at true values, which appear in the last line of equation (\ref{eq:measus_error}). Corrected moments can thus be constructed using
\begin{equation}\label{eq:correctedmoment}
    M^x_{it}(Z_{it})-\frac{\partial^2 M^x_{it}(Z_{it})}{\partial \mathbb{E}_{it}[y_{it+1}]^2}\frac{\mathbb{E}[\upsilon_y^2]}{2}-\frac{\partial^2 M^x_{it}(Z_{it})}{\partial \mathbb{E}_{it}[l_{it+1}]^2}\frac{\mathbb{E}[\upsilon_l^2]}{2}.
\end{equation}
Treating the variance terms $\mathbb{E}[\upsilon_y^2]$ and $\mathbb{E}[\upsilon_l^2]$ as additional parameters to estimate, one can therefore use corrected moments using \eqref{eq:correctedmoment} to recover estimates that are robust to moderate levels of measurement error in firms' expectations.

To understand the meaning of ``moderate'' in the preceding sentence, observe that the expectations data enter the moment in equation \eqref{eq:generalmoment} via the uni-dimensional $Z_{it}$ variable, which represents $\mathbb{E}_{it}[\omega_{it+1}|\Omega_{it}]$. The extent to which measurement errors in $\mathbb{E}_{it}[y_{it+1}|\Omega_{it}]$ and $\mathbb{E}_{it}[l_{it+1}|\Omega_{it}]$ both impede the performance of the NPR estimator and can be withstood by EZ-NPR therefore depends on how their combined variance (equal to $\mathbb{E}[\upsilon_y^2]+\beta_l^2\mathbb{E}[\upsilon_l^2]$) compares to the variance of the true $\mathbb{E}_{it}[\omega_{it+1}|\Omega_{it}]$. If persistent productivity follows an AR(1) process, for example, NPR and EZ-NPR can tolerate measurement error in firms' expectations of increasingly higher variance as the persistence of the process increases, since this implies greater variance in $\mathbb{E}_{it}[\omega_{it+1}|\Omega_{it}]$. We explore this quantitatively in Monte Carlo simulations discussed in Section \ref{sect:montecarlo}.

EZ-NPR may appear simpler to implement than the iterative NPR algorithm as consistent estimates can be obtained via Generalized Method of Moments (GMM) estimation using corrected moments of the form \eqref{eq:correctedmoment}. In practice, however, the non-linear high-dimensional objective function is hard to optimize numerically. We therefore use a nested optimization procedure that estimates $\Psi$ and the variance parameters in an ``inner'' optimization stage, while an ``outer''  optimization focuses on the two-dimensional production function parameter vector of interest. We discuss this nested approach alongside further implementation details in Appendix \ref{adix:EZNPR}. 

\subsection{Accommodating Biased Expectations and Imperfect Knowledge of the Production Technology}\label{subsect:method_bias}

In our baseline case, when firms know the true production technology (as in the other proxy methods) and their beliefs align with this, the law of motion for $\omega_{it}$ can be inverted to obtain equation (\ref{eq:gomega_inverse}). This inversion is crucial to the NPR estimation algorithm and is analogous to the invertibility condition that OP impose on the investment policy function and that LP and ACF impose on the material input policy function. We thus rely on a one-to-one mapping between firms' expectations and their current productivity to generate a proxy for the latter. 

Given the centrality of expectations to the NPR algorithm, it is important to consider whether and how expectational biases undermine the proposed approach. The first thing to note is that our suggested method can accommodate biased input expectations as long as such bias is also reflected in firms' expected output and vice versa. If, for example, a firm is systematically optimistic in its sales forecasts we would require it to be similarly optimistic in its employment forecasts. The precise meaning of `similarity' in this context is governed by the production function. Specifically, in the case of a firm with over-optimistic output expectations, we require bias in the firm's employment expectations such that the integral of the production function with respect to expected labor equals the biased output expectation.  

As an example, consider the case of Cobb-Douglas production in equation (\ref{eq:prodfunct_cd}).  When the bias in sales expectations, say $\texttt{bias}_{y,it}$, balances the bias in the employment expectations, say $\texttt{bias}_{l,it}$, so that $\texttt{bias}_{y,it}=\beta_l \texttt{bias}_{l,it}$, then one is still able to recover an unbiased expectation of firms' next-period productivity as the residual on the right hand side of equation (\ref{eq:gomega_inverse}). 
Furthermore, in this particular case, when production is Cobb-Douglas, it is possible to construct a `Wald'-type estimator of $\beta_l$ if one has the data to compare expectations of output and labor to their realized values. We do not pursue this in our main analysis because we present evidence that such bias is minimal in our empirical context (see the discussion of Table \ref{tab:mes_expectation_errors} in  subsection \ref{subsect:data_expectations} and also \cite{bloom_well_2021}). Nevertheless, we provide a further exposition in Appendix \ref{adix:bias}.

The ability of our method to accommodate these forms of input (and output) biases is encouraging. Biases in firms' productivity expectations, however, are more problematic. To see why, suppose firms' bias about their next-period productivity is captured by $\iota_{it}$ (positive values reflect optimism, negative values reflect pessimism), such that
\begin{equation*}
\mathbb{E}_{it}[\omega_{it+1}|\Omega_{it}]=g(\omega_{it})+\iota_{it}.
\end{equation*}
Even if firms' expectations about output and inputs align with the true production technology, the presence of bias means  
\begin{equation}\label{eq:eprodfunct_bias}
\mathbb{E}_{it}[y_{it+1}|\Omega_{it}] =\int f(k_{it+1},l_{it+1}; \theta)dF_{it}(l_{it+1})+g(\omega_{it})+\iota_{it}.
\end{equation}
The bias term $\iota_{it}$ precludes us from recovering $\omega$ since
\begin{equation*}
\begin{split}
g(\omega_{it})+\iota_{it} = \mathbb{E}_{it}[y_{it+1}|\Omega_{it}] - \int f(k_{it+1},l_{it+1}; \theta)dF_{it}(l_{it+1}) \\
\therefore \\
g^{-1}\left(\mathbb{E}_{it}[y_{it+1}|\Omega_{it}] - \int f(k_{it+1},l_{it+1}; \theta)dF_{it}(l_{it+1})\right) = g^{-1}\left(g(\omega_{it})+\iota_{it}\right) \neq \omega_{it}.
\end{split}
\end{equation*}  
Whether such bias is surmountable depends on its form. The simplest case is when the bias term $\iota_{it}$ is uncorrelated with all observables used by the NPR estimator (i.e.,$\{y_{it}, l_{it},$ $ k_{it}, \mathbb{E}_{it}[y_{it+1}], \mathbb{E}_{it}[l_{it+1}], k_{it+1}\}$). In this case, the bias term operates exactly as the measurement error discussed in Section \ref{subsect:measurement_error} and consistent estimates can be recovered using the EZ-NPR estimator.

In addition to this type of ``classical'' bias, we discuss in Appendix \ref{adix:bias} how the NPR estimator can be extended to withstand bias that is either time-invariant or a function of observables. In these cases, one can embed the NPR estimation algorithm in an outer iterative estimation loop to recover estimates of firms' expected productivity bias, which can subsequently be used to restore consistency of the NPR estimator. These methods require panel data on expectations which are very limited in the data of our empirical application -- we have a reasonable panel for outputs and inputs, but the MES subjective expectations data are two cross sections with limited longitudinal overlap. Hence, we do not pursue these extensions in the empirics.

Another factor that undermines consistency of the NPR estimator is if the firm has imperfect knowledge of the production technology. In this case, deviations between the production function parameters perceived by firms and the true values create a `wedge' between expected output, the true production function evaluated at $\left(k_{it+1},\mathbb{E}_{it}[l_{it+1}]\right)$ and $g(\omega_{it})$, similar to the bias term $\iota_{it}$ of equation (\ref{eq:eprodfunct_bias}). In Appendix \ref{adix:bias} we demonstrate an extension to the baseline NPR estimator in the presence of such imperfect knowledge when the production function is linear in parameters. However, similarly to the bias-robust extensions this requires a panel of firms' expectations. 

\section{Monte Carlo Simulations}\label{sect:montecarlo}
\subsection{Baseline Monte Carlo Results}
Our baseline Monte Carlo setup follows that of ACF (see more details in Appendix \ref{adix:mc_setup}). The production function specification is Leontief in the material input:
\begin{equation*}
Y_{it} = \min \{\beta_0 K^{\beta_k}_{it} L^{\beta_l}_{it} e^{\omega_{it}},\beta_m M_{it} \} e^{\epsilon_{it}},
\end{equation*}
where $\beta_0 = 1, \beta_k = 0.4, \beta_l = 0.6$ and $\beta_m = 1$. In our baseline analysis the productivity shock is assumed to follow an AR(1) process:
\begin{equation}
\omega_{it}= \rho \omega_{it-1} + \xi_{it},
\end{equation}
with $\rho = 0.7$. 
As pointed out by ACF, the LP estimator does not identify the production function parameters unless there is stochastic variation in firms' labor inputs, for example due to optimization error. We therefore focus on data generating processes featuring such variation, which we introduce in the same way as ACF by adding a mean-zero normally-distributed random variable to firms' optimal level of labor. In addition, we also consider the impact of optimization error in investment and materials (which ACF do not consider),\footnote{ACF's analysis considers the impact of \emph{measurement} error in materials but this is distinct from optimization error as it does not affect output. Optimization error, by contrast, will affect output via the assumption of Leontief technology.} which we incorporate in the same manner as labor.\footnote{Optimization errors in labor, investment and materials are simulated from a mean-zero normal distribution with standard deviation 0.37, which matches the distributional assumption ACF make regarding the labor optimization error in their DGPs.} 

For each DGP, we use the closed-form solutions of the model to simulate data for 1,000 firms over 100 periods. Capital is initialized at zero and we only use data from the last 10 periods for estimation purposes, as by this time the capital stock  appears to have reached steady state. Further details of the environment and the data generating process (DGP) -- including the simulation of firms' expectations -- are given in Appendix \ref{adix:mc_setup} and the Appendix in ACF. Following the procedure detailed in ACF's Appendix A.4, we concentrate out parameters in the ACF estimator's second stage such that it involves minimization of a nonlinear objective function over the two-dimensional parameter vector $(\beta_l,\beta_k)$. We use a BFGS algorithm implemented via the R \texttt{optimx} function to find the minimum of this objective, which requires the researcher to specify initial values of the parameters. Similarly to the NPR algorithm, we implement the ACF estimator using 16 initialization points (the interaction of two equally-spaced 4-point grids on a 0.1 to 0.9 range -- one for the initial labor coefficient value and one for the initial capital coefficient value), and retain the estimates that return the lowest value of the objective function subject to the constraint that both coefficients are strictly positive.\footnote{ \label{ACFfn16} We found that imposing the non-negativity condition when searching over initialization points eliminated implausible estimates that the ACF estimator returned due to a global identification issue in the moments employed by ACF: ``there is a `global' identification issue in that the moments have expectation zero not only at the true parameters, but also at one other point on the boundary of the parameter space where $\beta_k=0$ and $\beta_l=\beta_l+\beta_k$, and the estimated AR(1) coefficient on $\omega$ equals the AR(1) coefficient on the wage process'' (see footnote 16 in their paper).  This issue is also discussed in \citet{Rovigatti} and \citet{kimluosu2019}, for example.} Implementing a similar concentrating out approach in the second stage of both OP and LP estimators, by contrast, yields a uni-dimensional nonlinear optimization problem. The method we use to find the minimum is a combination of golden section search and successive parabolic interpolation as implemented in the R function \texttt{optimize} which, unlike \texttt{optimx}, is a uni-dimensional optimizer and instead requires the researcher to specify a search interval, which we set to [0,2]. Experimentation indicates that estimators are generally robust to varying these optimization settings. 

Table \ref{tab:mcrun_summtab} examines the performance of the estimators as various firm choices are subject to optimization error. Panel A has optimization error only in employment, $l$; Panel B adds in optimization error in investment ($l$,$i$); Panel C in materials ($l$,$m$) and Panel D in all three inputs ($l$,$i$,$m$). 

We highlight three salient points in Table \ref{tab:mcrun_summtab}. First, as expected, the OLS and OP parameter estimates are heavily biased across all DGPs. Bias in the OP estimates is due to the presence of firm-specific capital adjustment costs, added by ACF to obtain across-firm variation in capital similar to that observed in the data. 

Second, when optimization error affects labor only (Panel A), NPR, ACF and LP all perform well. As anticipated by ACF, LP (slightly) outperforms their proposed estimator in this environment in terms of precision. NPR improves on LP even more, achieving much greater precision on the capital coefficient. Given the iterative non-parametric structure of NPR, one may worry that this greater precision comes at the cost of additional computation requirements. While the precise time required for the NPR estimator varies somewhat across DGPs, the longest runtime is approximately 26 seconds, compared to approximately four seconds for ACF, and in some DGPs matches the ACF runtime.\footnote{Computations were performed on a desktop PC with an Intel Core i7-11700 processor (2.50 GHz, 8 cores, 16 threads) and 32 GB of RAM, running Windows 10 (64-bit).} Therefore, while NPR does indeed incur an additional computational burden, it is not prohibitive.

Third, when optimization error affects material inputs (Panels C and D), all estimators except NPR deteriorate. While ACF appears robust to errors in investment (Panel B), it is compromised by errors in materials which violate the monotonicity condition ACF require between material input choices and productivity. ACF returns very low labor coefficients and very high capital coefficients (e.g., 0.13 and 0.91 respectively in Panel D). The LP estimator also deteriorates, especially on the capital coefficient which is 0.49 on average, almost a quarter higher than its true value of 0.4. It is also the case that in alternative plausible DGPs discussed by ACF and replicated for our estimators in Table \ref{tab:mcrun_summtab_acftab1}, LP does substantially worse (see the discussion in Appendix \ref{subsec:Replication_ACF}).

In contrast to the other estimators, the NPR estimates remain accurate throughout, although optimization error in materials reduces the precision of the NPR estimates, particularly for the capital coefficient.\footnote{\label{fn:optm} Despite not relying on material input data, NPR is affected by materials optimization error because of the assumption of Leontief production. When material optimization error is negative, the Leontief assumption means output will be determined by sub-optimally low materials and hence the specification of output as a function of labor and capital will be incorrect.} This is a clear demonstration of the observation made in subsection \ref{subsect:methodcomp} that NPR is robust to optimization error in inputs, whereas the other proxy variable estimators are not.
    
\begin{table}[H]
   \caption{Input Optimization Error Monte Carlo Results}
   \label{tab:mcrun_summtab}
    \centering
    \begin{adjustbox}{max width=\textwidth}
        \begin{tabular}{l|cccc|cccc}
\hline\hline
&\multicolumn{4}{c|}{$\beta_{l}=0.6$} & \multicolumn{4}{c}{$\beta_{k}=0.4$} \\
& Mean & Median & S.D. & MSE & Mean & Median & S.D. & MSE \\
\hline
&\multicolumn{8}{c}{\textbf{A. Optimization error in} $ l $} \\
\hline
NPR &0.600 &0.600 &0.003 &0.000 &0.400 &0.400 &0.005 &0.000 \\
OLS &0.920 &0.920 &0.002 &0.103 &0.092 &0.093 &0.004 &0.095 \\
OP &0.842 &0.843 &0.004 &0.059 &0.002 &0.000 &0.025 &0.159 \\
LP &0.600 &0.600 &0.003 &0.000 &0.400 &0.400 &0.013 &0.000 \\
ACF &0.600 &0.599 &0.009 &0.000 &0.400 &0.400 &0.015 &0.000 \\
\hline
&\multicolumn{8}{c}{\textbf{B. Optimization error in} $ (l,i) $} \\
\hline
NPR &0.600 &0.600 &0.003 &0.000 &0.400 &0.400 &0.005 &0.000 \\
OLS &0.920 &0.920 &0.002 &0.103 &0.092 &0.092 &0.004 &0.095 \\
OP &0.913 &0.913 &0.002 &0.098 &0.044 &0.044 &0.018 &0.127 \\
LP &0.600 &0.600 &0.003 &0.000 &0.400 &0.400 &0.012 &0.000 \\
ACF &0.599 &0.600 &0.009 &0.000 &0.400 &0.399 &0.016 &0.000 \\
\hline
&\multicolumn{8}{c}{\textbf{C. Optimization error in} $ (l,m) $} \\
\hline
NPR &0.602 &0.600 &0.024 &0.001 &0.425 &0.401 &0.204 &0.042 \\
OLS &0.920 &0.920 &0.003 &0.103 &0.093 &0.093 &0.006 &0.094 \\
OP &0.842 &0.842 &0.005 &0.059 &0.245 &0.246 &0.016 &0.024 \\
LP &0.555 &0.555 &0.004 &0.002 &0.490 &0.490 &0.013 &0.008 \\
ACF &0.140 &0.141 &0.033 &0.213 &0.900 &0.901 &0.042 &0.252 \\
\hline
&\multicolumn{8}{c}{\textbf{D. Optimization error in} $ (l,i,m) $} \\
\hline
NPR &0.602 &0.600 &0.026 &0.001 &0.403 &0.401 &0.026 &0.001 \\
OLS &0.920 &0.920 &0.003 &0.103 &0.092 &0.092 &0.006 &0.095 \\
OP &0.913 &0.912 &0.004 &0.098 &0.047 &0.046 &0.030 &0.126 \\
LP &0.555 &0.555 &0.004 &0.002 &0.486 &0.486 &0.012 &0.007 \\
ACF &0.130 &0.130 &0.037 &0.222 &0.908 &0.906 &0.046 &0.261 \\
\hline
\end{tabular}
    \end{adjustbox}
    \medskip
    \caption*{{\scriptsize Note: The table contains summary statistics calculated over 500 Monte Carlo simulations. The true values of $\beta_l$ and $\beta_k$ are 0.6 and 0.4 respectively. $l$ is ln(labor), $m$ is ln(materials), $k$ is ln(capital) and $i$ is ln(investment).}}
\end{table}

The results of Table \ref{tab:mcrun_summtab} demonstrate the robustness of NPR to optimization error in contemporaneous firm decisions. Since NPR does not rely on any assumptions related to such decisions, unlike the other proxy variable estimators, this could be regarded as somewhat unsurprising. We now turn to tests of the NPR estimator relating to measurement error in expectations and to bias in expected productivity which, as discussed in Sections \ref{subsect:measurement_error} and \ref{subsect:method_bias} could be problematic for NPR.

\subsection{Monte Carlo Extensions to Measurement Error and Bias in Expectations}\label{sect:mc_extensions}

Table \ref{tab:mcrun_summtab_Emeaserror} shows the results from incorporating measurement error to the expectations measures as discussed in subsection \ref{subsect:measurement_error}. Each panel presents the NPR estimator in the first row and the EZ-NPR estimator, which can accommodate moderate levels of measurement error, in the second row. Panel A is our ``baseline'' and reproduces the results from Panel A in Table \ref{tab:mcrun_summtab}. EZ-NPR produces similar results to NPR, albeit with higher standard deviations due to correcting for measurement error when none is present.

The subsequent Panels in Table \ref{tab:mcrun_summtab_Emeaserror} add increasing amounts of measurement error in the two expectations variables, expressed as a percentage of the variance of the ``true'' $\mathbb{E}_{it}[\omega_{it+1}]$. This ranges from 2.5\% in Panel B to 15\% in Panel E. As one might expect, the NPR estimates deteriorate as measurement error increases. With 10\% measurement error (Panel D), for example, NPR returns a mean output elasticity of capital of 0.29 and one for labor of 0.73. In contrast EZ-NPR does much better with values of 0.35 and 0.64, respectively. Of course, with very large degrees of measurement error, even EZ-NPR eventually performs poorly although consistently better than standard NPR.

\begin{table}[H]
   \caption{Expectation Measurement Error Monte Carlo Results}
   \label{tab:mcrun_summtab_Emeaserror}
    \centering
    \begin{adjustbox}{max width=\textwidth}
        \begin{tabular}{l|cccc|cccc}
\hline\hline
&\multicolumn{4}{c|}{$\beta_{l}=0.6$} & \multicolumn{4}{c}{$\beta_{k}=0.4$} \\
& Mean & Median & S.D. & MSE & Mean & Median & S.D. & MSE \\
\hline
&\multicolumn{8}{c}{\textbf{A. Baseline - no measurement error}} \\
\hline
NPR &0.600 &0.600 &0.003 &0.000 &0.400 &0.400 &0.005 &0.000 \\
EZ-NPR &0.598 &0.599 &0.015 &0.000 &0.402 &0.401 &0.017 &0.000 \\
\hline
&\multicolumn{8}{c}{\textbf{B. Measurement error in} $ \mathbb{E}_{it}[\omega_{it+1}] $} \\
&\multicolumn{8}{c}{\textbf{variance = 2.5\% truth}} \\
\hline
NPR &0.625 &0.625 &0.003 &0.001 &0.377 &0.377 &0.008 &0.001 \\
EZ-NPR &0.610 &0.612 &0.023 &0.001 &0.388 &0.386 &0.024 &0.001 \\
\hline
&\multicolumn{8}{c}{\textbf{C. Measurement error in} $ \mathbb{E}_{it}[\omega_{it+1}] $} \\
&\multicolumn{8}{c}{\textbf{variance = 5\% truth}} \\
\hline
NPR &0.652 &0.652 &0.004 &0.003 &0.352 &0.352 &0.008 &0.002 \\
EZ-NPR &0.613 &0.622 &0.038 &0.002 &0.383 &0.374 &0.040 &0.002 \\
\hline
&\multicolumn{8}{c}{\textbf{D. Measurement error in} $ \mathbb{E}_{it}[\omega_{it+1}] $} \\
&\multicolumn{8}{c}{\textbf{variance = 10\% truth}} \\
\hline
NPR &0.731 &0.732 &0.010 &0.017 &0.290 &0.292 &0.015 &0.012 \\
EZ-NPR &0.643 &0.643 &0.006 &0.002 &0.345 &0.345 &0.010 &0.003 \\
\hline
&\multicolumn{8}{c}{\textbf{E. Measurement error in} $ \mathbb{E}_{it}[\omega_{it+1}] $} \\
&\multicolumn{8}{c}{\textbf{variance = 15\% truth}} \\
\hline
NPR &0.830 &0.842 &0.028 &0.054 &0.096 &0.118 &0.048 &0.094 \\
EZ-NPR &0.657 &0.657 &0.007 &0.003 &0.319 &0.319 &0.014 &0.007 \\
\hline
\end{tabular}
    \end{adjustbox}
    \medskip
    \caption*{{\scriptsize Note: The table contains summary statistics calculated over 500 Monte Carlo simulations. The true values of $\beta_l$ and $\beta_k$ are 0.6 and 0.4 respectively. $l$ is ln(labor), $y$ is ln(output) and $\omega$ is persistent productivity. All DGPs feature optimization error in labor.}}
\end{table}

Turning now to expectational biases, Table \ref{tab:mcrun_summtab_Eerror} shows moments of parameter estimates obtained by applying the NPR estimator to data simulated under the ``optimization error in $l$'' scenario (Panel A of Table \ref{tab:mcrun_summtab}) but with the addition of idiosyncratic shocks to firms' expectations. As explained in Section \ref{subsect:method_bias}, NPR is robust to biased expectations over labor because bias in expected inputs leads to bias in expected output according to the production technology, which means that a one-to-one mapping between expected outputs, inputs and productivity is preserved. The results in the first row confirm this. Panel B shows that NPR loses accuracy when there is bias in expected productivity (due to bias in expected output). In this DGP, the relationship between expected output and inputs is subject to two unobservables -- expected productivity and the bias shock -- and hence it is no longer possible to control for expected productivity using expectations data. Panel C considers a similar DGP to Panel B, although in this case the bias to firms' expected productivity is a function of a time-variant observable characteristic, $X_{it} \equiv \textrm{mgmt}_{it}$.\footnote{We use the label ${mgmt}_{it}$ because \citet{bloom_well_2021} find evidence that forecast biases are related to managerial quality. Management is simulated as a standard normal random variable drawn for each firm-period (i.e.,${mgmt}_{it}\sim N(0,1)$) and firms' expected productivity bias is simulated as -0.15 times management (i.e.,$\iota_{it}=-0.15{mgmt}_{it}$).} Although this type of expectation bias again compromises the performance of the basic NPR estimator, application of a modified version of the NPR estimator (``NPR (bias-robust extension)'') described in detail in Appendix \ref{adix:bias} under ``Case 3'' recovers consistency.

In summary, NPR is robust to bias in expected inputs and, while bias in expected productivity may undermine performance of the basic estimator, extensions to NPR can accommodate these biases under certain assumptions (see Appendix \ref{adix:bias} for details). We do not consider the bias-robust versions of NPR further since in our MES data, output expectations appear largely unbiased and we do not have many firms with multiple longitudinal data on firms' subjective expectations.

\begin{table}[H]
  \caption{Expectation Bias Monte Carlo Results}
\label{tab:mcrun_summtab_Eerror}
    \centering
    \begin{adjustbox}{max width=\textwidth}
        \begin{tabular}{l|cccc|cccc}
\hline\hline
&\multicolumn{4}{c|}{$\beta_{l}=0.6$} & \multicolumn{4}{c}{$\beta_{k}=0.4$} \\
& Mean & Median & S.D. & MSE & Mean & Median & S.D. & MSE \\
\hline
&\multicolumn{8}{c}{\textbf{A. Bias shocks to} $ \mathbb{E}_{it}[l_{it+1}] $} \\
\hline
NPR &0.600 &0.600 &0.003 &0.000 &0.401 &0.401 &0.008 &0.000 \\
\hline
&\multicolumn{8}{c}{\textbf{B. Bias shocks to} $ \mathbb{E}_{it}[\omega_{it+1}] $} \\
\hline
NPR &0.674 &0.674 &0.003 &0.005 &0.333 &0.333 &0.006 &0.004 \\
\hline
&\multicolumn{8}{c}{\textbf{C. Bias in} $ \mathbb{E}_{it}[\omega_{it+1}] $ f(mgmt)} \\
\hline
NPR &0.710 &0.685 &0.066 &0.017 &0.469 &0.302 &0.418 &0.179 \\
NPR (bias-robust extension) &0.607 &0.601 &0.037 &0.001 &0.431 &0.393 &0.243 &0.060 \\
\hline
\end{tabular}
    \end{adjustbox}
    \medskip
    \caption*{{\scriptsize Note: The table contains summary statistics calculated over 500 Monte Carlo simulations. The true values of $\beta_l$ and $\beta_k$ are 0.6 and 0.4 respectively. $l$ is ln(labor), $y$ is ln(output) and $\omega$ is persistent productivity. All DGPs feature optimization error in labor.}}
\end{table}

While the OP, LP and ACF control function estimators do not use subjective expectations data directly, they will be affected by expectational biases if such biases lead to changes in firm decisions. To illustrate this point, we focus on biased productivity expectations and run several simulations that change the variance of the bias component. In contrast to our baseline DGP, we eliminate heterogeneity in investment costs and set the degree of optimization error in labor to 0.1 instead of 0.37. These changes are made to reduce the bias in the OP estimator in a setting without expectational biases and thus highlight a distinct source of bias.\footnote{Results available from the authors on request show the performance of the LP, ACF and NPR estimators is very similar when productivity bias is varied in the context of the baseline DGP.} 

\begin{figure}[H]
    \caption{Estimator MSE Under Different Levels of Expected Productivity Bias}\label{fig:Eomgbias_mse}
    \begin{subfigure}[b]{0.5\textwidth}
        \centering
        \caption{Labor Coefficient MSE}\label{fig:Eomgbias_bl_mse}
        \includegraphics[width=\textwidth]{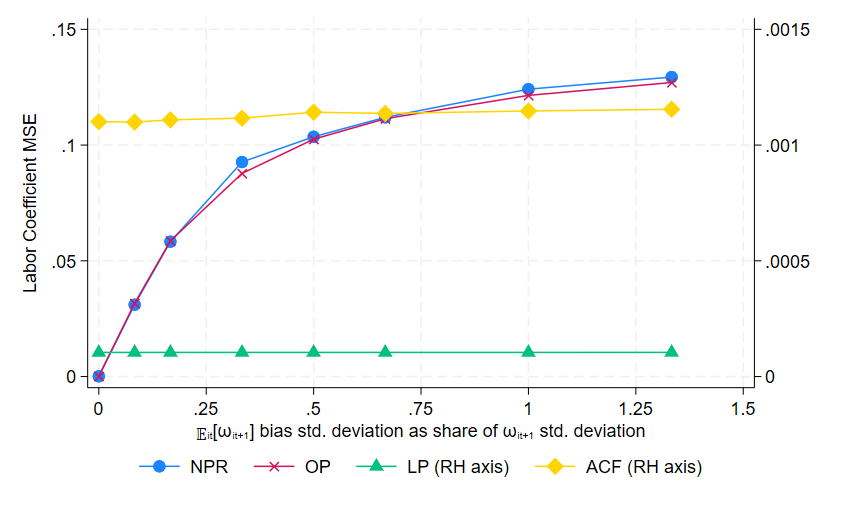}
    \end{subfigure}
    \begin{subfigure}[b]{0.5\textwidth}
        \centering 
        \caption{Capital Coefficient MSE}\label{fig:Eomgbias_bk_mse}
        \includegraphics[width=\textwidth]{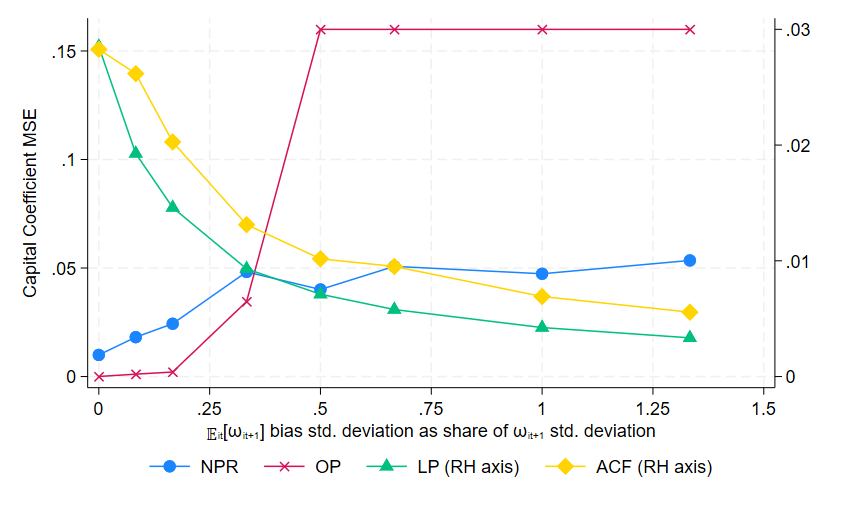}
    \end{subfigure}
 \caption*{{\scriptsize Note: In Panel (a) OP and NPR MSE are very similar, which is why they are hard to distinguish. The ACF series shows the mean squared error of ACF estimates across simulation runs after discarding runs where the estimator obtained $\hat{\beta}_l$ or $\hat{\beta}_k$ less than zero or greater than one.}}
\end{figure}

Figure \ref{fig:Eomgbias_mse} illustrates how mean squared errors of the control function estimates vary with the magnitude of bias in expected productivity. For small levels of expected productivity bias, the NPR estimator returns a small mean squared error but its performance deteriorates as the magnitude of the bias grows. This is expected following subsection \ref{subsect:method_bias}, which explains that expected productivity bias violates the scalar unobservable assumption on which the NPR estimator depends. The performance of the other control function estimators is perhaps more surprising. The OP estimator performs similarly to the NPR estimator, obtaining precise results when expected productivity is subject to low levels of bias but becoming increasingly biased as the magnitude of expected productivity bias grows. The LP and ACF estimators, by contrast, remain consistent at all levels of expected productivity bias, with the capital coefficient estimate of both estimators showing small gains in precision as the magnitude of the bias increases. 

The relative performance of the OP/LP/ACF estimators has a straightforward explanation. Firm-specific biases in expected next-period productivity will increase between-firm differences in current-period investment, leading to increased variation in next-period capital. Conditional on capital (and, in the case of ACF, labor), materials is still chosen optimally every period which, in the DGPs depicted in Figure \ref{fig:Eomgbias_mse}, generates a one-to-one mapping between materials (conditional on other inputs) and productivity. This preserves the monotonicity assumption that is crucial to both the LP and ACF estimators. Biased expectations of future productivity therefore leave the consistency of the LP/ACF estimators intact, while also increasing variation in capital, which translates into greater precision in these estimators. In contrast, the extra variation that biased productivity expectations introduce into investment violates the scalar unobservable assumption required by the OP estimator. The same logic applies when labor expectations are subject to bias, leading to similar results as those shown in Figure \ref{fig:Eomgbias_mse}. Thus, an additional benefit of the LP and ACF estimators over OP is that, by using conditional demand for a static input to control for persistent productivity, they are robust to firm-specific variation in expectation accuracy. Such firm-specific variation will likely influence dynamic decisions such as investment, violating the scalar unobservable assumption required for consistency of the OP estimator.

\section{Data}\label{sect:data}
\subsection{Sources and Sample Characteristics}\label{subsect:data_sourcessample}	

Our proposed methodology requires data on firms' log output, log inputs and one-period-ahead expectations of these quantities. While numerous firm surveys include questions related to expectations (see, for example, Table 2.1 in \cite{born_etal_handbook_2022}), the majority of these ask firms for single point-valued expectations of future levels or growth rates. On its own, this is not enough to obtain measures of firms' expectations of \emph{log} levels. To do so, one requires further information on firms' subjective expectation distributions, such as variance or, ideally, several points and associated probabilities from firms' subjective cumulative distribution functions (CDFs). 

Table \ref{tab:expectations_data_library} documents surveys that contain enough information to estimate firms' expected log quantities and thus implement our proposed estimator or that would require small modifications to do so. These include the Management and Organizational Practices Surveys (MOPS) by the US Census Bureau \citep{buffington_management_2016}, the Decision Maker Panel (DMP) by the Bank of England \citep{bloom_tracking_2017}, the Survey of Business Uncertainty (SBU) by the Atlanta Federal Reserve Bank and the China Employer-Employee Survey (CEES) \citep{altigetal2022}. 

Several points regarding Table \ref{tab:expectations_data_library} are worth highlighting. First, it is possible to implement our proposed estimator in three of the six largest economies, including the US and China. Small modifications to existing surveys would additionally include Japan, Germany and Italy. The estimator outlined above thus has broader relevance than one might initially presume. Second, the most common method of eliciting firms' expectations is to ask for expected outcomes in a number of quantitative `scenarios' (e.g.,`low', `medium' and `high'), along with perceived likelihoods of these scenarios occurring. Section \ref{subsect:data_expectations} discusses how to recover measures of firms' subjective expectation distributions from such information, which researchers using these datasets may find useful beyond the subject of production function estimation.

\begin{sidewaystable}[p]
\centering
\caption{Surveys Including Questions on Firms' Subjective Expectation Distributions}\label{tab:expectations_data_library}
\scriptsize
\setlength{\tabcolsep}{2pt}
\renewcommand{\arraystretch}{1.15}
\begin{adjustbox}{max totalsize={\textheight}{\textwidth},center}
\begin{tabular}{@{}
l
>{\raggedright\arraybackslash}p{0.17\linewidth}
>{\raggedright\arraybackslash}p{0.135\linewidth}
>{\raggedright\arraybackslash}p{0.085\linewidth}
>{\raggedright\arraybackslash}p{0.075\linewidth}
>{\raggedright\arraybackslash}p{0.075\linewidth}
>{\raggedright\arraybackslash}p{0.075\linewidth}
>{\raggedright\arraybackslash}p{0.09\linewidth}
>{\raggedright\arraybackslash}p{0.09\linewidth}
>{\raggedright\arraybackslash}p{0.11\linewidth}
>{\raggedright\arraybackslash}p{0.14\linewidth}
>{\raggedright\arraybackslash}p{0.13\linewidth}
@{}}
\toprule
Country & Survey & Documentation & Sectors & Freq. & Years avail. & Years NPR feas. & Sales & Employment & Expectations & Extra Qs for NPR & Example publication \\
\midrule
\multicolumn{12}{c}{\textbf{NPR feasible}} \\
\midrule
AU & Business Outlook Scenarios Survey (BOSS) & \url{https://tinyurl.com/5fx7manu} & Manuf. and services & Monthly & Jan 2025--present & Jan 2025--present$^{(1)}$ & $t$, $\mathbb{E}[\tfrac{t+1}{t}]$ & $t$, $\mathbb{E}[t+1]$ & 5 scenarios with probabilities & NA & \url{https://www.nber.org/papers/w34836} \\
CN & China Employer–Employee Survey & \url{https://tinyurl.com/2p8yp6xz} & Manuf. & Annual / Biannual & 2015, 2016, 2018 & 2018 & $t-1$, $t$, $\mathbb{E}[t+1]$ & $t-1$, $t$, $\mathbb{E}[t+1]$ & 5 scenarios with probabilities & NA & \\
UK & Management and Expectations Survey & \url{https://tinyurl.com/y5uvawdk} & Manuf. and services & Triannual & 2017, 2020, 2023 & 2017, 2020, 2023 & $t-1$, $t$, $\mathbb{E}[t+1]$ & $t-1$, $t$, $\mathbb{E}[t+1]$ & 5 scenarios with probabilities & NA & \\
UK & Decision Maker Panel & \url{https://tinyurl.com/y46e4wmk} & Manuf. and services & Monthly & 2016--present & 2016--present$^{(1)}$ & $t$, $\mathbb{E}[\tfrac{t+1}{t}]$ & $t$, $\mathbb{E}[\tfrac{t+1}{t}]$ & 5 scenarios with probabilities & NA & \url{https://www.bankofengland.co.uk/-/media/boe/files/working-paper/2024/the-decision-maker-panel-a-users-guide.pdf} \\
US & Survey of Business Uncertainty & \url{https://tinyurl.com/mv5vpnh8} & Manuf. and services & Monthly & 2014--present & 2014--present & $t-1$, $t$, $\mathbb{E}[t+1]$ & $t-1$, $t$, $\mathbb{E}[t+1]$ & 5 scenarios with probabilities & NA & \url{https://www.sciencedirect.com/science/article/pii/S0304407620302785\#sec1} \\
US & US Management and Organizational Practices Survey (MOPS) & \url{https://tinyurl.com/mt2aurz8} & Manuf. & Approx.\ every 5 years & 2010, 2015, 2021 & 2015, 2021 & $t-1$, $t$, $\mathbb{E}[t+1]$ & $t-1$, $t$, $\mathbb{E}[t+1]$ & 5 scenarios with probabilities & NA & \url{https://www.nber.org/papers/w28259} \\
\midrule
\multicolumn{12}{c}{\textbf{NPR feasible with slight modifications}} \\
\midrule
DE & Bundesbank Firm Expectations Survey & \url{https://tinyurl.com/y4dssbp5} & Manuf. and services & Monthly & 2020--2024 & & $t$, $\mathbb{E}[t+1]$ (2021--2025) & $t$, $\mathbb{E}[t+1]$ (2024Q3 only) & Sales: 5 scenarios with probabilities; Employment: mode with bounds & Add scenario-based employment expectations & \\
IT & Survey of Industrial and Service Firms (INVIND) & \url{https://tinyurl.com/5n6z54av} & Manuf. and services & Annual & 1996--present & & $t-1$, $t$, $\mathbb{E}[t+1]$ & $t-1$, $t$, $\mathbb{E}[t+1]$ & Sales (2017--2020): mean with probability bins; Employment: point forecast & Add subjective employment distribution & \url{https://link.springer.com/article/10.1007/s00181-025-02725-0} \\
JP & Japanese MOPS (JP MOPS) & \url{https://tinyurl.com/nhka5723} & Manuf. & Variable & 2017, 2021 & & $t-1$, $\mathbb{E}[t]^{(1)}$, $\mathbb{E}[t+1]$ & $t-1$, $\mathbb{E}[t]^{(2)}$, $\mathbb{E}[t+1]$ & $\mathbb{E}[t]$: point; $\mathbb{E}[t+1]$: 3 scenarios with probabilities & Add distribution over current $\mathbb{E}[t]$ or administer survey closer to year end & \url{https://www.sciencedirect.com/science/article/pii/S0889158321000319?via\%3Dihub} \\
\bottomrule
\end{tabular}

\end{adjustbox}
\caption*{\footnotesize{Notes. $(1)$: questionnaire has a rotating structure, with expectations questions for sales and employment asked in distinct months. Implementing NPR would therefore require an assumption about the stability of employment across months. $(2)$: the JP MOPS was administered between January and March of the survey year. This is unlike the other surveys, which were typically administered towards the end of the survey year. Current-year values are therefore subject to greater uncertainty in the JP MOPS than in other datasets and it is therefore inappropriate to use point-valued reports to infer log quantities.}}
\end{sidewaystable}

A final remark on Table \ref{tab:expectations_data_library} is that there are numerous surveys eliciting information on \emph{point-valued expectations} as opposed to subjective \emph{distributions}. \citet{born_etal_handbook_2022}, for example, document a further 13 surveys that contain questions on point-valued expectations. This suggests data collectors have a strong interest in measuring firms' expectations, yet the manner in which they conventionally do so constrains the analytic utility of such information. Our paper is one example of the econometric possibilities enabled by data on subjective distributions and we imagine that such data will become more common given the interest by data collectors in measuring firms' expectations noted above. Point-valued expectations can be insightful, but far greater insight can be gained from more detailed expectation measures, even if the additional detail is only some quantification of uncertainty/variance. Furthermore, this paper demonstrates how questions on expectations enable analysis that typically demands panel data to be conducted using a cross section; a very attractive feature given the costs of panel data collection. Including questions on subjective expectation distributions in cross-sectional surveys, such as the World Bank Enterprise Survey, would therefore enable production function estimation in contexts where prominent methods are currently infeasible due to an absence of firm-level panel data. 

Our empirical application uses data from the Management and Expectations Survey (MES), one of the datasets in Table \ref{tab:expectations_data_library}. The MES is a survey administered by the UK's statistical authority that is sent to a stratified random sample of non-financial private sector firms.\footnote{Formally, these are ``Reference Units'' which are groups of establishments (e.g., manufacturing plants or retail stores). Further details on the MES' sampling design are provided in \citet{bloom_well_2021}.} The survey was designed to have broadly the same set of questions as US MOPS (\citet{bloom_what_2019}) and was administered in 2017 and 2020, creating two ``waves'' of data. It can be linked to other administrative business datasets and we exploit this property to match MES respondents to the Annual Business Survey (ABS, \citet{office_for_national_statistics_absannual_2023}) enabling us to compare firms' subjective expectations to outturns.\footnote{Establishments with 250 or more employees are surveyed each year by the ABS. Businesses below this threshold are surveyed on a multi-year basis with surveyed businesses in any one year chosen as a stratified random sample from the population.} Both the 2017 and 2020 versions of the MES ask firms for their sales revenue, employment, capital expenditure and expenditure on intermediates (purchases of energy, materials and services) in the year of the survey and the previous year. The survey specifies it should be completed by `the most senior person responsible for day-to-day operations', as a senior member of staff who is likely to have adequate knowledge of these quantities (this will correspond to the plant-manager or COO in most firms). We use firms' reported investment in these years to build capital stocks via the perpetual inventory method, imputing base period capital from national accounts data on industry-specific capital stocks, apportioning industry totals among firms according to within-industry intermediate input shares.

The latest 2023 wave of the MES omitted questions on previous-year quantities. Although, as indicated in Table \ref{tab:expectations_data_library}, it is feasible to implement our proposed estimator using this data, it is not possible to implement the OP/LP/ACF control-function estimators owing to the absence of previous-year quantities. We therefore exclude this data from our empirical application to ease the comparison of results across estimators. 

Inclusion in the subsample of MES data we use in the majority of our analysis is conditioned on responding to all MES expectations questions and having adequate observations of sales, capital, employment, intermediates and investment to implement the OP, LP and ACF estimators.\footnote{Both the 2017 and 2020 MES contain a limited number of what appear to be data entry errors in sales and employment. We identify these by calculating ratios of sales to employment and comparing these with equivalent ratios observed in the ABS. We identify spurious observations as those whose sales-employment ratios differ across the MES and the ABS by a factor of two \emph{and} with year-to-year growth in their MES sales-employment ratio in the top 5\% of the distribution. These observations are dropped from our analysis.}  To allow for parameter heterogeneity by sector, we confine attention to three industries defined using the UK's Standard Industrial Classification (SIC): ``electronics'' consists of firms that manufacture computer electronic and optical products or electrical equipment (SIC groups 26 and 27 respectively); ``retail'' consists of firms in the wholesale and retail trade except motor vehicles and motorcycles (SIC groups 46 and 47 respectively); and ``restaurants'' consists of firms who conduct food and beverage serving activities (SIC group 56). These industries were selected as they are among the largest in the MES and provide examples of manufacturing as well as service activities. 

Table \ref{tab:mes_sumstats} summarizes the characteristics of the MES analysis subsample in terms of means, standard deviations and (fuzzy) medians. We show these for the whole MES across all industries (first panel), and the three industries we focus on (subsequent panels). Relative to the combined sample, electronics firms are smaller in terms of inputs and output. Firms in the retail sector are relatively similar in size to the overall non-financial private sector economy, whereas restaurants are larger than average in terms of employment but smaller in terms of sales and value added, indicating this sector is relatively labor intensive. 

\begin{table}[H]
       \caption{MES Analysis Sample Characteristics}
       \label{tab:mes_sumstats}
        \centering
        \begin{adjustbox}{max width=\textwidth}
            \begin{tabular}{l|ccc|cc}
\hline\hline
 &  Mean  &  S.D.  &  p50  &  N obs.  &  N firms \\ \hline 
 &  \multicolumn{5}{c}{\textbf{(A) Full MES} } \\\hline
Sales (\pounds k)  & 23160 & 55758 & 6000 & 29314 & 12831 \\ 
Employment  & 162 & 298 & 58 & 29314 & 12831 \\ 
Capital (\pounds k)  & 10889 & 34959 & 1512 & 35417 & 12831 \\ 
Investment (\pounds k)  & 1238 & 4606 & 100 & 31310 & 12831 \\ 
Intermediates (\pounds k)  & 15171 & 40454 & 2902 & 29314 & 12831 \\ 
Value added (\pounds k)  & 7556 & 20015 & 1659 & 29314 & 12831\\ \hline 
 &  \multicolumn{5}{c}{\textbf{(B) Electronics 2016-2020}} \\\hline
Sales (\pounds k)  & 14418 & 23406 & 6265 & 848 & 374 \\ 
Employment  & 88 & 101 & 49 & 848 & 374 \\ 
Capital (\pounds k)  & 6335 & 11682 & 2167 & 1013 & 374 \\ 
Investment (\pounds k)  & 358 & 784 & 100 & 908 & 374 \\ 
Intermediates (\pounds k)  & 9657 & 20445 & 3350 & 848 & 374 \\ 
Value added (\pounds k)  & 5203 & 9718 & 2173 & 848 & 374\\ \hline 
 &  \multicolumn{5}{c}{\textbf{(C) Retail 2016-2020}} \\\hline
Sales (\pounds k)  & 30790 & 68665 & 8088 & 3668 & 1633 \\ 
Employment  & 124 & 262 & 44 & 3668 & 1633 \\ 
Capital (\pounds k)  & 5735 & 22291 & 725 & 4543 & 1633 \\ 
Investment (\pounds k)  & 791 & 2856 & 97 & 3931 & 1633 \\ 
Intermediates (\pounds k)  & 23377 & 55718 & 4990 & 3668 & 1633 \\ 
Value added (\pounds k)  & 6708 & 16958 & 1599 & 3668 & 1633\\ \hline 
 &  \multicolumn{5}{c}{\textbf{(D) Restaurants 2016-2020}} \\\hline
Sales (\pounds k)  & 12172 & 20091 & 3749 & 726 & 338 \\ 
Employment  & 321 & 500 & 91 & 726 & 338 \\ 
Capital (\pounds k)  & 10284 & 21212 & 2399 & 878 & 338 \\ 
Investment (\pounds k)  & 1254 & 3519 & 99 & 756 & 338 \\ 
Intermediates (\pounds k)  & 8690 & 17654 & 1816 & 726 & 338 \\ 
Value added (\pounds k)  & 4154 & 8372 & 648 & 726 & 338\\ \hline 
\hline\hline\end{tabular}
        \end{adjustbox}
        \caption*{{\scriptsize Note: `Mean' and `S.D.' columns show the mean and standard deviation calculated over the sample indicated in the `N obs.' column which consists of a number of distinct firms indicated in the `N firms' column. While the table is calculated using data from two MES waves, `N obs.' is often more than twice as large as `N firms' as the statistics are calculated from firms' reports of current and previous year values, yielding at most two observations per firm-wave. The `p50' column contains the mean value among the 50 observations closest to the median owing to data disclosure requirements. Monetary values are given in current prices.}}
\end{table}

\subsection{Subjective Expectations}\label{subsect:data_expectations}
 
To elicit firms' expectations for the following year, the MES asks firms to report on five scenarios ranging from ``lowest'' to ``highest''. Firms are asked for the value they expect each variable separately to take under each scenario in the next year and the likelihood of the scenario occurring. In the 2017 MES, for example, the exact wording of the question regarding sales expectations was: ``Looking ahead to the 2018 calendar year, what is the approximate pound sterling value of turnover you would anticipate for this business in the following scenarios [Lowest, Low, Medium, High, Highest], and what likelihood do you assign to each scenario?'' (note that the question uses ``turnover'', which is the British term for sales revenue). This wording is very similar to that used in the US MOPS and in the other surveys listed in Table \ref{tab:expectations_data_library}. An image showing the expectations question and its position in relation to questions on current and previous year sales is given in Appendix \ref{adix:supp_results}. 

The 2017 MES used these questions to elicit firms' expectations of sales, employment, capital expenditure and expenditure on intermediate  inputs (energy, goods and services which we label ``materials''), whereas the 2020 MES only asked firms about their expectations of sales and employment in order to limit survey length. In both years, expectations over monetary quantities were asked in nominal terms. We therefore conduct all analysis on a nominal basis and include a time dummy to allow for industry-year specific shocks (like output prices).
	
The 2017 MES was administered as a paper survey and, although firms were instructed to ensure the likelihoods assigned to the five scenarios summed to 100, some responses did not meet this criterion. In these cases the reported likelihoods were rescaled to sum to 100 if the total likelihood across the five scenarios was between 90 and 110. A small number of responses with a total reported likelihood outside of this window were discarded. This issue does not appear in the 2020 data, as this wave of the MES was administered online and required respondents' reported likelihoods to sum to 100 before they could proceed to subsequent questions.
	
The MES respondents report point-values for each of the five scenarios and their related probabilities. This is in contrast to the survey design implemented in, for example, \citet{dominitz_using_1997}, which recovers households' subjective CDFs of one-year-ahead income by asking for the perceived likelihood that income next year will fall below a number of thresholds, where the thresholds are determined by first asking households for the minimum and maximum income they expect next year and splitting the interval into ``bins''. Because of this discrepancy, it is not obvious how to use the MES' questions on subjective expectations to recover firms' subjective CDFs. One approach is to treat the scenario values as points on either a corresponding CDF or survival function on the stated support. In the CDF approach, for example, the cumulative likelihood for the Medium scenario would be taken as the sum of the likelihoods a firm reports against the Lowest, Low and Medium scenarios. In the survival function approach, by contrast, the cumulative distribution function value for the Medium scenario would be taken as one minus the survival function, which is the sum of the likelihoods a firm reports against the Highest, High and Medium scenarios. 

Experimentation with both approaches found the former (CDF) created a lower mean expectation relative to a simple weighted sum across scenarios, while the latter (survival function) created a higher mean expectation compared with a simple weighted sum across scenarios. We therefore estimate lognormal parameters for both approaches by choosing mean and variance parameters to minimize the sum of squared deviations between the fitted distribution and firms' reported scenario values and their corresponding CDF or survival function points. Firms' subjective CDFs are then characterised as $F_{it}(l_{it+1}) = \mathcal{N}(\bar{\mu_{i}},\overline{\sigma_i^{2}})$, where bars denote the averages across the CDF and survival function estimates.\footnote{In the Cobb-Douglas specification, one only requires expectations rather than the subjective CDF, which could alternatively be estimated as a weighted sum across (log) scenario values using the reported likelihoods as weights. These weighted sums are very similar to those of the fitted lognormals. An alternative would be to rely on ``bounds'' defined by the CDF and survival functions.} Finally, to reduce the degree of measurement error in the expectations variables, we exclude observations in the top and bottom 1\% of expected sales or employment calculated within each of our focus industries.

\begin{figure}[H]
    \centering 
    \caption{Example of Fitting a Subjective CDF to Scenario-based Expectations}\label{fig:subjdistexample}
    \begin{subfigure}[b]{0.7\textwidth}
        \centering
        \caption{Reported Scenario Values and Probabilities}\label{fig:subjdistexample_pdf}
        \includegraphics[width=\textwidth]{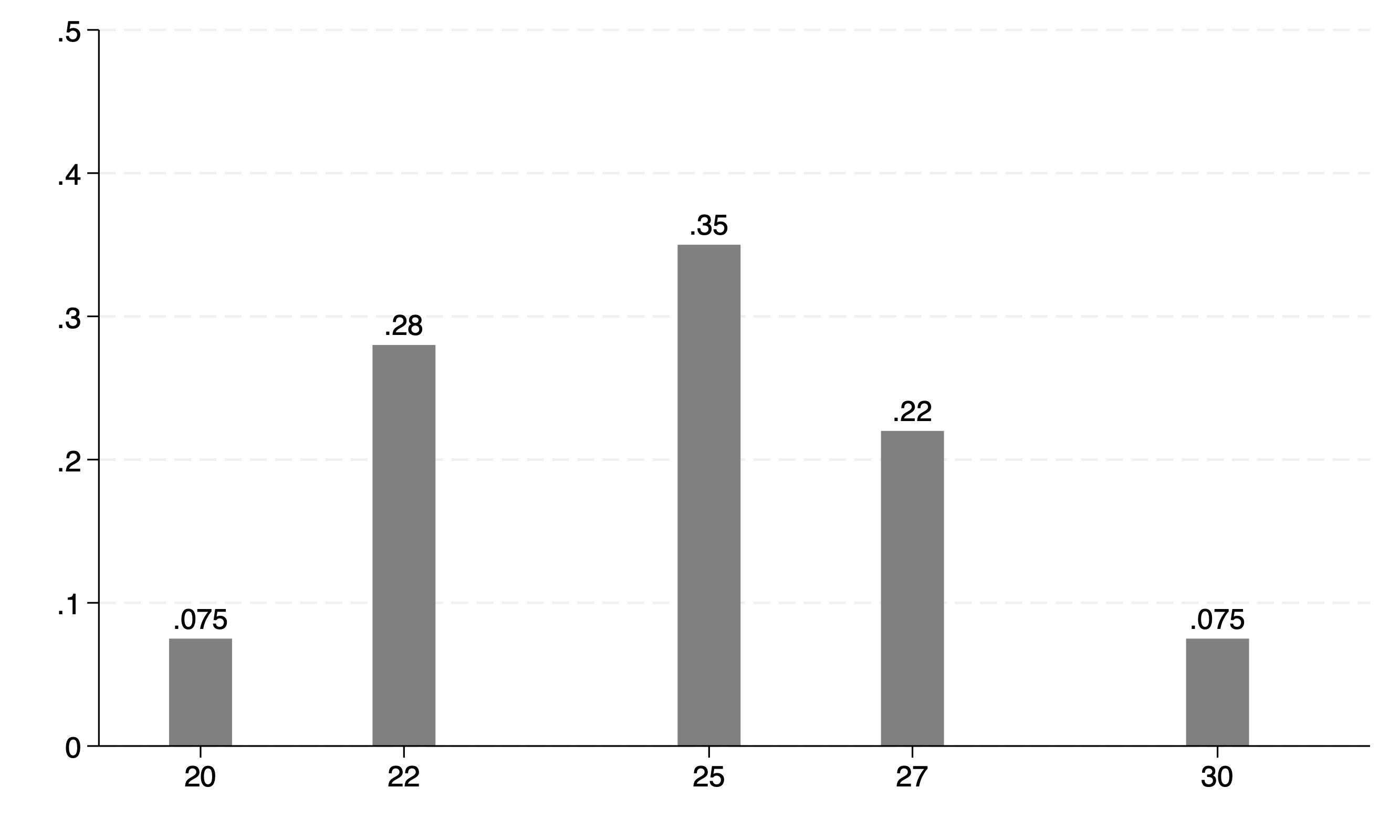}
    \end{subfigure} \\
    \begin{subfigure}[b]{0.7\textwidth}
        \centering 
        \caption{Subjective Distribution Functions}\label{fig:subjdistexample_cdfs}
        \includegraphics[width=\textwidth]{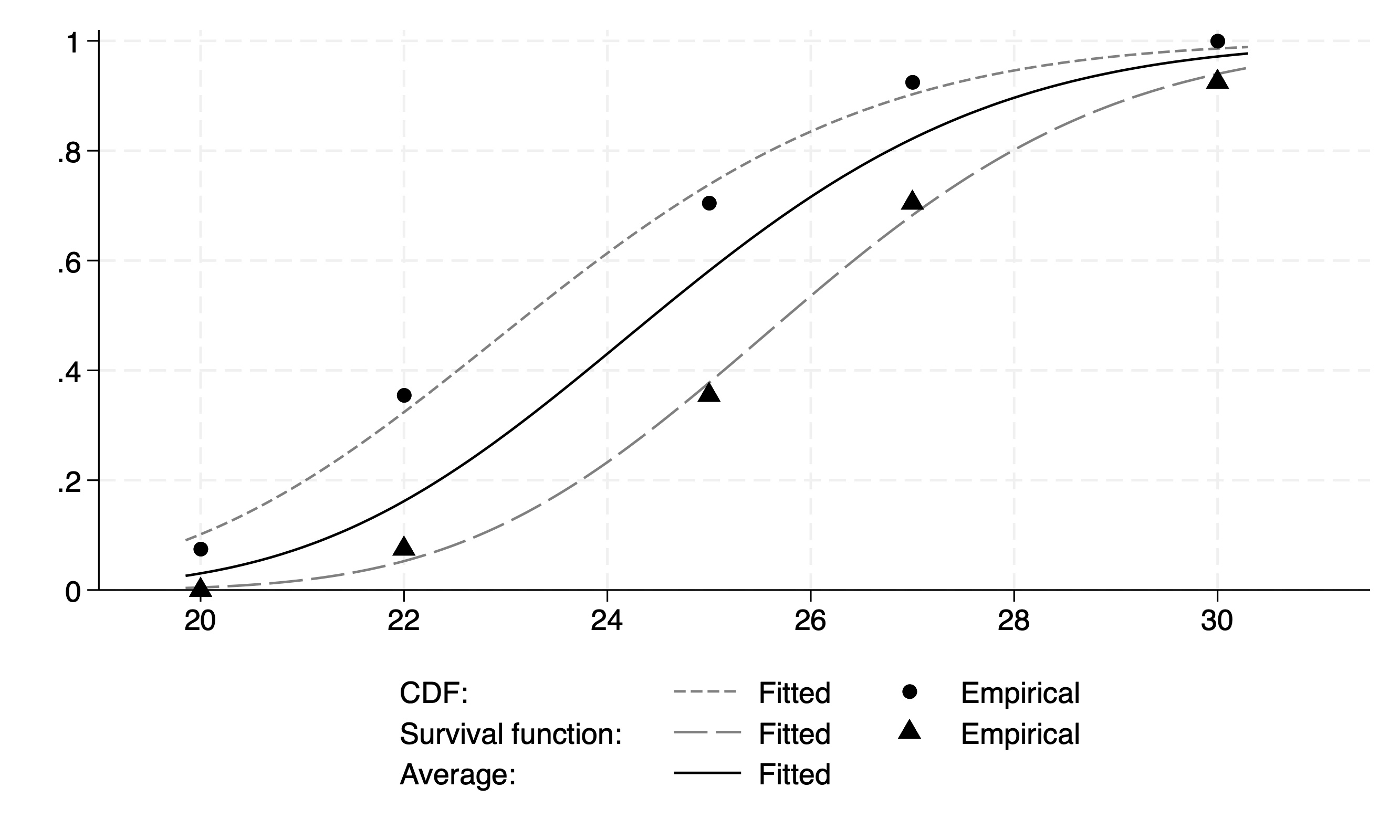}
    \end{subfigure}
            \caption*{{\scriptsize Note: This is an example of how we create subjective distribution functions from the reported scenario values and probabilities from the MES data. Panel (a) has the data where a respondent estimates there is, for example, a 35\% probability that next year's sales will be \pounds 25,000, a 28\% chance of \pounds 22,000, etc. The lower panel shows the two methods of translating these responses into a fitted lognormal distribution (the CDF vs. survivor function approaches are dashed lines). We calculate the solid line as a lognormal CDF using the average of the two means and variances. See text for details.}}
\end{figure}

Figure \ref{fig:subjdistexample} provides an example of our approach. Panel (a) is a hypothetical response to a scenario-based expectation question. The values on the x-axis represent the expected values in each of the five scenarios (ranked lowest to highest), and the bars represent the reported subjective likelihoods of these scenarios occurring. The markers in Panel (b) show the empirical CDF and survival function implied by these responses, calculated using the method described above. The dashed lines in the panel are lognormal CDFs with mean and variance chosen to minimise the sum of square residuals between the fitted CDF and the empirical distribution. The solid line is a lognormal CDF with mean and variance taken as the average mean and variance underlying the two dashed CDFs and is the subjective CDF we use to calculate expected values in the MES data.

A legitimate concern with this approach is that it imposes symmetry in logs on firms' subjective expectations, which may introduce non-negligible measurement error if firms' expectations exhibit skewness. Table \ref{tab:skewness} in Appendix \ref{adix:expectation_robustness} shows that the majority of firms in our estimation sample report scenario values that are symmetric in logs, with a small number exhibiting negative skewness due to reporting very low values for the ``lowest'' expectations scenario. Table \ref{tab:backfit_results_ecomp} in the same Appendix shows our main estimates are broadly similar if one fits the parameters of a beta distribution instead of those of a lognormal, which suggests these firms are not biasing our results. The table additionally shows the NPR estimates are stable if one constructs firms' expectations using the CDF or the survival function approach in isolation, and if one constructs expectations as a weighted sum across scenarios where weights are taken as firms' reported scenario probabilities. 

Table \ref{tab:l2_av_ssestimates_lnorm} summarises absolute deviations between firms' reported point values and those implied by the fitted lognormal CDFs. These differences are small across all variables for all industries implying the fitted lognormal distributions provide a good continuous approximation of the subjective distributions underlying firms' responses to the discrete MES expectations questions and compares favorably with the fit obtained in related works such as \citet{dominitz_using_1997}.\footnote{See the discussion in their Section 3.5. Note that we confine analysis of firms' expectations here to sales and employment as these are the key expectations variables required by the baseline NPR estimator. Similar results for firms' expectations of investment and intermediate inputs as reported in the 2017 MES are available from the authors on request.}

\begin{table}[H]
       \caption{Summary Statistics of Average Absolute Deviations Between Reported and Fitted Probabilities}
       \label{tab:l2_av_ssestimates_lnorm}
        \centering
        \begin{adjustbox}{max width=\textwidth}
            \begin{tabular}{l|ccccc}
\hline \hline
& Mean & p25 & p50 & p75 & N \\
\hline
& \multicolumn{5}{c}{\textbf{(A) Electronics}} \\\hline
\hspace{.4cm} Sales &   0.031 &   0.020 &   0.027 &   0.038 &     424 \\ 
\hspace{.4cm} Employment &   0.030 &   0.018 &   0.027 &   0.038 &     424 \\ \hline
& \multicolumn{5}{c}{\textbf{(B) Retail}} \\\hline
\hspace{.4cm} Sales &   0.031 &   0.019 &   0.028 &   0.038 &    1834 \\ 
\hspace{.4cm} Employment &   0.030 &   0.019 &   0.027 &   0.038 &    1834 \\ \hline
& \multicolumn{5}{c}{\textbf{(C) Restaurants}} \\\hline
\hspace{.4cm} Sales &   0.032 &   0.020 &   0.030 &   0.040 &     363 \\ 
\hspace{.4cm} Employment &   0.033 &   0.020 &   0.030 &   0.040 &     363 \\ 
\hline \hline
\end{tabular}

        \end{adjustbox}
        \caption*{{\scriptsize Note: The table contains means and quantile-group averages of the mean absolute deviations across firms' reported scenario likelihoods and those implied by the fitted lognormal distributions. The column titled `Mean' contains the arithmetic mean calculated across the number of observations given in the column titled `N'. Columns titled `p25'/`p50'/`p75' contain the mean values calculated across the 50 observations that are closest to the 25th/50th/75th percentile respectively owing to data disclosure requirements.}}
\end{table}

\begin{figure}[H]
    \caption{Expected growth rates}\label{fig:dist_e_gr}
    \begin{subfigure}[b]{0.5\textwidth}
        \centering
        \caption{Electronics: Sales}\label{fig:dist_e_gr_turnover_electronics}
        \includegraphics[width=\textwidth]{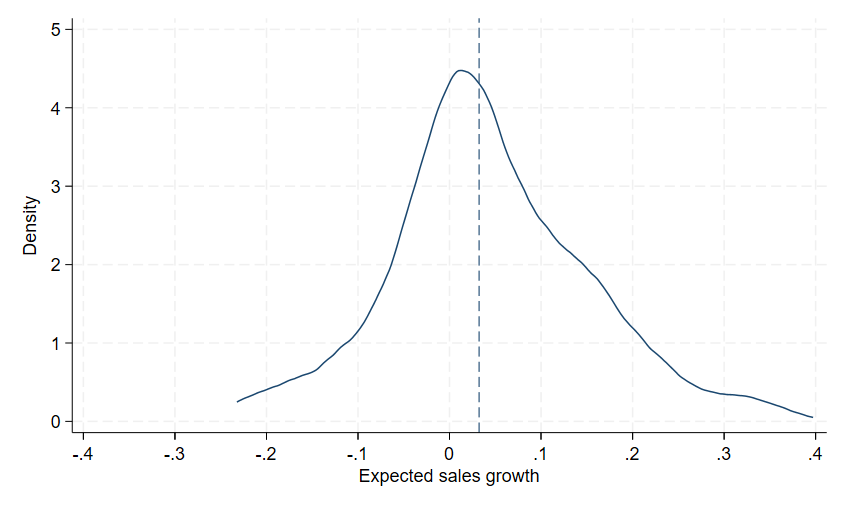}
    \end{subfigure}
    \begin{subfigure}[b]{0.5\textwidth}
        \centering 
        \caption{Electronics: Employment}\label{fig:dist_e_gr_emp_electronics}
        \includegraphics[width=\textwidth]{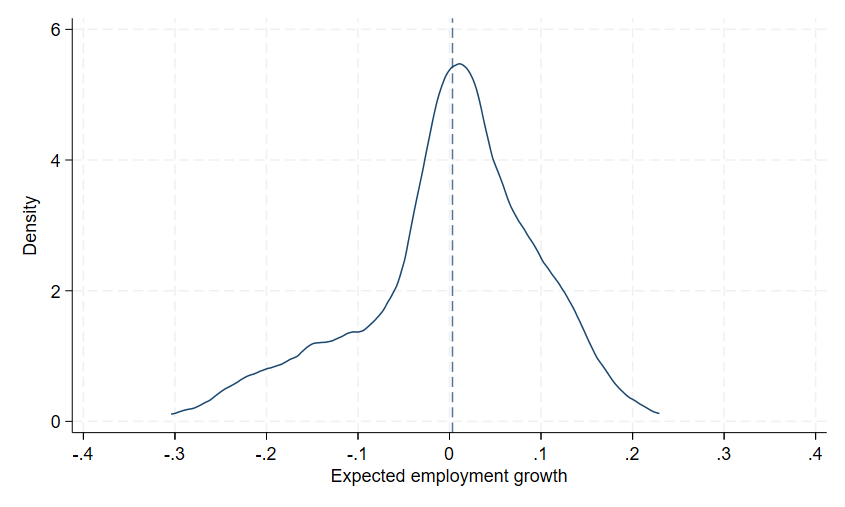}
    \end{subfigure}
    \begin{subfigure}[b]{0.5\textwidth}
        \centering
        \caption{Retail: Sales}\label{fig:dist_e_gr_turnover_wsale_retail}
        \includegraphics[width=\textwidth]{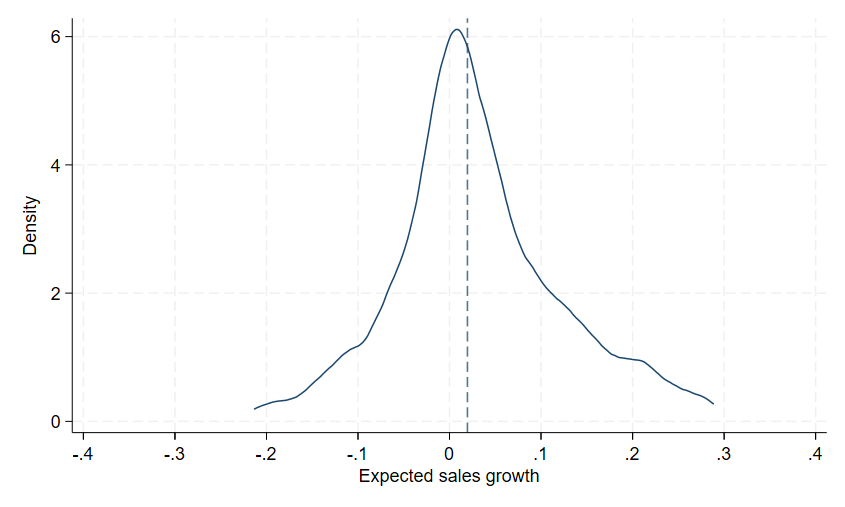}
    \end{subfigure}
    \begin{subfigure}[b]{0.5\textwidth}
        \centering 
        \caption{Retail: Employment}\label{fig:dist_e_gr_emp_wsale_retail}
        \includegraphics[width=\textwidth]{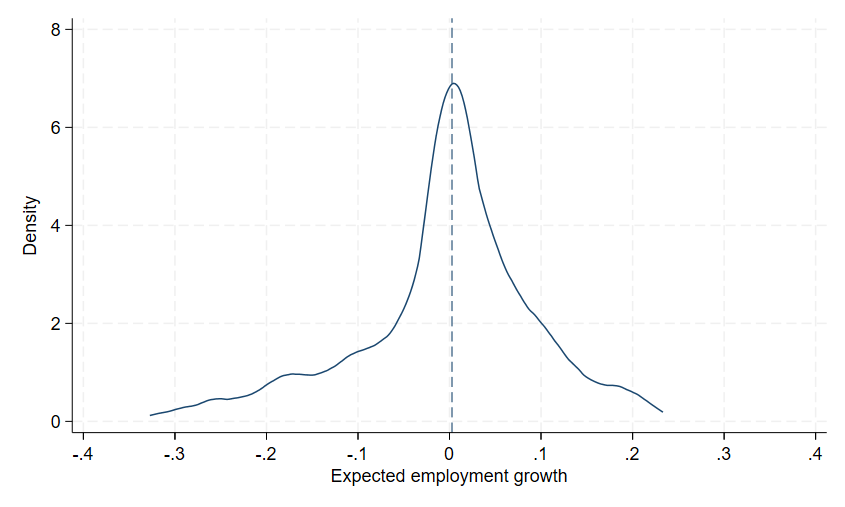}
    \end{subfigure}
    \begin{subfigure}[b]{0.5\textwidth}
        \centering
        \caption{Restaurants: Sales}\label{fig:dist_e_gr_turnover_foodbev_serv}
        \includegraphics[width=\textwidth]{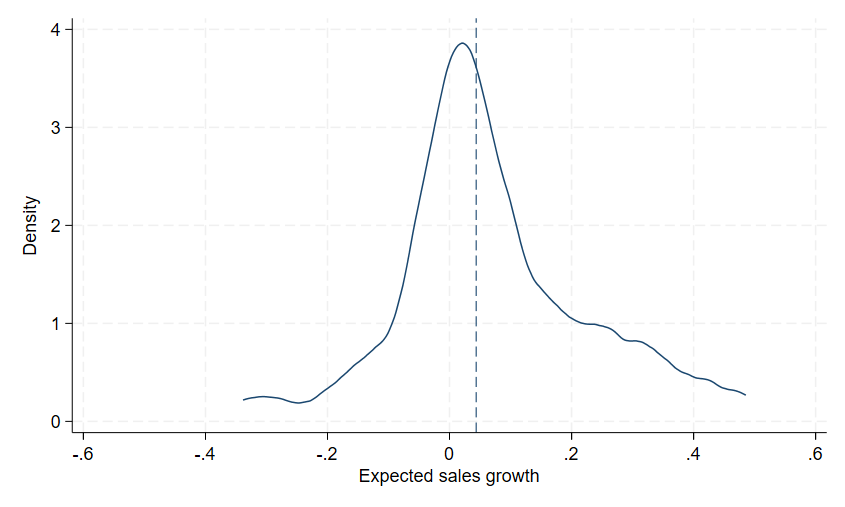}
    \end{subfigure}
    \begin{subfigure}[b]{0.5\textwidth}
        \centering 
        \caption{Restaurants: Employment}\label{fig:dist_e_gr_emp_foodbev_serv}
        \includegraphics[width=\textwidth]{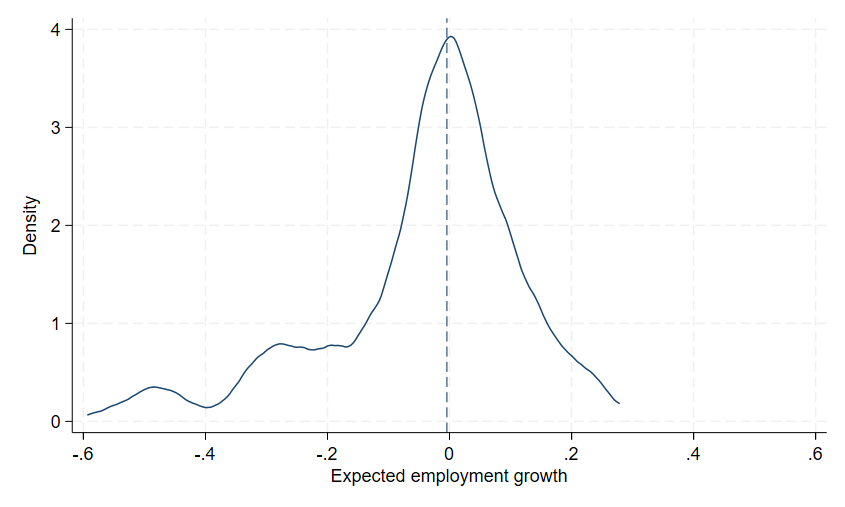}
    \end{subfigure}
 \caption*{{\scriptsize Note: Subfigures show kernel densities of $(\mathbb{E}_{it}[\text{ln(X)}_{it+1}]-\text{ln(X)}_{it})$ where X is either sales or employment as indicated in the subfigure captions. Dashed vertical lines indicate medians. The top and bottom 5\% of growth rates are included in the calculation of the medians but are not plotted in the figure.}}
\end{figure}

Our method of obtaining subjective distributions from the MES responses yields a distribution of lognormal parameters and implied moments across MES respondents. Table \ref{tab:estdists_medians_lognormal_inds} summarizes the medians of these parameters and a selection of moments across variables and samples.\footnote{Percentile statistics are prohibited from being exported from the secure server through which we access the MES and ABS data. Table \ref{tab:estdists_medians_lognormal_inds} therefore reports `fuzzy medians' of the firm-specific lognormal subjective distribution parameters, calculated as the mean value across the 50 observations closest to the median.} The average $\sigma$ parameter of the fitted subjective distributions indicates firms' uncertainty and shows that restaurants are slightly more uncertain on average about year-ahead sales than firms in electronics or retail, and more substantially so regarding year-ahead employment. As well as greater within-firm uncertainty, the restaurant sector also exhibits greater across-firm variation in mean expectations. This can be seen in Figure \ref{fig:dist_e_gr}, which plots kernel densities of firms' expected mean growth in sales and employment. 

\begin{table}[H]
       \caption{Median Fitted Subjective Distribution Characteristics}
       \label{tab:estdists_medians_lognormal_inds}
        \centering
        \begin{adjustbox}{max width=\textwidth}
            \begin{tabular}{l|ccccccc}
\hline \hline
& $ \mu $ & $ \sigma $ & Mean & S.D. & Median & IQR & N \\
\hline
\hline
& \multicolumn{7}{c}{\textbf{(A) Electronics}}  \\ \hline
\hspace{.4cm} Sales  &    8.78 &    0.06 &    6469 &    3939 &    6405 &     518 &     424 \\ 
\hspace{.4cm} Employment  &    3.90 &    0.04 &      50 &      30 &      49 &       3 &     424 \\ 
\hline
& \multicolumn{7}{c}{\textbf{(B) Retail}}  \\ \hline
\hspace{.4cm} Sales  &    9.06 &    0.06 &    8639 &    5272 &    8602 &     661 &    1834 \\ 
\hspace{.4cm} Employment  &    3.80 &    0.05 &      45 &      27 &      45 &       3 &    1834 \\ 
\hline
& \multicolumn{7}{c}{\textbf{(C) Restaurants}}  \\ \hline
\hspace{.4cm} Sales  &    8.16 &    0.08 &    3224 &    1992 &    3211 &     274 &     363 \\ 
\hspace{.4cm} Employment  &    4.44 &    0.08 &      81 &      50 &      81 &       8 &     363 \\ 
\hline \hline
\end{tabular}

        \end{adjustbox}
        \caption*{{\scriptsize Note: The table contains mean values among the 50 observations closest to the median of the across-firm distribution of the parameter or moment indicated in the column title. Units for employment are for number of workers and for all other variables are thousands of pounds. The $\mu$ and $\sigma$ columns give log-normal distribution parameters and therefore are interpreted as logged values of the variable given in the row title. The column `N' indicates the number of observations for which we are able to fit a subjective distribution.}}
\end{table}

A subsample of firms in the MES were also surveyed in the ABS the year following their MES response, which allows us to compare their subjective expectations to actual outcomes. Table \ref{tab:mes_expectation_errors} shows the log difference between firms' expectations and outcomes separately by variable and year. On average, firms' expectations are generally good, with the forecast error being insignificantly different from zero. For employment, the forecast error was on average zero for retail, -2\% for electronics and -5\% for restaurants (and 1\%, 2\% and 3\% when medians are considered, respectively).  Electronics and retail industries are also reasonably accurate at forecasting sales (-7\% and -6\% respectively on average and -4\% and 3\% for the median errors). Restaurants, however, underestimated sales by 18 log points (and 12 log points at the median).  An explanation for this pattern relates to difficulties faced by the sector, especially during 2020.  As noted earlier, the specification of the production function we use in estimation includes a time dummy to allow for industry-year specific shocks, which should (at least partially) accommodate the pessimism in the sector for that year.

\begin{table}[H]
       \caption{Log Deviation Between Expected Levels and Outcomes}
       \label{tab:mes_expectation_errors}
        \centering
        \begin{adjustbox}{max width=\textwidth}
            \begin{tabular}{l|ccc|cc}
\hline\hline
 &  Mean  &  S.D.  &  p50  &  N obs.  &  N firms \\ \hline 
 &  \multicolumn{5}{c}{\textbf{(A) Electronics}} \\\hline
Sales  & -0.07 & 0.45 & -0.04 & 202 & 186\\ 
Employment  & -0.02 & 0.35 & 0.02 & 202 & 186\\ \hline 
 &  \multicolumn{5}{c}{\textbf{(B) Retail}} \\\hline
Sales  & -0.06 & 0.3 & -0.03 & 654 & 589\\ 
Employment  & 0.00 & 0.21 & 0.01 & 656 & 591\\ \hline 
 &  \multicolumn{5}{c}{\textbf{(C) Restaurants}} \\\hline
Sales  & -0.18 & 0.46 & -0.12 & 139 & 122\\ 
Employment  & -0.05 & 0.44 & 0.03 & 139 & 122\\ 
\hline\hline\end{tabular}
        \end{adjustbox}
        \caption*{{\scriptsize Note: The table contains summary statistics of the log deviation between the expected value and outcome of the variable denoted in the row title. Columns titled `Mean' and `S.D.' contain the mean and standard deviation respectively calculated across the number of firms in the column titled `N'. The column titled `p50' contains the mean value among the 50 observations closest to the median of the across-firm distribution denoted in the column title as data disclosure requirements prevent us from reporting the median value itself.}}
\end{table}

\section{Results}\label{sect:results}
\subsection{Main Results}\label{subsection:results_main}
To evaluate the empirical performance of the NPR estimator, we implement a number of other popular production function estimators on the MES data separately for each of our industries. In Table \ref{tab:backfit_results_summary}, estimates for the electronics sector are in Panel A, for the retail sector in Panel B and for the restaurant sector in Panel C.\footnote{All results discussed in this section are from a Cobb-Douglas gross output production function (i.e.,one that takes sales as the measure of output). Appendix \ref{adix:prodfunct_va} contains an explanation of how one can construct expectations of (log) value added using expectations questions of the type contained in the MES for sales and intermediates. We do not pursue estimation of a value added (net) production function since expectations of intermediates were only asked in the 2017 MES, which would result in prohibitively small analysis samples. The number of observations is less than twice the number of firms because not all firms were surveyed in both the 2017 and 2020 MES.}

Across all industries in Table \ref{tab:backfit_results_summary}, the first-differenced OLS specification (contained in column `OLS FD') returns capital coefficients that are near zero and statistically insignificant. The labor coefficient is also generally smaller in magnitude than the other estimators. The large attenuation indicates that exploiting some of the cross sectional variation across firms is necessary to yield credible estimates (as emphasized \textit{inter alia} by \cite{strom_production_1999}). The results in the first column show that NPR generates plausible coefficients throughout the table that are generally similar to ACF and OLS levels, with exceptions discussed below. OP and LP return lower labor coefficients than the OLS, NPR and ACF estimators, with OP yielding an insignificant capital coefficient across all industries. The EZ-NPR estimator returns results very similar to the baseline NPR results, suggesting that measurement error is not a major source of bias in our context.

More specific comparisons between the various estimators differ by industry. In  electronics (Panel A), NPR returns a labor coefficient of 0.83, slightly higher than ACF at 0.76 and OLS at 0.79. The capital coefficient of 0.24 is similar to ACF and only slightly lower than OLS (0.28).
The labor coefficient for NPR is again only slightly higher than ACF and OLS (0.75 vs. 0.71) in retail (Panel B). The capital coefficient for NPR is two-thirds of OLS (0.20 vs. 0.29) and ACF is lower (0.17) and insignificant. 

The largest differences between NPR and ACF are in the restaurant sector. The capital coefficient is 0.11 for NPR and only 0.02 for ACF, while the labor coefficients are 0.90 and 1.01, respectively. It is \textit{ex ante} implausible to believe that capital does not matter at all in this sector. The very low value for $\hat \beta_k$, together with the high value obtained for $\hat \beta_l$ is suggestive of the identification issue highlighted earlier in Section \ref{sect:montecarlo}. It is also reassuring that EZ-NPR has a similar capital (and labor) coefficient to NPR of 0.13, which is statistically significant.\footnote{The NPR capital coefficient in restaurants attains significance when we relax the condition restricting the sample to firms for which the other proxy variable estimators can be implemented (it is 0.15 with a standard error of 0.07). This suggests the statistical insignificance is primarily due to small sample size.}

\begin{table}[H]
     \caption{Production Function Coefficient Estimates}
      \label{tab:backfit_results_summary}
       \centering
       \begin{adjustbox}{max width=\textwidth}
           \begin{tabular}{l|ccccccc}
\hline
& NPR &ACF& OLS & OLS FD & OP & LP & EZ-NPR \\
\hline
&\multicolumn{7}{c}{\textbf{(A) Electronics}} \\
\hline
 $ \beta_l $  & 0.83$ ^{\ast\ast\ast} $ & 0.76$ ^{\ast\ast\ast} $ & 0.79$ ^{\ast\ast\ast} $ & 0.42$ ^{\ast\ast\ast} $ & 0.61$ ^{\ast\ast\ast} $ & 0.62$ ^{\ast\ast\ast} $ & 0.87$ ^{\ast\ast\ast} $  \\
 &(0.07)&(0.21)&(0.06)&(0.11)&(0.05)&(0.08)&(0.24) \\
 $ \beta_k $  & 0.24$ ^{\ast\ast\ast} $ & 0.23$ ^{\ast\ast\ast} $ & 0.28$ ^{\ast\ast\ast} $ &0.02&0.23& 0.34$ ^{\ast\ast\ast} $ &0.26 \\
 &(0.06)&(0.08)&(0.05)&(0.01)&(0.14)&(0.09)&(0.21) \\
 \hline
 $ \beta_l+\beta_k $ &1.08&0.99&1.07&0.44&0.84&0.97&1.13 \\
 CRS  &0.1&0.95&0.01&0.00&0.27&0.79&0.12 \\
 \hline
 N obs.  &424&424&424&424&424&424&424 \\
 N firms  &374&374&374&374&374&374&374 \\
\hline
&\multicolumn{7}{c}{\textbf{(B) Retail}} \\
\hline
 $ \beta_l $  & 0.75$ ^{\ast\ast\ast} $ & 0.71$ ^{\ast\ast\ast} $ & 0.71$ ^{\ast\ast\ast} $ & 0.50$ ^{\ast\ast\ast} $ & 0.61$ ^{\ast\ast\ast} $ & 0.54$ ^{\ast\ast\ast} $ & 0.78$ ^{\ast\ast\ast} $  \\
 &(0.07)&(0.20)&(0.04)&(0.05)&(0.04)&(0.05)&(0.08) \\
 $ \beta_k $  & 0.20$ ^{\ast\ast\ast} $ &0.17& 0.29$ ^{\ast\ast\ast} $ &0.00&0.03& 0.16$ ^{\ast\ast\ast} $ & 0.14$ ^{\ast\ast\ast} $  \\
 &(0.04)&(0.13)&(0.03)&(0.01)&(0.06)&(0.06)&(0.05) \\
 \hline
 $ \beta_l+\beta_k $ &0.95&0.88&1.00&0.50&0.64&0.70&0.92 \\
 CRS  &0.29&0.63&0.91&0.00&0.00&0.00&0.23 \\
 \hline
 N obs.  &1834&1834&1834&1834&1834&1834&1834 \\
 N firms  &1633&1633&1633&1633&1633&1633&1633 \\
\hline
&\multicolumn{7}{c}{\textbf{(C) Restaurants}} \\
\hline
 $ \beta_l $  & 0.90$ ^{\ast\ast\ast} $ & 1.01$ ^{\ast\ast\ast} $ & 0.86$ ^{\ast\ast\ast} $ & 0.75$ ^{\ast\ast\ast} $ & 0.73$ ^{\ast\ast\ast} $ & 0.74$ ^{\ast\ast\ast} $ & 0.89$ ^{\ast\ast\ast} $  \\
 &(0.07)&(0.10)&(0.05)&(0.08)&(0.05)&(0.06)&(0.08) \\
 $ \beta_k $  &0.11&0.02& 0.15$ ^{\ast\ast\ast} $ &0.03&0.04&0.05& 0.13$ ^{\ast\ast} $  \\
 &(0.07)&(0.11)&(0.06)&(0.03)&(0.09)&(0.14)&(0.06) \\
 \hline
 $ \beta_l+\beta_k $ &1.01&1.03&1.01&0.78&0.77&0.79&1.02 \\
 CRS  &0.84&0.67&0.49&0.00&0.01&0.12&0.66 \\
 \hline
 N obs.  &363&363&363&363&363&363&363 \\
 N firms  &338&338&338&338&338&338&338 \\
\hline\hline\end{tabular}

       \end{adjustbox}
       \caption*{\footnotesize{Note: dependent variable is log sales. Parentheses contain standard errors. NPR standard errors calculated from 100 bootstrap replications. Column titles indicate estimation methods. `OLS' and `OLS FD' denote OLS estimation in levels and first-differences, respectively. Row `CRS' contains the p-value from a test that the labor and capital coefficient sum to 1. All specifications include survey year dummies. $^{\ast}$/$^{\ast\ast}$/$^{\ast\ast\ast}$ denote significance at the 10/5/1 percent level respectively.}}
\end{table}

NPR is most similar to OLS levels in Table \ref{tab:backfit_results_summary}, which may be surprising in light of the simultaneity issues with OLS. Note however that in retail the capital elasticity is about 50\% larger in OLS and also that there are more marked differences when a translog production function is estimated (see Appendix \ref{adix:prodfunct_results}). 
 
Why should NPR and ACF deliver such different results for restaurants? One explanation is that material input invertibility is less likely to hold in the restaurant industry than in either electronics or retail. Evidence to this point comes from a subsample of firms in the MES that were surveyed by the ABS. Unlike the ABS, which asks firms for annual values retrospectively after the year is complete, the MES was dispatched to firms in Fall-Winter of the survey year. This means the ``current-year'' values that firms were asked for (i.e.,2017 values for firms surveyed in MES 2017 and 2020 values for firms surveyed in MES 2020), were estimates made before the year was complete. Comparing levels of inputs and output that firms report in the MES survey year to the equivalent values observed in the ABS therefore gives an indication of how unpredictable inputs are, and thus how hard they are to optimize. 

Table \ref{tab:mes_errors} contains moments of the distribution of MES-ABS log differences separately by industry and variable. This shows sales was relatively easy to forecast prior to year-end, with a median difference between MES reported values and the ABS equivalents of zero across all industries. For electronics and retail, employment was equally as predictable; it was only slightly less so in the restaurant sector with a median MES-ABS deviation of 4\%. The predictability of material inputs, however, varies more markedly across industries. Focusing on the median to avoid the influence of outliers, the median MES-ABS difference for intermediates is small for electronics and retail firms at 2\% and 1\% respectively. For restaurants, however, it is much larger at 34 log points. Under the assumption of Leontief technology, the fact that employment was relatively unchanged from the date at which restaurants were surveyed by the MES whereas intermediate inputs changed substantially implies intermediate inputs cannot be at their intended level for the year. As shown in Section \ref{sect:montecarlo}, this optimization error in intermediate inputs undermines ACF, owing to violation of the monotonicity condition they require, and we therefore view Table \ref{tab:mes_errors} as further reason to doubt the ACF production function estimates for the restaurants industry in favour of the NPR estimates.\footnote{A third more technical point is to first note that the returns to scale implied by the ACF estimates are similar to those implied by the NPR estimates. This suggests that the ACF estimates may be affected by the global identification issue noted above in footnote \ref{ACFfn16}.}

\begin{table}[H]
       \caption{Moments of Within-Firm MES-ABS Log Differences}
       \label{tab:mes_errors}
        \centering
        \begin{adjustbox}{max width=\textwidth}
            \begin{tabular}{l|ccc|cc}
\hline\hline
 &  Mean  &  S.D.  &  p50  &  N obs.  &  N firms \\ \hline 
\hline
 &  \multicolumn{5}{c}{\textbf{(A) Electronics}} \\\hline
Sales  & 0.00 & 0.15 & 0.00 & 245 & 218\\ 
Employment  & 0.02 & 0.17 & 0.01 & 245 & 218\\ 
Intermediates  & -0.18 & 1.5 & 0.02 & 245 & 218\\ \hline
 &  \multicolumn{5}{c}{\textbf{(B) Retail}} \\\hline
Sales  & 0.01 & 0.15 & 0.00 & 982 & 906\\ 
Employment  & 0.02 & 0.19 & 0.00 & 982 & 906\\ 
Intermediates  & -0.28 & 1.33 & 0.01 & 981 & 905\\ \hline
 &  \multicolumn{5}{c}{\textbf{(C) Restaurants}} \\\hline
Sales  & 0.03 & 0.29 & 0.00 & 189 & 171\\ 
Employment  & -0.01 & 0.34 & 0.04 & 189 & 171\\ 
Intermediates  & 0.22 & 1.12 & \textbf{0.34} & 189 & 171\\ \hline
\hline\hline\end{tabular}
        \end{adjustbox}
	\caption*{\footnotesize{Note: The table shows summary statistics of $\left(\text{ln}(X^{MES}_{it})-\text{ln}(X^{ABS}_{it})\right)$ where superscripts denote survey name. X is either sales, employment or intermediates as indicated in the row title. $t$ is either 2017 or 2020 according to which year firms were surveyed by the MES. Sample is restricted to firms in the MES analysis sample that are also observed in the wave of the ABS that records values for the same period as the MES survey year. The `p50' column contains the mean values calculated across the 50 observations that are closest to the 50th percentile owing to data disclosure requirements.}}
\end{table}

While we have focused here on Cobb-Douglas production technology, the NPR estimator can also estimate parameters of a translog production function. Appendix \ref{adix:prodfunct_results} describes the additional data preparation necessary for this alternate specification and presents estimates for our three industries along with alternative estimates obtained via various OLS estimators and the ACF method. Testing whether the additional nonlinear translog parameters are jointly zero, the NPR estimates fail to reject Cobb-Douglas technology for the retail and restaurant sectors, but do for electronics. 

\subsection{Productivity Decompositions}\label{subsect:results_productivity}

An implication of the estimators relates to the relative importance of persistent and transitory components in explaining overall TFP variation. Given estimates of the production function parameters, one can recover an estimate of TFP as
\begin{equation*}
	\hat{a}^{m}_{it}=y_{it}-\hat{\beta}^{mj}_ll_{it}-\hat{\beta}^{mj}_kk_{it},
\end{equation*}
where superscripts $m$ and $j$ on the $\hat{\beta}$ estimates refer to estimator and industry, respectively. TFP estimates constructed in this manner comprise both the persistent $\omega$ component and the transitory $\epsilon$ component. While OLS estimates do not allow one to decompose the overall residual into these separate elements, both the control function estimators and NPR do, albeit through different mechanisms. The residual from the first-stage of the OP/LP/ACF control functions acts as an estimate of the $\epsilon$ term, which can then be subtracted from the overall TFP residual to recover $\omega$. For NPR, the estimated nonparametric $\Psi$ function combined with firms' subjective expectations yields an estimate of $\omega$, which can then be subtracted from the overall TFP estimate to recover an estimate of $\epsilon$. 

This decomposition is informative because the persistent and transitory components of TFP have distinct economic interpretations and policy implications. The persistent component $\omega$ reflects durable firm-level characteristics such as managerial quality or technology, whereas the transitory component $\epsilon$ captures measurement error and short-lived shocks. The relative importance of these two components in explaining overall TFP dispersion therefore has direct bearing on how one interprets cross-firm productivity differences and the extent to which they are amenable to policy intervention. A higher share of persistent productivity in the overall variance of TFP, for example, would imply greater scope for policies that target structural firm capabilities, whereas a dominant transitory component would suggest that much of the observed cross-firm productivity dispersion is ephemeral.

Since the NPR decomposition relies on firms' subjective expectations, we implement the TFP decomposition using the MES data from 2017 and 2020. After obtaining separate estimates of the persistent component of TFP ($\omega$) and the transitory component ($\epsilon$), we compare their variances to that of overall TFP to quantify the extent to which productivity dispersion reflects persistent rather than transitory differences.

Table \ref{tab:omega_decomp} summarizes the results of this decomposition. Across all three industries, TFP estimates obtained using NPR exhibit similar total variance to both OLS and ACF, but the share of this variance attributable to persistent productivity is substantially higher under NPR. In the electronics industry, the NPR estimates imply that 64\% of TFP variance is due to persistent productivity, compared to 43\% under ACF. The differences are even more pronounced in the service sectors: in retail, NPR attributes 77\% of TFP variance to persistent productivity versus 49\% from ACF, while in restaurants the corresponding figures are 52\% and 33\%. The OP and LP estimators present a less consistent picture across industries. In some cases they attribute a similar share of total TFP variance to persistent productivity as NPR, but alongside substantially greater total TFP variance, which likely reflects the lower production function coefficient estimates these estimators return.

\begin{table}[H]
        \centering
       \caption{TFP Variance Decomposition}
       \label{tab:omega_decomp}
       \begin{adjustbox}{max width=\textwidth}
            \begin{tabular}{l|ccccc}
\hline\hline
&OLS&NPR&ACF&OP&LP\\
\hline
&\multicolumn{5}{c}{\textbf{(A) Electronics}}\\
\hline
TFP Variance&0.218&0.240&0.231&0.264&0.411\\
Omega Variance&--&0.153&0.100&0.084&0.263\\
Omega Variance Share&--&0.638&0.431&0.319&0.640\\
\hline
N Obs. &424&424&424&424&424\\
N Firms &374&374&374&374&374\\
\hline
&\multicolumn{5}{c}{\textbf{(B) Retail}}\\
\hline
TFP Variance&0.807&0.882&0.859&1.155&1.675\\
Omega Variance&--&0.679&0.419&0.390&1.209\\
Omega Variance Share&--&0.769&0.488&0.338&0.722\\
\hline
N Obs. &1834&1834&1834&1834&1834\\
N Firms &1633&1633&1633&1633&1633\\
\hline
&\multicolumn{5}{c}{\textbf{(C) Restaurants}}\\
\hline
TFP Variance&0.213&0.259&0.239&0.387&0.378\\
Omega Variance&--&0.134&0.080&0.203&0.193\\
Omega Variance Share&--&0.518&0.334&0.523&0.511\\
\hline
N Obs. &363&363&363&363&363\\
N Firms &338&338&338&338&338\\
\hline\hline\end{tabular}
        \end{adjustbox}
        \caption*{\footnotesize{Note: The table shows variances of TFP and its persistent component in levels and ratio. TFP variances are shown for different industries and estimators, as indicated in panel and column titles, respectively.}}
\end{table}

In summary, the variance decomposition analysis shows that the NPR estimator attributes a larger share of TFP dispersion to persistent productivity than conventional control function methods. If this decomposition is accurate, existing methods may overstate the role of transitory shocks in driving cross-firm productivity variation and correspondingly understate the contribution of persistent structural factors. This finding has implications for productivity analysis more broadly, as it suggests that the degree of persistent productivity dispersion across firms -- the component most relevant for understanding long-run differences in firm performance -- may be larger than estimates based on conventional proxy variable methods would indicate.

\section{Conclusion}\label{sect:conclusion}

In this paper we have proposed a new production function estimation methodology that leverages data on firms' observable expectations, data which are becoming increasingly available across a range of countries. We show that such information enables one to relax assumptions on firms' input choices (strict monotonicity and invertibility with respect to a single scalar unobservable productivity term), which underpin currently-used proxy variable approaches in the \citet{olley_dynamics_1996} tradition. Moreover, our method can be implemented on a single cross-section of data. This is an attractive feature relative to both proxy variable methods, which require two or more periods, and dynamic panel methods, such as those of \citet{blundell_gmm_2000}, which assume linear productivity dynamics and require three or four consecutive time period observations per firm.  

We present Monte Carlo simulations similar to those in \citet{ackerberg_identification_2015} featuring forward-looking firms with heterogeneous quadratic adjustment costs in capital. These show that our proposed NPR estimator is robust to optimization error in firms' input choices, whereas other methods are biased when firms make optimization errors in their material input choices. We also demonstrate that the NPR estimator can be extended to accommodate certain forms of bias in firms' expectations.

Implementing our proposed estimator on UK data, we show it recovers parameter estimates that are either comparable to or more credible than those recovered by conventionally-used estimators. Decomposing TFP into its persistent and transitory components, we find that NPR consistently attributes a substantially larger share of productivity dispersion to persistent factors than the ACF estimator. This suggests that existing approaches may understate the degree of structural productivity differences across firms.

\printbibliography

\newpage

\appendix 

\noindent \LARGE \textbf{Appendices}

\normalsize 

\section{Expectation Formation Processes}\label{adix:expectations}

As noted in the main text, we mirror the notation used in \cite{pesaranweale2006} and use the operator $\mathbb{E}_{it}[ \cdot | \Omega_{it} ]$ to denote a firm's expectation given its information set when applied to output $y_{it}$ and inputs (e.g., $l_{it}$): $\mathbb{E}_{it}[ \mathbf{x}_{it+1} | \Omega_{it} ] = \int \mathbf{x}_{t+1} h_{it}(\mathbf{x}_{t+1}| \Omega_{it})d\mathbf{x}_{t+1}$, where $\mathbf{x}_{t+1}$ is a vector and $h_{it}(\cdot)$ represents firm $i$'s distribution for it in period $t$ given their information $\Omega_{it}$.   As long as they satisfy the identification conditions in Theorem \ref{thm:ident} and Corollary \ref{cor:ident} (see below), we do not take a stand on the expectation formation process for output or inputs and treat those as data.  One may nonetheless envision different scenarios on the expectation process for productivity.   Pesaran and Weale discuss several expectation formation protocols (see their Section 2) and we borrow from their discussion here.   
\newline \newline
\textbf{Rational Expectations:} This case corresponds to: $\mathbb{E}_{it}[ \omega_{it} | \Omega_{it-1} ] = \mathbb{E}[\omega_{it}|\Omega_{it-1}]$, which is used to obtain equation (\ref{eq:eprodfunct_general}) in the main text:
\begin{equation*}
\begin{split}
\mathbb{E}_{it}[y_{it+1}|\Omega_{it}] & =\int f(k_{it+1},l_{it+1}; \beta)dF_{it}(l_{it+1})+\mathbb{E}_{it}[\omega_{it+1}|\Omega_{it}]+\mathbb{E}_{it}[\epsilon_{it+1}|\Omega_{it}]\\
& =\int f(k_{it+1},l_{it+1}; \beta)dF_{it}(l_{it+1})+g(\omega_{it}), \\
\end{split}
\end{equation*}
where $F_{it}(l_{it+1})$ represents firm $i$'s subjective probability distribution over their next-period labor input given its information $\Omega_{it}$ and the second equality follows from the assumptions that $\mathbb{E}_{it}[\epsilon_{it}|\Omega_{it-1}]=\mathbb{E}[\epsilon_{it}|\Omega_{it-1}]$ and $\mathbb{E}_{it}[\xi_{it}|\Omega_{it-1}]=\mathbb{E}[\xi_{it}|\Omega_{it-1}]$ are equal to zero.
\newline \newline
\textbf{Static Expectations:} This is the simplest form of expectation formation protocol dating back to \citet{keynes1936} and amounts to $\mathbb{E}_{it}[ \mathbf{x}_{it+1} | \Omega_{it} ] = \mathbf{x}_{it}$.  One can envision then a process by which $\mathbb{E}_{it}[\omega_{it+1}|\Omega_{it}] = \tilde g(\omega_{it})$, where $\tilde g(\cdot)$ is potentially different from $g(\cdot)$.  In this case,
\begin{equation*}
\begin{split}
\mathbb{E}_{it}[y_{it+1}|\Omega_{it}] & =\int f(k_{it+1},l_{it+1}; \beta)dF_{it}(l_{it+1})+\mathbb{E}_{it}[\omega_{it+1}|\Omega_{it}]+\mathbb{E}_{it}[\epsilon_{it+1}|\Omega_{it}]\\
& =\int f(k_{it+1},l_{it+1}; \beta)dF_{it}(l_{it+1})+\tilde g(\omega_{it}), \\
\end{split}
\end{equation*}
where once again we assume that $\mathbb{E}_{it}[\epsilon_{it}|\Omega_{it-1}]=0$, which seems reasonable since, by definition, $\epsilon_{it}$ is either measurement error or unanticipated by the firm.  In this case,  as long as $\tilde g(\cdot)$ is invertible, $\Psi(\cdot)$ is not $g^{-1}(\cdot)$ but $\tilde g^{-1}(\cdot)$.
\newline \newline
\textbf{Return to normality models:}  \citet{pesaranweale2006} discuss this case in terms of the average or consensus expectation.  For an individual firm, expectations are presumed to follow: $\mathbb{E}_{it}[ \mathbf{x}_{it+1} | \Omega_{it} ] = (\mathbf{I} - \mathbf{\Lambda}_i) \mathbf{x}_{it} + \mathbf{\Lambda}_i \mathbf{x}^*_{it}$ where $\mathbf{x}^*_{it}$ represents the ``normal'' or ``long run equilibrium'' level of $\mathbf{x}_{it}$.  Here,  $\mathbb{E}_{it}[\omega_{it+1}|\Omega_{it}] = (1-\lambda) \omega_{it} + \lambda \omega^*_i$ where we impose homogeneity in $\lambda$ across firms and a time-invariant normal or long run equilibrium level.   Then,
\begin{equation*}
\begin{split}
\mathbb{E}_{it}[y_{it+1}|\Omega_{it}] & =\int f(k_{it+1},l_{it+1}; \beta)dF_{it}(l_{it+1})+\mathbb{E}_{it}[\omega_{it+1}|\Omega_{it}]+\mathbb{E}_{it}[\epsilon_{it+1}|\Omega_{it}]\\
& =\int f(k_{it+1},l_{it+1}; \beta)dF_{it}(l_{it+1})+(1-\lambda) \omega_{it} + \lambda \omega^*_i, \\
\end{split}
\end{equation*}
where once again we assume that $\mathbb{E}_{it}[\epsilon_{it}|\Omega_{it-1}]=0$.   In this case,  one ends up with $\Psi(x) = x/(1-\lambda)$ and $-\lambda \omega^*_i / (1-\lambda)$ in equation (\ref{eq:prodfunct_cd}).  If $-\lambda \omega^*_i / (1-\lambda)$ is homogeneous across firms within an industry, then an industry-specific estimate as we perform here will be adequate. If it is a firm ``fixed effect'' then we need a panel as discussed below in Appendix \ref{adix:bias} (Case 2).
\newline \newline
\textbf{Adaptive Expectations:} This case amounts to $\mathbb{E}_{it}[ \mathbf{x}_{it+1} | \Omega_{it} ] = \mathbb{E}_{it-1}[ \mathbf{x}_{it} | \Omega_{it-1} ] + \mathbf{\Gamma}_i ( \mathbf{x}_{it} - \mathbb{E}_{it-1}[ \mathbf{x}_{it} | \Omega_{it-1} ])$ (once again specialising this to the individual firm) and the expectations can also be represented as an infinite distributed lag process.  For a unidimensional case considering $\omega_{it}$, this means that $\mathbb{E}_{it}[\omega_{it+1}|\Omega_{it}] = \gamma_i \omega_{it} + (1-\gamma_i) \mathbb{E}_{it-1}[\omega_{it}|\Omega_{it-1}]=\gamma_i [ \omega_{it}+(1-\gamma_i) \omega_{it-1} + (1-\gamma_i)^2 \omega_{it-2} + \dots ]$.   Assuming that $\gamma_i = \gamma$ for any $i$, we then have that
\begin{equation}\label{eq:Ey_infinitesum}
\begin{split}
\mathbb{E}_{it}[y_{it+1}|\Omega_{it}] & =\int f(k_{it+1},l_{it+1}; \beta)dF_{it}(l_{it+1})+\mathbb{E}_{it}[\omega_{it+1}|\Omega_{it}]+\mathbb{E}_{it}[\epsilon_{it+1}|\Omega_{it}]\\
& =\int f(k_{it+1},l_{it+1}; \beta)dF_{it}(l_{it+1})+\gamma [ \omega_{it}+(1-\gamma) \omega_{it-1} + (1-\gamma)^2 \omega_{it-2} + \dots ]. \\
\end{split}
\end{equation}
While this breaks the Markov assumption we rely on, we can adapt usual protocols in the estimation of geometric (Koyck) models.  Namely, equation \eqref{eq:Ey_infinitesum} implies that
$$\gamma [ (1-\gamma) \omega_{it-1} + (1-\gamma)^2 \omega_{it-2} + \dots] = (1-\gamma) \{ \mathbb{E}_{it-1}[y_{it}|\Omega_{it-1}]-\int f(k_{it},l_{it}; \beta)dF_{it-1}(l_{it}) \}.$$
Consequently, we obtain that
\begin{eqnarray*}
\mathbb{E}_{it}[y_{it+1}|\Omega_{it}] & = & \int f(k_{it+1},l_{it+1}; \beta)dF_{it}(l_{it+1}) \\
 & & +(1-\gamma) \{ \mathbb{E}_{it-1}[y_{it}|\Omega_{it-1}]-\int f(k_{it},l_{it}; \beta)dF_{it-1}(l_{it}) \}+ \gamma \omega_{it}. 
\end{eqnarray*}
Thus, we get that $\Psi(x) = x/\gamma$ and it takes as argument the original expression minus $(1-\gamma) \{ \mathbb{E}_{it-1}[y_{it}|\Omega_{it-1}]-\int f(k_{it},l_{it}; \beta)dF_{it-1}(l_{it}) \}$ in equation (\ref{eq:prodfunct_cd}), for example.  Notice that this would require that observations on expectations be collected in $t$ and $t-1$.
\newline \newline
\textbf{Conclusions on Expectations Processes:} The discussion here shows that NPR (or extensions of it) can handle a wide range of canonical expectations processes. Although we have developed our formal proofs for NPR under rational expectations, it encompasses a wider class of firm forecasting protocols.

\section{Convergence of the NPR Estimator} \label{adix:convergence}

The demonstration here follows Theorem 2 in \citet{dominitzsherman2005} (DS), and we adopt the notation in Theorem 1 and Corollary 1 of our paper. Suppose that $f(\cdot)$ is linear and focus on a cross-section for simplicity:
\begin{equation} \label{eq:dgp}
y_{i} = x_{i}^\top \beta_0 + h_0(z_{i})+\epsilon_{i} = x_{i}^\top \beta_0 + \Psi_0(z_{1i}-z_{2i}^\top \beta_0)+\epsilon_{i},
\end{equation}
where $x_i \in \mathbb{R}^k$ (e.g., $(\ln k_i, \ln l_i) \in \mathbb{R}^2$), $z_i = [z_{1i} ~ z_{2i}^\top]^\top$ with $z_{1i} \in \mathbb{R}$ (i.e., expected log-production next period) and $z_{2i} \in \mathbb{R}^k$ (i.e.,  log-capital and expected log-labor next period).  In what follows, we let $\mathcal{N} \subset \mathbb{R}^k$ denote the closure of an open convex neighbourhood of $\beta_0$. 

We define two objects as in DS.  For $\phi \in \mathbb{R}^k$, let
\begin{equation} \label{eq:Mn} 
\mathcal{M}_n(\phi) = \arg \min_\beta \sum_{i=1}^n (y_{i}(\phi) - x_{i}^\top \beta)^2 = (\mathbf{X}^\top \mathbf{X})^{-1}  \mathbf{X}^\top \mathbf{Y}(\phi) \in \mathbb{R}^k,
\end{equation}
where $y_i(\phi) \equiv y_{i} - \Psi_0(z_{1i}-\phi^\top z_{2i}) - \epsilon_{i}$, $\mathbf{X}_{n \times k} = [ x_1 ~  \cdots ~ x_n ]^\top$ and $\mathbf{Y}(\phi)_{n \times 1} = [ y_1(\phi) \dots y_n(\phi)]^\top.$  Note that $\mathcal{M}_n(\beta_0)=(\mathbf{X}^\top \mathbf{X})^{-1}  \mathbf{X}^\top \mathbf{Y}(\beta_0)=(\mathbf{X}^\top \mathbf{X})^{-1}  \mathbf{X}^\top \mathbf{X} \beta_0 = \beta_0$.   We later also use the matrix $\mathbf{Z}_{2 ~ (n \times k)} = [ z_{21} ~  \cdots ~ z_{2n} ]^\top$.  Furthermore, let
\begin{equation} \label{eq:Mnhat} 
\widehat{\mathcal{M}}_n(\phi) = \arg \min_\beta \sum_{i=1}^n (\hat y_{i}(\phi) - x_{i}^\top \beta)^2 = (\mathbf{X}^\top \mathbf{X})^{-1}  \mathbf{X}^\top \widehat{\mathbf{Y}}(\phi)  \in \mathbb{R}^k,
\end{equation}
where $\hat y_{i}(\phi) \equiv y_i - \hat  \Psi(z_{1i}-\phi^\top z_{2i})$ and $\widehat{\mathbf{Y}}(\phi)_{n \times 1} = [ \hat y_{1}(\phi) ~ \cdots ~ \hat y_{n}(\phi)]^\top$.   Here, $\hat  \Psi(\cdot)$ is an estimate for $\Psi_0(\cdot)$,  which is constructed using splines in our application.  $\mathcal{M}_n(\cdot)$ and $\widehat{\mathcal{M}}_n(\cdot)$ relate to the steps in our algorithm.  As noted in DS, for $i \ge 1$,  $\hat \beta_i = \widehat{\mathcal{M}}_n(\hat \beta_{i-1})$ and $\beta_i = \mathcal{M}_n(\beta_{i-1})$ correspond to sample and population iterates in the iterative protocol.  

Following the notation in DS, we also define $\nabla_\phi$ as the differential operator with respect to $\phi$.  Consequently, $\nabla_\phi \mathcal{M}_n(\phi)_{k \times k}$ is the matrix of first-order derivatives for $\mathcal{M}_n(\phi)$ and similarly for $\nabla_\phi \widehat{\mathcal{M}}_n(\phi)_{k \times k}$.  In particular, notice that $\nabla_\phi \mathcal{M}_n(\phi)_{k \times k} = (\mathbf{X}^\top \mathbf{X})^{-1}  \mathbf{X}^\top \nabla_\phi \mathbf{Y}(\phi)$ and $\nabla_\phi \widehat{\mathcal{M}}_n(\phi)_{k \times k} = (\mathbf{X}^\top \mathbf{X})^{-1}  \mathbf{X}^\top \nabla_\phi \widehat{\mathbf{Y}}(\phi)$, where $\nabla_\phi \mathbf{Y}(\phi)_{n \times k}$ and $\nabla_\phi \widehat{\mathbf{Y}}(\phi)_{n \times k}$ are derivative matrices for $\mathbf{Y}(\phi)$ and $\widehat{\mathbf{Y}}(\phi)$, respectively.

To demonstrate convergence for our protocol, we use the following \emph{sufficient} conditions:

\begin{ass}[DGP and Sampling] \label{ass:dgp}
The data is generated according to Equation (\ref{eq:dgp}), where 
\begin{enumerate}
\item[i.] $x_i, z_i \indep \epsilon_i$; 
\item[ii.] $\mathbb{E}[x_i x_i^\top]$ is non-singular and $\sum_{i=1}^n | x_{ik_1} z_{2i,k_2} z_{2i,k_3} | / n = O_p(1)$;
\item[iii.] $(y_i,x_i,z_i)_{i=1}^n$ is i.i.d.;
\item[iv.] $\beta_0$ belongs to a compact subset of $\mathbb{R}^k$; and
\item[v.] $\Psi_0(\cdot) \in C_2$ with bounded second-order derivative.
\end{enumerate}
\end{ass}

In the above, $x_{ik_1}$ is the $k_1$th entry in the vector $x_i$ and $z_{2i,k_2}$ and $z_{2i,k_3}$ are the $k_2$th and $k_3$th entries in the vector $z_{2i}$ where $1 \le k_1, k_2, k_3 \le k$.  Assumption 1 (i), (ii), (iii) and (iv) are conventional.  We suppose above that $\Psi_0(\cdot)$ is sufficiently smooth (i.e.,  twice continuously differentiable) with a bounded second-order derivative in Assumption 1 (v).  If, for example, the law of motion for $\omega$ is a linear autoregressive process with coefficient $\rho_0$, $\Psi_0: x \mapsto x/\rho_0$ and the assumption holds.   In addition, we assume:

\begin{ass}[Estimation Protocol]  \label{ass:estimation} The estimation is such that
\begin{enumerate}
\item[i.] $\mathbb{P}( \sigma_{\textrm{max}}(\nabla_\phi \mathcal{M}_n(\beta_0)) < 1 ) \rightarrow 1$; and
\item [ii.]$\hat \Psi(\cdot)$ converges to $\Psi_0(\cdot)$ uniformly in probability.
\end{enumerate}
\end{ass}

Assumption 2 (i) requires that the maximum singular value for $\nabla_\phi \mathcal{M}_n(\beta_0)$ be strictly less than 1 with probability approaching one as the sample since increases.   If the law of motion for $\omega$ is a linear autoregressive process with coefficient $\rho_0$,  the assumption is satisfied if $\sigma_{\textrm{max}}(\nabla_\phi \mathcal{M}_n(\beta_0)) = \sigma_{\textrm{max}}( (\mathbf{X}^\top \mathbf{X})^{-1}  \mathbf{X}^\top \mathbf{Z}_2 )/\rho_0 < 1$ with probability approaching 1 when $n \rightarrow \infty$.   Since the maximum singular value of a matrix is a continuous function of the matrix
,  by the Continuous Mapping Theorem the assumption is satisfied as long as $\sigma_{\textrm{max}}( \mathbb{E}[ x_i x_i^\top ]^{-1} \mathbb{E}[ x_i z_{2i}^\top ] )/\rho_0 < 1$.   For more general laws of motion, one can in principle enforce this latter assumption on the estimation by constraining $\sup |\widehat \Psi'(\cdot)| < \sigma_{\textrm{max}}( (\mathbf{X}^\top \mathbf{X})^{-1}  \mathbf{X}^\top)^{-1}  \sigma_{\textrm{max}}(\mathbf{Z}_2 )^{-1}$.\footnote{Notice that $\nabla_\phi \mathbf{Y}(\phi)$ defined in the proof for Lemma 1 is $\textrm{diag}(\Psi'(\mathbf{Z})) \mathbf{Z}_2$ where $\Psi'(\mathbf{Z})_{n \times 1} = [ \Psi'(z_{11}-\phi^\top z_{21}) ~ \cdots ~  z_{1n}-\phi^\top \Psi'(z_{2n}) ]^\top$.  Then,  $\sigma_{\textrm{max}}(\nabla_\phi \mathcal{M}_n(\beta_0)) = \sigma_{\textrm{max}}((\mathbf{X}^\top \mathbf{X})^{-1}  \mathbf{X}^\top \textrm{diag}(\Psi'(\mathbf{Z})) \mathbf{Z}_2 )$.  By Theorem 3.3.4 in Horn and Johnson (Topics in Matrix Analysis),  $\sigma_{\textrm{max}}((\mathbf{X}^\top \mathbf{X})^{-1}  \mathbf{X}^\top \textrm{diag}(\Psi'(\mathbf{Z})) \mathbf{Z}_2 ) \le \sigma_{\textrm{max}}((\mathbf{X}^\top \mathbf{X})^{-1}  \mathbf{X}^\top ) \sigma_{\textrm{max}}(\textrm{diag}(\Psi'(\mathbf{Z}))) \sigma_{\textrm{max}}( \mathbf{Z}_2 ) = \max_i | \Psi'(z_{1i}-\phi^\top z_{2i})| \times  \sigma_{\textrm{max}}((\mathbf{X}^\top \mathbf{X})^{-1}  \mathbf{X}^\top )\sigma_{\textrm{max}}( \mathbf{Z}_2 ) \le \sup | \Psi'(\cdot) |  \times \sigma_{\textrm{max}}((\mathbf{X}^\top \mathbf{X})^{-1}  \mathbf{X}^\top )\sigma_{\textrm{max}}( \mathbf{Z}_2 )$.}

Assumption 2 (ii) in turn requires that  $\hat \Psi(\cdot)$ converge to $\Psi_0(\cdot)$ uniformly in probability.  Conditions for uniform consistency for (non-penalized) splines can be found, for example, in \cite{horowitzmammen2004} (see Theorem 1c) and for penalized splines in  \cite{wangshenruppert2011} or  \cite{xiao2019} (see Section 6).  For kernel methods, results can be found, for example, in \cite{bierens1983}.

As previously defined, let $\hat \beta_i = \widehat{\mathcal{M}}_n(\hat \beta_{i-1})$ denote the sample iterate in the iterative protocol.  As in DS, we allow $i$ to depend on $n$ and denote this by the function $i(n)$ mapping $n \in \mathbb{Z}^+$ into $i \in \mathbb{Z}^+$.  Below, we use $\| \cdot \|$ to denote the Euclidean norm.  We can then state the following convergence result:

\begin{prop}\label{prop:convergence}
Suppose $i(n) \rightarrow \infty$ as $n \rightarrow \infty$.  If Assumptions \ref{ass:dgp} and \ref{ass:estimation} hold, then $\| \hat \beta_{i(n)} - \beta_0 \| = o_p(1)$ as $n \rightarrow \infty$.
\end{prop}

This result follows from Theorem 2 in DS.  This theorem relies on three conditions.  The first condition requires that $\mathcal{M}_n(\phi)$ be an Asymptotic Contraction Mapping (ACM), which we define below, and the third condition requires that $\| \widehat{\mathcal{M}}_n(\phi) - \mathcal{M}_n(\phi) \|$ converge to zero uniformly in probability.  The second condition is implied by the first one if $i(n) \rightarrow \infty$ as $n \rightarrow \infty$.   In order to demonstrate convergence, we thus establish two lemmas.  Lemma \ref{lem:acm} obtains that $\mathcal{M}_n(\phi)$ is an Asymptotic Contraction Mapping (ACM) and Lemma \ref{lem:unifconv} obtains that $\| \widehat{\mathcal{M}}_n(\phi) - \mathcal{M}_n(\phi) \|$ converges to zero uniformly in probability.  Proposition \ref{prop:convergence} thus follows.

The following notation is used below.  For a matrix $A_{k \times k}$, let $\| A \|_2$ denote the matrix norm induced by the Euclidean norm.   It coincides with the Euclidean norm for vectors (which we denote by $\| \cdot \|$) and is also known as the operator norm or the spectral norm since $\| A \|_2 = \sigma_{\textrm{max}}(A)$, where $ \sigma_{\textrm{max}}(A)$ is the largest singular value for $A$.  For Lemma \ref{lem:acm}, we define an Asymptotic Contraction Mapping (ACM) as (see DS, p.841):

\begin{definition}[ACM] 
Let $(\Omega, \mathcal{A}, \mathbb{P})$ be a probability space.  For $\omega \in \Omega$, let $W_1(\omega,\beta_0),\dots,$ $W_n(\omega,\beta_0)$ denote a sample of size $n$ from $\mathbb{P}$, where $\beta_0 \in \mathbb{R}^k$ denotes a fixed parameter of interest.  For each $n \ge 1$ and $\omega \in \Omega$, let $K^\omega(\cdot)$ be a function defined on a set $\mathcal{X}$ where $(\mathcal{X},d)$ is a metric space.  The collection $\{ K^\omega_n(\cdot) : n \ge 1, \omega \in \Omega \}$ is an asymptotic contraction mapping (ACM) on $(\mathcal{X},d)$ if there exists a constant $c \in [0,1)$ that does not depend on $n$ or $\omega$, and sets $\{ A_n \}$ with $A_n \subset \Omega$ and $\mathbb{P}(A_n) \rightarrow 1$ as $n \rightarrow \infty$, such that for each $\omega \in A_n$, $K^\omega_n(\cdot)$ maps $\mathcal{X}$ to itself and for all $x, y \in \mathcal{X}$, $d(K^\omega_n(x),K^\omega_n(y)) \le c \times d(x,y)$.
\end{definition}

We can now establish that:

\begin{lem} \label{lem:acm}
If Assumptions \ref{ass:dgp} and \ref{ass:estimation}(i) hold, there exists an open ball centered at $\beta_0$ with closure $B_0 \subset \mathcal{N}$ such that $\{ \mathcal{M}_n(\phi) \}$ is an ACM on $(B_0, \| \cdot \|)$.
\end{lem}

\begin{proof}
Take $\phi$ and $\gamma \in B_0 \subset \mathcal{N}$.  By the Mean Value Theorem we have:
$$\mathcal{M}_n(\phi)  = \mathcal{M}_n(\gamma) + \nabla_\phi \mathcal{M}_n(\phi^*) (\phi - \gamma),$$
where $\phi^*$ is (componentwise) between $\phi$ and $\gamma$.  Recall that
$$\nabla_\phi \mathcal{M}_n(\phi) = (\mathbf{X}^\top \mathbf{X})^{-1}  \mathbf{X}^\top \nabla_\phi \mathbf{Y}(\phi),$$
where $ \nabla_\phi \mathbf{Y}(\phi)_{n \times k} = [ \nabla_\phi y_1(\phi) \dots \nabla_\phi y_n(\phi)]^\top$ with $\nabla_\phi y_i(\phi) \equiv \Psi_0'(z_{1i}-\phi^\top z_{2i})z_{2i}$.   Then, adding and subtracting $(\nabla_\phi \mathcal{M}_n(\beta_0)-\mathbb{E}[x_i x_i^\top]^{-1}\mathbb{E}[x_i \nabla y_i(\beta_0)^\top] + \mathbb{E}[x_i x_i^\top]^{-1}\mathbb{E}[x_i \nabla y_i(\phi^*)^\top])$ $\times (\phi - \gamma)$, we can obtain:
\begin{equation} \label{eq:lem1}
\begin{array}{l}
\| \mathcal{M}_n(\phi)  - \mathcal{M}_n(\gamma) \| =  \| \nabla_\phi \mathcal{M}_n(\phi^*) (\phi - \gamma) \|
\le \| \nabla_\phi \mathcal{M}_n(\beta_0) (\phi - \gamma) \| \\ \qquad + \| \mathbb{E}[x_i x_i^\top]^{-1} \left\{ \mathbb{E}[x_i \nabla y_i(\phi^*)^\top]  -  \mathbb{E}[x_i \nabla y_i(\beta_0)^\top] \right\} (\phi - \gamma) \| + o_p( \| \phi - \gamma \| ),
\end{array}
\end{equation}
where $\| \cdot \|$ is the Euclidean norm.   

The $o_p$ term in (\ref{eq:lem1}) corresponds to $\| \{ - \left( \nabla_\phi \mathcal{M}_n(\beta_0) - \mathbb{E}[x_i x_i^\top]^{-1}\mathbb{E}[x_i \nabla y_i(\beta_0)^\top] \right) + \nabla_\phi \mathcal{M}_n(\phi^*)$ $- \mathbb{E}[x_i x_i^\top]^{-1}\mathbb{E}[x_i \nabla y_i(\phi^*)^\top] \} (\phi - \gamma) \|$.  This can be shown to converge to zero in probability if $\nabla_\phi \mathcal{M}_n(\cdot) - \mathbb{E}[x_i x_i^\top]^{-1}\mathbb{E}[x_i \nabla y_i(\cdot)^\top]$ converges uniformly to zero in probability.   This is established using Lemma 2.9 in \cite{neweymcfadden1994}.  

To employ this lemma, first note that the parameter space is compact by Assumption \ref{ass:dgp}.   Also by Assumption \ref{ass:dgp}, $\| \nabla_\phi \mathcal{M}_n(\phi) - \mathbb{E}[x_i x_i^\top]^{-1}\mathbb{E}[x_i \nabla y_i(\phi)^\top]  \|_2$ converges to zero in probability for given $\phi$ by the Law of Large Numbers and the Continuous Mapping Theorem.   Moreover,  we can also obtain that 
$$\| \nabla_\phi \mathcal{M}_n(\phi) - \nabla_\phi \mathcal{M}_n(\gamma) ]  \|_2 \le \hat B_n \| \phi - \gamma \|,$$
where $\hat B_n = O_p(1)$.  For notational simplicity,  assume that $k = 1$.  In this case, by the Mean Value Theorem,
\begin{eqnarray*}
\nabla_\phi \mathcal{M}_n(\phi) - \nabla_\phi \mathcal{M}_n(\gamma) & = & \left(\sum_{i=1}^n x_i^2\right)^{-1} \times \sum_{i=1}^n x_i z_{2i} \left( \Psi_0'(z_{1i}-\phi  z_{2i})-\Psi_0'(z_{1i}-\gamma  z_{2i}) \right) \\
 & = & - \left(\sum_{i=1}^n x_i^2\right)^{-1} \times \sum_{i=1}^n x_i z_{2i}^2 \Psi_0''(z_{1i}-\phi^*  z_{2i})(\phi-\gamma),
\end{eqnarray*}
where $\phi^*$ is (componentwise) between $\phi$ and $\gamma$.  By the triangle inequality and Assumption \ref{ass:dgp} (v) (i.e., $| \Psi_0''(\cdot) | \le c$ for some $c$), we have that 
\begin{eqnarray*}
| \nabla_\phi \mathcal{M}_n(\phi) - \nabla_\phi \mathcal{M}_n(\gamma) | & \le & \underbrace{c \left(\sum_{i=1}^n x_i^2 / n \right)^{-1} \times \sum_{i=1}^n \left| x_i z_{2i}^2 \right| / n }_{\equiv \hat B_n}  \times | \phi-\gamma |.
\end{eqnarray*}

Given Assumption \ref{ass:dgp} (ii),  $\hat B_n = O_p(1)$.  Finally, continuity of $ \Psi_0'(\cdot)$ implies continuity of $\mathbb{E}[x_i x_i^\top]^{-1} \times$ $\mathbb{E}[x_i \nabla y_i(\phi)^\top]$ for $\phi \in \mathcal{N}$.   These suffice for uniform convergence by Lemma 2.9 in \cite{neweymcfadden1994}.  

Since $\Psi_0(\cdot) \in C_2$,  $\mathbb{E}[x_i x_i^\top]^{-1} \mathbb{E}[x_i \nabla y_i(\cdot)^\top]$ is continuously differentiable, which allows one to control the second term in (\ref{eq:lem1}) by the diameter of $B_0$.  To see this, and again using $k=1$ for notational simplicity, note that
\begin{eqnarray*}
\left| \mathbb{E}[x_i \nabla y_i(\phi^*)^\top]  -  \mathbb{E}[x_i \nabla y_i(\beta_0)^\top] \right| & = & \left| \mathbb{E}[x_i z_{2i}^2 \Psi_0''(z_{1i}-\tilde \phi  z_{2i}) ] \right| \times |(\phi^*-\beta_0)| \\
 & \le & c  \times \mathbb{E}[ | x_i z_{2i}^2 | ] \times |(\phi^*-\beta_0)|. 
\end{eqnarray*}
where $\tilde \phi$ is a value between $\phi^*$ and $\beta_0$ and the second line uses the triangle inequality and Assumption \ref{ass:dgp} (v) (i.e., $| \Psi_0''(\cdot) | \le c$ for some $c$).  The second term in (\ref{eq:lem1}) can thus be seen to be $o(\sup_{\phi,\gamma \in B_0}| \phi - \gamma |)$.

Then, as noted in the proof for Lemma 5 in DS, it suffices to show that $\mathbb{P}(\| \nabla_\phi \mathcal{M}_n(\beta_0) (\phi - \gamma) \| < C \| (\phi - \gamma) \|) \rightarrow 1$ as $n \rightarrow \infty$, where $C$ is a positive scalar such that $C < 1$.  Since the matrix norm induced by the Euclidean distance is submultiplicative, we have that:
$$\| \nabla_\phi \mathcal{M}_n(\beta_0) (\phi - \gamma) \|  \le \| \nabla_\phi \mathcal{M}_n(\beta_0) \|_2 \times \| (\phi - \gamma) \| = \sigma_{\textrm{max}}(\nabla_\phi \mathcal{M}_n(\beta_0)) \times \| (\phi - \gamma) \|.$$
Since $\mathbb{P}( \sigma_{\textrm{max}}(\nabla_\phi \mathcal{M}_n(\beta_0)) < 1 ) \rightarrow 1$ as $n \rightarrow \infty$, we obtain the desired result.
\end{proof}

For the uniform convergence condition, we obtain that:

\begin{lem} \label{lem:unifconv}
If Assumptions \ref{ass:dgp} and \ref{ass:estimation}(ii) hold, $\sup_\phi \| \widehat{\mathcal{M}}_n(\phi) - \mathcal{M}_n(\phi) \| = o_p(1)$ as $n \rightarrow \infty$.
\end{lem}

\begin{proof}
First notice that 
\begin{eqnarray*}
y_i(\phi) & = & y_{i} - \Psi_0(z_{1i}-\phi^\top z_{2i}) - \epsilon_{i} \\
 & = & (x_{i}^\top \beta_0 +\Psi_0(z_{1i}-z_{2i}^\top \beta_0)+\epsilon_{i}) - \Psi_0(z_{1i}-\phi^\top z_{2i}) - \epsilon_{i} \\
 & = & x_{i}^\top \beta_0 + \left(\Psi_0(z_{1i}-z_{2i}^\top \beta_0) -  \Psi_0(z_{1i}-\phi^\top z_{2i})  \right)
\end{eqnarray*}
and
\begin{eqnarray*}
\hat y_{i}(\phi) & = & (x_{i}^\top \beta_0 +\Psi_0(z_{1i}-z_{2i}^\top \beta_0)+\epsilon_{i}) - \hat  \Psi(z_{1i}-\phi^\top z_{2i}) \\
 & = & x_{i}^\top \beta_0 + (\Psi_0(z_{1i}-z_{2i}^\top \beta_0)- \Psi_0(z_{1i}-\phi^\top z_{2i}))+\epsilon_{i} + ( \Psi_0(z_{1i}-\phi^\top z_{2i})- \hat  \Psi(z_{1i}-\phi^\top z_{2i})) \\
 & = & y_i(\phi)+\epsilon_{i} + ( \Psi_0(z_{1i}-\phi^\top z_{2i})- \hat  \Psi(z_{1i}-\phi^\top z_{2i})).
\end{eqnarray*}
Consequently, we can write: 
$$\widehat{\mathbf{Y}}(\phi)_{n \times 1} = \mathbf{Y}(\phi)_{n \times 1} + \mathbf{\epsilon}_{n \times 1} + \tilde{\Psi}(\phi,\mathbf{Z})_{n \times 1},$$
where $ \mathbf{\epsilon} = [\epsilon_1 ~ \cdots ~ \epsilon_n ]^\top$ and $\tilde{\Psi}(\phi,\mathbf{Z}) = [ ( \Psi_0(z_{11}-\phi^\top z_{21})- \hat \Psi(z_{11}-\phi^\top z_{21})) ~ \cdots ~ ( \Psi_0(z_{1n}-\phi^\top z_{2n})- \hat \Psi(z_{1n}-\phi^\top z_{2n})) ]^\top$.  ($\widehat{\mathbf{Y}}(\phi)$ and $\mathbf{Y}(\phi)$ were defined earlier.)  Using the definitions for $\mathcal{M}_n(\phi)$ and $\widehat{\mathcal{M}}_n(\phi)$ in equations (\ref{eq:Mn}) and (\ref{eq:Mnhat}), we then obtain:
$$
\widehat{\mathcal{M}}_n(\phi) - \mathcal{M}_n(\phi) =  \underbrace{(\mathbf{X}^\top \mathbf{X})^{-1}  \mathbf{X}^\top \mathbf{\epsilon}}_{\equiv A} + \underbrace{(\mathbf{X}^\top \mathbf{X})^{-1}  \mathbf{X}^\top \tilde{\Psi}(\phi,\mathbf{Z})}_{\equiv B(\phi)}.
$$
The first term above (A) does not depend on $\phi$ and converges in probability to 0 given Assumption \ref{ass:dgp}.  (This obtains from $A = (\sum_{i=1}^n x_i x_i^\top /n)^{-1} \times (\sum_{i=1}^n x_i \epsilon_i^\top /n) \rightarrow_p \mathbb{E}[x_i x_i^\top]^{-1}  \mathbb{E}[x_i \epsilon_i] = 0$.) 

For the second term,  let $B_1 \equiv \mathbf{X} (\mathbf{X}^\top \mathbf{X})^{-1} \sqrt{n}$ and $B_2(\phi) \equiv \tilde{\Psi}(\phi,\mathbf{Z}) / \sqrt{n}$ so that $B(\phi)_{k \times 1} = B_1^\top B_2(\phi)$.  Then,
it can be deduced that
$$\| B(\phi) \|^2 = \| B_1^\top B_2(\phi) \|^2 = (B_1(:,1)^\top B_2(\phi))^2 + \dots + (B_1(:,k)^\top B_2(\phi))^2,$$
where $B_1(:,j)$ is the $j$th column of $B_1$.  Applying the Cauchy-Schwarz inequality to each of the $k$ terms in the sum, we get:
\begin{eqnarray*}
 \| B(\phi) \|^2 & \le & \| B_1(:,1) \|^2 \| B_2(\phi) \|^2 + \dots + \| B_1(:,k) \|^2 \| B_2(\phi) \|^2 \\
 & = & (\| B_1(:,1) \|^2 + \dots + \| B_1(:,k) \|^2 ) \| B_2(\phi) \|^2.   
\end{eqnarray*}
Notice that $\| B_1(:,1) \|^2 + \dots + \| B_1(:,k) \|^2 = \| B_1 \|^2_F$, where $\| \cdot \|_F$ is the Frobenius norm.   Since $\| B_1 \|^2_F = tr( B_1^\top B_1 )$, where $tr(\cdot)$ is the trace operator, we have that
$$\| B_1 \|^2_F = tr( (\mathbf{X}^\top \mathbf{X})^{-1}  \mathbf{X}^\top \mathbf{X} (\mathbf{X}^\top \mathbf{X})^{-1} n ) = tr( (\mathbf{X}^\top \mathbf{X})^{-1} n ) = tr( (\sum_{i=1}^n x_i x_i^\top /n)^{-1} ).$$
Given that $\sum_{i=1}^n x_i x_i^\top /n \rightarrow_p \mathbb{E}[x_i x_i^\top]$, continuity of $tr(\cdot)$ and non-singularity of $\mathbb{E}[x_i x_i^\top]$, $\| B_1 \|^2_F \rightarrow_p tr(\mathbb{E}[x_i x_i^\top]^{-1})$.

In addition,  notice that: 
$$ \| B_2(\phi) \|^2 = B_2(\phi)^\top B_2(\phi) = \tilde{\Psi}(\phi,\mathbf{Z})^\top \tilde{\Psi}(\phi,\mathbf{Z}) / n = \sum_{i=1}^n \left(  \Psi_0(z_{1i}-\phi^\top z_{2i})- \hat  \Psi(z_{1i}-\phi^\top z_{2i}) \right)^2 / n.$$  Since $\left(  \Psi_0(z_{1i}-\phi^\top z_{2i})- \hat  \Psi(z_{1i}-\phi^\top z_{2i}) \right)^2 \le  \sup_\phi |  \Psi_0(z_{1i}-\phi^\top z_{2i})- \hat  \Psi(z_{1i}-\phi^\top z_{2i}) |^2 = \sup | \Psi_0(\cdot)- \hat \Psi(\cdot) |^2$,  $ \| B_2(\phi) \|^2 \le \sup | \Psi_0(\cdot)- \hat \Psi(\cdot) |^2$ which converges to zero in probability if $\hat \Psi(\cdot)$ converges to $\Psi_0(\cdot)$ uniformly in probability which holds by Assumption 2(ii).  Thus,  $ \| B_2(\phi) \|^2 \rightarrow_p 0$ and consequently
$$\| B(\phi) \|^2 \le \| B_1 \|^2_F \times \| B_2(\phi) \|^2 \rightarrow_p 0$$
uniformly in $\phi$.   
\end{proof}

\section{Implementing the NPR Estimator}\label{adix:implementation}

As explained in subsection \ref{subsect:estimation}, the NPR estimator (assuming a Cobb-Douglas production function) consists of the following steps:
\begin{enumerate}
    \item Pick initial parameter values $(\hat{\beta}_{k0},\hat{\beta}_{l0})$.
    \item\label{algo:repeat} For iteration $j$, calculate $Z_{it}=\mathbb{E}_{it}[y_{it+1}|\Omega_{it}] - \hat \beta_{kj-1}k_{it+1}-\hat \beta_{lj-1}\mathbb{E}_{it}[l_{it+1}|\Omega_{it}]$.
    \item Fit the model $y_{it}=\beta_kk_{it}+\beta_ll_{it}+\Psi\left(Z_{it}\right)+\epsilon_{it}$ using the shape constrained estimation protocol of \citet{pya_shape_2015} to obtain $(\hat{\beta}_{j},\hat{\Psi}_j)$.
    \item Calculate the Euclidean distance between $(\hat{\beta}_{kj},\hat{\beta}_{lj})$ and $(\hat{\beta}_{kj-1},\hat{\beta}_{lj-1})$. If the distance is below some tolerance level, stop and treat $\hat{\beta}_{j}$ as the model's parameter estimates. If not then update the iteration number $j\leftarrow j+1$ and repeat from step \ref{algo:repeat}.
\end{enumerate}

\noindent For more general production functions, one instead should use $Z_{it} = \mathbb{E}_{it}[y_{it+1}|\Omega_{it}] - \int f(k_{it+1},l_{it+1}; \hat \beta)dF_{it}(l_{it+1})$ in step 2 and  $y_{it}=f(k_{it},l_{it}; \beta)+\Psi\left(Z_{it}\right)+\epsilon_{it}$ in step 3.  

Implementing this iterative algorithm requires a number of decisions. First, one must decide on the initialization values $(\hat{\beta}_{k0},\hat{\beta}_{l0})$. \citet{ichimura_chapter_2007} recommend that, when fitting a generalized additive model, initial values be obtained from a linear regression of the outcome variable on all independent (explanatory) variables (i.e.,both those that enter the model linearly and those that are arguments of non-parametric smooth model components). In our context this would involve regressing $y_{it}$ on a constant, $k_{it},l_{it}$ and $Z_{it} \equiv \mathbb{E}_{it}[y_{it+1}|\Omega_{it}] - \beta_{k0}k_{it+1}-\beta_{l0}\mathbb{E}[l_{it+1}|\Omega_{it}]$, which nonetheless depends on the unknown parameters ($\beta_{k0}$ and $\beta_{l0}$).  One alternative is to instead regress $y_{it}$ on a constant and $X_{it} = \left[ k_{it},l_{it},\mathbb{E}_{it}[y_{it+1}|\Omega_{it}],k_{it+1},\mathbb{E}[l_{it+1}|\Omega_{it}] \right]$ and set $(\hat{\beta}_{k0},\hat{\beta}_{l0})$ as the coefficient estimates on current-period capital and labor respectively. However, in both Monte Carlo simulations and our empirical context, this approach frequently initialized the capital coefficient at a negative value due to high correlations between $y_{it}$ and $\mathbb{E}_{it}[y_{it+1}]$ and between $k_{it}$ and $k_{it+1}$. 

Rather than specify a rule for selecting a single initialization parameter vector, we therefore recommend a grid search approach. Under this method, one specifies a grid of $N$ initial values for $(\hat{\beta}_{k0},\hat{\beta}_{l0})$ and obtains $N$ parameter estimates by implementing the NPR estimator initialised at each point on the grid. The optimal parameter estimates are then chosen as those associated with the lowest least squares objective function from the shape constrained estimation protocol. All results in this paper were calculated using this initialization approach, with an additional non-negativity condition explained below, implemented on a 16-point initialization grid consisting of all combinations of $\hat{\beta}_{k0} \in [0.1,0.333,0.617,0.9]$ and $\hat{\beta}_{l0} \in [0.1,0.333,0.617,0.9]$. Experimentation showed our Monte Carlo and empirical results are generally robust to other grid choices. We recommend that users of the NPR estimator test the sensitivity of their results to the choice of initial values, just as one should do when using any estimator that relies on numerical optimization.

The selection of initial values of production function input coefficients is also required by the ACF estimator. To make our estimates comparable across ACF and NPR, we adopt the same approach as we use with NPR. This involves running ACF 16 times using the $\hat{\beta}_{k0} \in [0.1,0.333,0.617,0.9]$ and $\hat{\beta}_{l0} \in [0.1,0.333,0.617,0.9]$ initialization grids and storing the results that return the optimal ACF objective value among initializations. ACF note in their Monte Carlo results that they initialize at the true values ``to ease non-linear search''. They note their results are ``fairly robust to other nearby starting values (e.g.,OLS estimates), though with further away starting values, one can sometimes end up at the spurious minimum around ($\beta_k = 0, \beta_l = \beta_l+\beta_k$)'', as described in their footnote 16. We also noticed this problem in our Monte Carlo simulations of ACF, so when selecting the values that minimize the objective function, we report the best estimates where $\hat{\beta}_l>0$ and $\hat{\beta}_k>0$.  In other words, when the optimum is attained at negative estimates, we instead use the non-negative estimates that best fit the data.\footnote{ACF only report their Monte Carlo estimates of $\beta_k$ and $\beta_l$ that are bounded between zero and one, discarding Monte Carlo runs where this was not the case.} 

To make things comparable, we follow the same procedure with NPR, keeping only positive values of the input elasticities. We found that in all simulated DGPs where the assumptions for consistency of NPR hold, the optimal value of the objective function was always attained at positive values for the input coefficients. Hence, the restriction on positive values was unnecessary. But for DGPs where NPR assumptions did \emph{not} fully hold (e.g., when expectations are biased or measured with error), it was sometimes the case that the minimum of the objective function could return a negative capital coefficient. Hence, these parameter restrictions are likely useful in empirical applications.  As explained in Section \ref{sect:montecarlo}, the implementation of both LP and OP estimators used in our Monte Carlo simulations involves a one-dimensional search over the capital coefficient. To mirror the non-negativity constraint used for both NPR and ACF, we set the search grid of this optimization to $\left[0,2\right]$. 

We use the same initialization procedure for NPR in both our Monte Carlo simulations and our empirical analysis. For ACF, we found the grid search approach to initialization sometimes caused the ACF estimator to return implausible parameter estimates similar to the ``spurious minimum'' noted in ACF's footnote 16. The results reported for ACF in our empirical application therefore use the default initialization in the popular \texttt{prodest} package in \texttt{Stata} (\cite{Rovigatti}), which initializes at OLS estimates plus a small random disturbance. Using Nelder-Mead minimization, the default algorithm in the \texttt{prodest} package, yielded estimates that were almost unchanged from the initial values. Using the gradient-based BFGS minimization routine, however, achieved results that differed more from the OLS estimates and achieved a lower value of the objective function. The ACF, LP and OP results we report in our empirical application were therefore obtained using OLS initialization and BFGS minimization.

A second implementation decision regards the non-linear component of the estimator. The protocol outlined by \citet{pya_shape_2015} (PW) allows one to estimate a generalized additive model of the form 
\begin{equation}\label{eq:scam_1}
    y_i = \theta^\top x_i+\sum_jf_j(z_{ji})+\epsilon_i,
\end{equation}
where $y$ is a univariate response variable, $x$ is a vector of linear independent variables and $\theta$ a vector of unknown parameters. The non-linear part of the model is represented by unknown, monotonically increasing smooth predictor functions $f_j$, with predictor variable $z_j$. The variable $\epsilon$ is an unobserved mean-zero disturbance.\footnote{While we focus this description of \citet{pya_shape_2015} on our particular application for parsimony, it should be noted that their framework allows for more general models.} To estimate this model, PW follow convention in the literature on generalized additive model estimation and approximate the unknown smooth functions $f_j$ with penalized B-splines (i.e.,P-splines). Under this simplification, the model becomes 
\begin{equation}\label{eq:scam_2}
    y_i = \theta^\top x_i+\sum_{j=1}^q\gamma_jB_j(z_i)+\epsilon_i,
\end{equation}
where $q$ is the number of basis functions, $B_j$ are B-spline basis functions of at least second order that represent smooth functions over interval $[a,b]$ based on evenly-spaced knots and $\gamma_j$ are the corresponding spline coefficients. Ensuring the smooth part of the model is monotonically increasing amounts to imposing restrictions on the spline parameters $\gamma$ and PW show how the model in equation (\ref{eq:scam_2}) can be reparameterized to guarantee such restrictions are satisfied. Following this reparameterization, estimates of the original model parameters can be obtained via minimising the squared difference between the response variable and the observable components on the RHS of equation (\ref{eq:scam_2}). To avoid overfitting during this minimization, one may also include a penalty term that controls the `wiggliness' of the B-splines. Specifically, PW account for this concern by penalizing the squared difference between adjacent B-spline parameters. While it is possible to pre-specify the `smoothing parameter', which scales the `wiggliness' term in the minimization and thereby controls the smoothness of the estimated functions, PW propose to estimate the `optimal' smoothing parameter via an outer estimation algorithm. The outer part of the estimation uses the generalized cross validation prediction error criterion to evaluate the performance of the model estimated using a particular value of the smoothing parameter and finds the optimal smoothing parameter according to this criterion using the Newton-Raphson method. 

In our application of the PW estimator, we specify the smooth term $\Psi$ using monotone increasing penalized B-splines (i.e., P-splines, using the \texttt{`bs="mpi"'} option), with 10 basis functions. In practice, we found that the `optimal' smoothing parameter estimated using the PW approach was near-zero in all our focus industries. We therefore set it to zero, as pre-specification considerably reduces the computational burden of the estimator (and the bootstrap replications required for inference). After specifying these settings, coefficients are estimated using the BFGS algorithm.

\section{Extensions to the NPR Estimator}\label{adix:bias}

\subsection{Value Added Production Functions}\label{adix:prodfunct_va}

Adapting the NPR estimator to a value added production function is trivial if one has data on firms' expected year-ahead value added. In this case, one can simply change $y$ in the estimation algorithm from log sales to log value added and the resultant estimates will have a value added interpretation. If such data is unavailable, but one instead possesses data on firms' expected material inputs in addition to expected sales and employment, one can use copula methods to estimate expected log value added. Specifically, firms' subjective \emph{joint} distribution between expected sales and expected materials can be estimated by applying a parametric copula to firms' subjective marginal distributions.\footnote{One could, for example, fit the normal copula to match the observed correlation between observed sales and materials.} One can then take an adequately large number of draws of expected sales and intermediates from this joint subjective distribution, calculate $\text{ln}(sales-intermediates)$ for each draw and recover expected log value added as the mean over the random draws where this quantity is defined (i.e.,when $(sales-intermediates)$ is greater than zero). Equipped with an estimate of expected value added, one can again simply set $y$ as log value added rather than log sales and proceed with the iterative NPR estimation algorithm detailed in subsection \ref{subsect:estimation}. Although theoretically possible, we do not explore this extension in our empirical application because expectations of material inputs were only elicited in the 2017 wave of the MES, leading to small industry-specific samples.\footnote{Copula methods can also be employed in (non-linear) gross output production functions with both labor and intermediates where the joint distribution for both variables is required.}

\subsection{NPR Estimation with Measurement Error in Expectations Data (``EZ-NPR'')}\label{adix:EZNPR}

The mis-measurement of inputs is well known to be a common problem for production function estimation, and can in principle cause bias for proxy methods (e.g., \cite{collard-wexler_productivity_2021}) and other approaches (e.g., \cite{strom_production_1999}). A related concern for the NPR estimator is that firms' subjective expectations are elicited with error. 

As discussed in subsection \ref{subsect:measurement_error} of the main text, NPR may be vulnerable in the presence of measurement error in firms' subjective expectations data. To address this, we propose an alternative estimation strategy (EZ-NPR) that combines NPR with the insights of \cite{evdokimov_zeleneev_2025}. This is based on moments using the functions given by expression \eqref{eq:correctedmoment}, which we repeat here for reference:

\begin{equation*}
    M^x_{it}(Z_{it})-\frac{\partial^2 M^x_{it}(Z_{it})}{\partial \mathbb{E}_{it}[y_{it+1}]^2}\frac{\mathbb{E}[\upsilon_y^2]}{2}-\frac{\partial^2 M^x_{it}(Z_{it})}{\partial \mathbb{E}_{it}[l_{it+1}]^2}\frac{\mathbb{E}[\upsilon_l^2]}{2},
\end{equation*}
where
\begin{equation*}
    M^x_{it}(Z_{it})=\left(y_{it}-\beta_k k_{it}-\beta_l l_{it} -\Psi\left(Z_{it}\right)\right)x_{it},
\end{equation*}
and
\begin{equation*}
    Z_{it} = \mathbb{E}_{it}[y_{it+1}|\Omega_{it}] -\beta_k k_{it+1}-\beta_l \mathbb{E}_{it}[l_{it+1}|\Omega_{it}].
\end{equation*}

Below, we derive the corrected EZ-NPR moments and discuss how one can use them to recover parameter estimates using GMM.

\subsubsection{Deriving the corrected moment functions}
Constructing the corrected moments requires derivation and calculation of the second partial derivatives of $M^x_{it}$ with respect to $\mathbb{E}_{it}[y_{it+1}]$ and $\mathbb{E}_{it}[l_{it+1}]$. 

First, consider the case where the instrument $x_{it}$ does not depend on either of the variables measured with error (e.g.,$x_{it}\in\left\{1,l_{it},k_{it},k_{it+1}\right\}$). From equation \eqref{eq:correctedmoment} and the definition of $Z_{it}$, the product rule implies
\begin{equation*}
    \begin{split}
        \frac{\partial^2 M^x_{it}}{\partial \mathbb{E}_{it}[y_{it+1}]^2}&=-\Psi''(Z_{it})x_{it} \quad \textrm{and} \quad
        \frac{\partial^2 M^x_{it}}{\partial \mathbb{E}_{it}[l_{it+1}]^2}=-\beta_l^2\Psi''(Z_{it})x_{it},
    \end{split}
\end{equation*}
where $\Psi''(Z_{it})$ denotes the second derivative of the B-spline function with respect to the argument $Z_{it}$. 

Now, consider the case where the instrument $x_{it}$ \emph{does} depend on the variables measured with error (e.g.,$x_{it}=Z_{it}$). In this case we have:
\begin{equation*}
    \begin{split}
        \frac{\partial^2 M^x_{it}}{\partial \mathbb{E}_{it}[y_{it+1}]^2}&=-\Psi''(Z_{it})x_{it}-2\frac{\partial x_{it}}{\partial \mathbb{E}_{it}[y_{it+1}]}\Psi'(Z_{it}) +\frac{\partial^2 x_{it}}{\partial \mathbb{E}_{it}[y_{it+1}]^2}\left(y_{it} -\beta_kk_{it}-\beta_l l_{it}-\Psi(Z_{it})\right) \\
        \frac{\partial^2 M^x_{it}}{\partial \mathbb{E}_{it}[l_{it+1}]^2}&=-\beta_l^2\Psi''(Z_{it})x_{it}+2\beta_l\frac{\partial x_{it}}{\partial \mathbb{E}_{it}[l_{it+1}]}\Psi'(Z_{it}) +\frac{\partial^2 x_{it}}{\partial \mathbb{E}_{it}[l_{it+1}]^2}\left(y_{it} -\beta_kk_{it}-\beta_l l_{it}-\Psi(Z_{it})\right).
    \end{split}
\end{equation*}
For $x_{it}=Z_{it}$, $\partial x_{it}/\partial \mathbb{E}_{it}[y_{it+1}]=1$, $\partial x_{it}/\partial \mathbb{E}_{it}[l_{it+1}]=-\beta_l$ and the second-order partial derivatives $\partial^2 x_{it}/\partial \mathbb{E}_{it}[y_{it+1}]^2$ and $\partial^2 x_{it}/\partial \mathbb{E}_{it}[l_{it+1}]^2$ are both equal to zero. 

The precise form of the correction terms evidently depends on how one models $\Psi$. Following the baseline NPR implementation, we estimate $\Psi$ using B-splines with monotonicity enforced via coefficient ordering constraints
\begin{equation*}
    \Psi(Z_{it}) = c_0 + \sum_{j=1}^{K} c_j \, b_j(Z_{it}), \qquad c_1 \leq c_2 \leq \cdots \leq c_K,
\end{equation*}
where $b_1, \ldots, b_K$ are b-spline basis functions of degree $p$ and $c_0$ is an unrestricted intercept.\footnote{The ordering constraint $c_{j+1} \geq c_j$ is sufficient for monotonicity because the derivative of a degree-$p$ b-spline can be written as
\begin{equation*}
    \Psi'(Z_{it}) = p \sum_{j} \frac{c_{j+1} - c_j}{\tau_{j+p} - \tau_j} \tilde{b}_j(Z_{it}),
\end{equation*}
where $\tilde{b}_j$ are b-splines of degree $p-1$ (which are everywhere non-negative) and $\tau_j$ are the knot positions. Since $c_{j+1} - c_j \geq 0$ and $\tau_{j+p} - \tau_j > 0$, we have $\Psi'(Z_{it}) \geq 0$ everywhere.}

The first and second derivatives of $\Psi$ with respect to $Z_{it}$ are therefore
\begin{equation*}
         \Psi'(Z_{it})=\sum_{j=1}^{K}c_j \, b'_j(Z_{it}) \quad \textrm{and} \quad
         \Psi''(Z_{it})=\sum_{j=1}^{K}c_j \, b''_j(Z_{it}),
\end{equation*}
which can be computed directly from the derivatives of the B-spline basis functions.

Putting all these components together, the corrected moment functions, which we denote $\tilde{M}^x_{it}$ are therefore
\begin{equation}\label{eq:correctedmoments_specific1}
    \tilde{M}^x_{it}=\left(y_{it}-\beta_k k_{it}-\beta_l l_{it}-c_0 -\sum_{j=1}^{K}c_j \, b_j(Z_{it})\right)x_{it}+x_{it}\left(\frac{\mathbb{E}[\upsilon_y^2]}{2}+\beta_l^2 \frac{\mathbb{E}[\upsilon_l^2]}{2}\right)\sum_{j=1}^{K}c_j \, b''_j(Z_{it}), 
\end{equation}
for $x\in\left\{1, k_{it},l_{it},k_{it+1}\right\}$ and 

\begin{equation}\label{eq:correctedmoments_specific2}
\begin{split}
      \tilde{M}^x_{it}&=\left(y_{it}-\beta_k k_{it}-\beta_l l_{it}-c_0-\sum_{j=1}^{K}c_j \, b_j(Z_{it})\right)x_{it} \\
      & \qquad +\left(\frac{\mathbb{E}[\upsilon_y^2]}{2}+\beta_l^2 \frac{\mathbb{E}[\upsilon_l^2]}{2}\right)\left(x_{it}\sum_{j=1}^{K}c_j \, b''_j(Z_{it}) + 2\sum_{j=1}^{K}c_j \, b'_j(Z_{it})\right),  
\end{split}
\end{equation}
for $x=Z_{it}$. 

\subsubsection{Implementing EZ-NPR}

To implement the EZ-NPR estimator, we take $\left\{1, k_{it},l_{it},k_{it+1},Z_{it}\right\}$ as the instrument set giving $M = 5$ moment conditions $\mathbb{E}[\tilde{M}^x_{it}]=0$, where $\tilde{M}^x_{it}$ is given by equations \eqref{eq:correctedmoments_specific1} and \eqref{eq:correctedmoments_specific2}. We denote the corresponding
$M \times 1$ sample moment vector by $\bar{\bm{g}}$ and the GMM weighting
matrix by~$W$, which is $M \times M$.

The EZ-NPR moments feature $K+4$ parameters: the two production function factor elasticities $(\beta_l,\beta_k)$, the intercept $c_0$ and $K$ spline coefficients $(c_1, \ldots, c_K)$ of the $\Psi$ function, and the combined measurement error variance $\gamma \equiv \tfrac{\mathbb{E}[\upsilon_y^2]}{2}+\beta_l^2 \tfrac{\mathbb{E}[\upsilon_l^2]}{2}$. In Monte Carlo experiments, we found that a Generalized Method of Moments (GMM) estimator was not numerically robust in recovering the true parameters when using a wider set of moments than those outlined above and optimizing over all $K + 4$ parameters simultaneously. This is unsurprising: the flexibility of the B-spline function introduces local optima in the spline coefficients, and these interact with the measurement error parameter, whose identification relies on the shape of $\Psi$. We therefore reduce the dimension of the ``outer optimization'' to two by (numerically) computing $c_0$, $(c_1, \ldots, c_K)$, and $\gamma$ at each candidate $(\beta_l, \beta_k)$ by constrained least-squares in an ``inner optimization'' step.

We now describe the full estimation algorithm, which proceeds in three stages: (i) knot selection, (ii) corrected GMM with identity weighting and (iii) efficient two-step GMM.

\paragraph{Stage 1: Knot selection.}
The B-spline basis functions require a set of interior knots $\kappa_1 < \cdots < \kappa_J$, which determine where the basis functions are centred. Since the argument $Z_{it}$ depends on $(\beta_l, \beta_k)$, the distribution of $Z_{it}$ --- and hence the appropriate knot locations --- changes with the candidate parameters. Recomputing knots at each evaluation of the objective would introduce discontinuities, which affected numerical stability in Monte Carlo simulations. We therefore fix the knots once, prior to the corrected estimation, using the following procedure. For each point on a $(\beta_l, \beta_k)$ grid,\footnote{Similar to the baseline NPR estimator, the range of this grid is 0.05 to 0.9. Unlike the baseline NPR estimator, however, MC simulations indicated the importance of having a sufficiently fine grid. We therefore use a $20\times 20$ grid, which consists of 20 equally-spaced points between 0.1 and 0.9.} we evaluate the uncorrected GMM objective (i.e.\ setting $\gamma = 0$) using the identity weighting matrix. The grid point achieving the lowest objective is then ``polished'' using Nelder-Mead optimization and the resulting estimates $(\hat{\beta}_l^{(0)}, \hat{\beta}_k^{(0)})$ are used to construct
$\hat{Z}_{it}^{(0)} = \mathbb{E}_{it}[y_{it+1}] - \hat{\beta}_k^{(0)} k_{it+1} - \hat{\beta}_l^{(0)} \mathbb{E}_{it}[l_{it+1}]$. The interior knots are set at equally spaced quantiles of
$\hat{Z}_{it}^{(0)}$ and remain fixed for all subsequent stages of the estimation.

\paragraph{Stage 2: Corrected GMM.}
For a given candidate $(\beta_l, \beta_k)$, the intercept $c_0$ and spline coefficients $(c_1, \ldots, c_K)$ can be recovered by regressing $r_{it} \equiv y_{it}-\beta_k k_{it}-\beta_l l_{it}$ on the B-spline basis functions $b_1(Z_{it}), \ldots, b_K(Z_{it})$ and a constant, subject to the monotonicity constraints $c_1 \leq c_2 \leq \cdots \leq c_K$. When $Z_{it}$ is subject to classical measurement error, however, the resultant $c$ estimates will be attenuated, since the B-spline basis functions are evaluated at the noisy $Z_{it}$ rather than the true $Z_{it}^\ast$. The regression must therefore be adjusted using the second-order Taylor expansion principle of EZ, which corrects the normal equations
underlying the constrained regression to account for measurement error in the argument of the basis functions. This creates a circularity: if the measurement error variance $\gamma$ were known, the correction would have a closed-form solution, but to concentrate out the measurement error variance one requires unbiased estimates of $c_0$ and $(c_1, \ldots, c_K)$. For given $(\beta_l, \beta_k)$, we therefore confine estimation of $c_0$, $(c_1, \ldots, c_K)$ and $\gamma$ to an ``inner'' iterative optimization procedure:

\begin{enumerate}
    \item Initialise the measurement error variance at
    $\gamma^{(0)} = 0$ and set the iteration counter $\iota = 0$.

    \item\label{step:qp} Given $\gamma^{(\iota)}$, solve the
    corrected constrained regression for $c_0$ and
    $(c_1, \ldots, c_K)$. This involves solving the quadratic
    programme
    \begin{equation*}
        \min_{c_0, c_1, \ldots, c_K} \;
        \frac{1}{2} \bm{c}^\top A^{(\iota)} \bm{c}
        - \bm{f}^{(\iota)\top} \bm{c}
        \qquad \text{s.t.} \quad
        c_1 \leq c_2 \leq \cdots \leq c_K,
    \end{equation*}
    where $\bm{c} = (c_0, c_1, \ldots, c_K)^\top$ and the corrected
    normal equation components are
    \begin{equation*}
        A^{(\iota)} = \frac{B^\top B}{n}
        + \gamma^{(\iota)}
        \frac{B_2^\top B + 2B_1^\top B_1 + B^\top B_2}{n},
        \qquad
        \bm{f}^{(\iota)} = \frac{B}{n}
        + \gamma^{(\iota)} \frac{B_2}{n}.
    \end{equation*}
    Here $B$ denotes the matrices of B-spline basis
    function values evaluated at the observed $Z_{it}$ augmented with a column of ones for the intercept, and $B_1$ and $B_2$ denote the corresponding matrices of first and second derivatives (with a leading column of zeros for the intercept).\footnote{The
    correction terms arise from applying the EZ second-order Taylor
    expansion to each normal equation
    $\mathbb{E}[(r - \Psi(Z)) \, b_k(Z)] = 0$. The product rule
    applied to $\Psi(Z) \cdot b_k(Z)$ generates three terms ---
    $B_2'B$ from the curvature of $\Psi$ times the level of the
    instrument, $2B_1'B_1$ from the slope of $\Psi$ times the slope
    of the instrument, and $B'B_2$ from the level of $\Psi$ times
    the curvature of the instrument.} When $\gamma^{(0)} = 0$, this
    reduces to an uncorrected constrained regression at putative values for $\beta_l$ and $\beta_k$.

    \item\label{step:psi_derivs} Using the solution from
    step~\ref{step:qp}, compute the first and second derivatives of
    the estimated $\Psi$ function:
    \begin{equation*}
        \hat{\Psi}'(Z_{it})^{(\iota)} = \sum_{j=1}^{K} \hat{c}_j^{(\iota)}
        \, b'_j(Z_{it}),
        \qquad
        \hat{\Psi}''(Z_{it})^{(\iota)} = \sum_{j=1}^{K} \hat{c}_j^{(\iota)}
        \, b''_j(Z_{it}).
    \end{equation*}

    \item\label{step:moments} Compute the residuals
    $\hat{\epsilon}_{it}^{(\iota)} = r_{it} - \hat{c}_0^{(\iota)}
    - \sum_{j=1}^{K} \hat{c}_j^{(\iota)} b_j(Z_{it})$, where $r_{it} \equiv y_{it}-\beta_k k_{it}-\beta_l l_{it}$. Use the estimated residuals to calculate the observed (uncorrected) moment vector
    $\bar{\bm{g}}_{\text{obs}}^{(\iota)} = n^{-1} \sum_i \hat{\epsilon}_{it}^{(\iota)}
    \cdot \bm{h}_{it}$, where $\bm{h}_{it}$ is the vector of
    instruments. Using $\hat{\Psi}'^{(\iota)}$ and $\hat{\Psi}''^{(\iota)}$ from
    step~\ref{step:psi_derivs}, construct the correction vector
    $\bar{\bm{g}}_{\gamma}^{(\iota)}$, whose $m$-th element is the sample mean
    of the second-derivative correction term from equations
    \eqref{eq:correctedmoments_specific1}--\eqref{eq:correctedmoments_specific2}
    corresponding to instrument $x_{it}^m$.

    \item\label{step:profile} Profile $\gamma$ by minimising the GMM
    objective over $\gamma$ alone, holding the spline coefficients
    fixed. Since the corrected moment vector is linear in~$\gamma$,
    \begin{equation*}
        \bar{\bm{g}}_{\text{corr}}^{(\iota)}
        = \bar{\bm{g}}_{\text{obs}}^{(\iota)}
        - \bar{\bm{g}}_{\gamma}^{(\iota)} \, \gamma,
    \end{equation*}
    minimizing
    $\bar{\bm{g}}_{\text{corr}}^{(\iota)\top} W \,
    \bar{\bm{g}}_{\text{corr}}^{(\iota)}$
    with respect to~$\gamma$ yields a univariate weighted
    least-squares problem with closed-form solution
    \begin{equation*}
        \gamma^{(\iota+1)}_{\text{raw}} = \max\!\left(
        \frac{\bar{\bm{g}}_{\gamma}^{(\iota) \top} \, W \,
        \bar{\bm{g}}^{(\iota)}_{\text{obs}}}
        {\bar{\bm{g}}_{\gamma}^{(\iota) \top} \, W \,
        \bar{\bm{g}}_{\gamma}^{(\iota)}},
        \; 0 \right),
    \end{equation*}
    where $W$ is the GMM weighting matrix and the non-negativity
    constraint reflects the fact that $\gamma$ is a variance.

    \item\label{step:damp} Update with dampening:
    $\gamma^{(\iota+1)} = (1-\lambda)\gamma^{(\iota)}
    + \lambda \, \gamma^{(\iota+1)}_{\text{raw}}$, where
    $\lambda \in (0,1)$ is a dampening factor. Increment $\iota$ and
    return to step~\ref{step:qp}. Repeat until
    $|\gamma^{(\iota+1)} - \gamma^{(\iota)}| < \tau$ for some
    tolerance $\tau$, or until a maximum number of iterations is
    reached.
\end{enumerate}

Upon convergence, the corrected moments can be evaluated at the
final estimates of $\hat{c}_0$, $(\hat{c}_1, \ldots, \hat{c}_K)$ and
$\hat{\gamma}$ using equations
\eqref{eq:correctedmoments_specific1}--\eqref{eq:correctedmoments_specific2},
and the GMM objective $Q(\beta_l, \beta_k) = n \,
\bar{\bm{g}}_{\text{corr}}^\top W \, \bar{\bm{g}}_{\text{corr}}$
evaluated, where
$\bar{\bm{g}}_{\text{corr}} = \bar{\bm{g}}_{\text{obs}}
- \hat{\gamma} \, \bar{\bm{g}}_{\gamma}$.

We implement this procedure using the identity
weighting matrix $W = I$ on each point of the $20\times 20$ grid, evaluate the GMM objective function and then further refine the initialization point that achieved the lowest objective value using Nelder-Mead optimization. 

\paragraph{Stage 3: Efficient two-step GMM.}
Using the estimates recovered in Stage 2, the efficient weighting matrix is then constructed as $W^\ast = \hat{S}^{-1}$, where $\hat{S} = n^{-1} \sum_i \hat{\bm{g}}_{it} \hat{\bm{g}}_{it}^\top$ is estimated from the corrected residuals. This efficient weighting matrix places greater weight on the precisely estimated predetermined moments relative to the noisier $Z_{it}$ moment, thereby improving the identification of $(\beta_l, \beta_k)$. The full procedure described in Stage 2 is then repeated using $W^\ast$ rather than $W = I$ and the estimates it returns are taken as the EZ-NPR estimates.

\subsection{NPR Estimation with Biased Expectations}\label{app:npr_biased}

In this section we consider how biases in firms' expectations affect the NPR estimator. We show that a particular type of bias in expected inputs and outputs provides an additional moment that can be used for identification and present extensions to the baseline NPR estimation algorithm that can accommodate biases in firms' productivity expectations. We also show how similar extensions can accommodate imperfect knowledge of the production technology among firms. 

\subsection*{Case 1: Expected Input Bias and Technology-Consistent Output Bias}

As stated in Section \ref{subsect:method_bias}, the baseline NPR estimator can accommodate biased output expectations as long as such bias is also reflected in firms' expected inputs. Specifically, in the case of a firm with over-optimistic output expectations, for example, we require bias in the firm's employment expectations such that the integral of the production function with respect to expected labor equals the biased output expectation. In this particular case, and under the assumption of Cobb-Douglas technology, the difference between expected log output and realized log output gives:

\begin{equation}\label{eq:eq1}
    \begin{split}
        \mathbb{E}_{it-1}\left[y_{it}|\Omega_{it-1} \right]-y_{it}&= \beta_0+\beta_kk_{it}+\beta_l\mathbb{E}_{it-1}\left[l_{it}|\Omega_{it-1} \right] + \mathbb{E}_{it-1}\left[\omega_{it}|\Omega_{it-1} \right] \\
        & \qquad -\beta_0-\beta_kk_{it}-\beta_ll_{it}-g(\omega_{it-1})-\xi_{it}-\epsilon_{it} \\
        &=\beta_l\left(\mathbb{E}_{it-1}\left[l_{it}|\Omega_{it-1} \right]-l_{it}\right) + \mathbb{E}_{it-1}\left[\omega_{it}|\Omega_{it-1} \right]-g(\omega_{it-1})-\xi_{it}-\epsilon_{it} \\
        &=\beta_l\left(\mathbb{E}_{it-1}\left[l_{it}|\Omega_{it-1} \right]-l_{it}\right) + g(\omega_{it-1})-g(\omega_{it-1})-\xi_{it}-\epsilon_{it} \\
        &=\beta_l\left(\mathbb{E}_{it-1}\left[l_{it}|\Omega_{it-1} \right]-l_{it}\right) -\xi_{it}-\epsilon_{it}.
    \end{split}
\end{equation}

Taking expectations of equation (\ref{eq:eq1}) and maintaining the assumption that firms' productivity expectations are unbiased, the errors $(\xi,\epsilon)$ cancel to zero:

\begin{equation*}\label{eq:eq2}
    \begin{split}
        \mathbb{E}\left[\mathbb{E}_{it-1}\left[y_{it}|\Omega_{it-1} \right]-y_{it}\right]&=\mathbb{E}\left[\beta_l\left(\mathbb{E}_{it-1}\left[l_{it}|\Omega_{it-1} \right]-l_{it}\right) -\xi_{it}-\epsilon_{it}\right] \\ 
        & =\beta_l\mathbb{E}\left[\mathbb{E}_{it-1}\left[l_{it}|\Omega_{it-1} \right]-l_{it}\right] -\mathbb{E}\left[\xi_{it}+\epsilon_{it}\right] \\
        & =\beta_l\mathbb{E}\left[\mathbb{E}_{it-1}\left[l_{it}|\Omega_{it-1} \right]-l_{it}\right].
    \end{split}
\end{equation*}

Assuming $\mathbb{E}\left[\mathbb{E}_{it-1}\left[x_{it}|\Omega_{it-1} \right]-x_{it}\right] \neq 0$ for $x\in(y,l)$, the labor coefficient is identified as the average of population mean expectation errors:

\begin{equation}\label{eq:eq3}
    \beta_l = \frac{\mathbb{E}\left[\mathbb{E}_{it-1}\left[y_{it}|\Omega_{it-1} \right]-y_{it}\right]}{\mathbb{E}\left[\mathbb{E}_{it-1}\left[l_{it}|\Omega_{it-1} \right]-l_{it}\right]}.
\end{equation}

This shows that, in the Cobb-Douglas technology case, a simple Wald estimator for $\beta_l$ may therefore be obtained by replacing the population mean expectation errors on the right hand side of equation (\ref{eq:eq3}) with their sample equivalents. Since the capital coefficient remains unidentified, equation (\ref{eq:eq3}) is not, on its own, enough to identify the entire production function. Rather it can be used either as a method to test the labor coefficient estimated by the baseline NPR algorithm or as an additional moment in estimation. Nonetheless, it is worth reiterating that equation (\ref{eq:eq3}) is only informative if firms' expectations over inputs and output exhibit bias that is consistent with the production technology and that their expectations over productivity are unbiased. In our application using MES data, the expectations of sales and employment seem reasonably well aligned with outturns (see Table \ref{tab:mes_expectation_errors}).

\subsection*{Case 2: Time-invariant Expected Productivity Bias}
Suppose firms' productivity bias was time-invariant, so that $\iota_{it}=\iota_i$ in (\ref{eq:eprodfunct_bias}). If one had panel observations of firms' forecast errors, it would be possible to estimate firms' time-invariant bias and then use $\hat{\iota}_i$ to recover consistent estimates of $\beta$. In practice, this would lead to an ``outer'' estimation loop and the NPR algorithm would become
\begin{enumerate}
   \item Pick an initial vector of bias terms (one for each firm), $\hat{\iota}_0$.
   \item \label{item:starthere} For iteration $r$, implement the NPR estimation algorithm using \\
   $\left(\mathbb{E}_{it}[y_{it+1}|I_{it}] - \int f(k_{it+1},l_{it+1}; \beta)dF_{it}(l_{it+1})-\hat{\iota}_{r-1}\right)$ as the argument of the smooth function $\Psi$, to obtain coefficient estimates $\hat{\beta}_r$.
   \item \label{item:calciota} Use $\hat{\beta}_r$ to calculate $\textrm{a}_{it} = \left(\mathbb{E}_{it-1}[y_{it}|I_{it-1}] - \int f(k_{it},l_{it}; \hat{\beta}_r)dF_{it-1}(l_{it})\right)$ and $\textrm{b}_{it} = y_{it}-f(k_{it},l_{it}; \hat{\beta}_r)$ and their difference for all periods for which both expectations (needed to calculate a), and outcomes (needed to calculate b), are observed.
   \item Recover an updated estimate of the bias terms $\hat{\iota}_r$ as the firm-level mean of $(\textrm{a}_{it}-\textrm{b}_{it})$ and compare to $\hat{\iota}_{r-1}$. If the difference is sufficiently small then stop, if not then repeat from step \ref{item:starthere}.
\end{enumerate}
To understand how the calculation in step \ref{item:calciota} is used to recover an estimate of $\iota_i$, observe that
\begin{enumerate}
\item $\textrm{a}_{it}= \mathbb{E}_{it-1}[y_{it}|I_{it-1}] - \int f(k_{it},l_{it}; \beta)dF_{it-1}(l_{it}) = g(\omega_{it-1})+\iota_{i}$
\item $\textrm{b}_{it}= y_{it} - f(k_{it},l_{it}; \beta) = g(\omega_{it-1})+\xi_{it}+\epsilon_{it}$
\end{enumerate}
therefore  
\begin{align*}
(\textrm{a}_{it}-\textrm{b}_{it}) &= \left(g(\omega_{it-1})+\iota_{i}\right) - \left(g(\omega_{it-1})+\xi_{it}+\epsilon_{it}\right) \\
&= \iota_{i}-\xi_{it}-\epsilon_{it}
\end{align*}
Since the shocks $\xi$ and $\epsilon$ are each assumed i.i.d. and mean-zero, multiple observations of productivity forecast errors $(\textrm{a}_{it}-\textrm{b}_{it})$ contain information on the time-invariant bias term $\iota$.

\subsection*{Case 3: Time-varying Expected Productivity Bias}
If firms' productivity bias is assumed to vary over time, one can proceed by specifying the bias as a function of observable firm attributes. Suppose firms' productivity bias can be expressed as 
\begin{equation*}
\iota_{it}=\Lambda\left(X_{it-1};\lambda\right)
\end{equation*}
where $X$ is a vector of observable firm attributes. Using productivity forecast errors, one can estimate the $\lambda$ parameters of the productivity bias function and thereby recover consistent estimates of $\theta$. Again, this leads to an additional estimation loop outside the main NPR estimation routine and the NPR algorithm becomes
\begin{enumerate}
\item Pick an initial vector of bias function parameters, $\hat{\lambda}_0$.
\item \label{item:starthere2} For iteration $r$, implement the NPR estimation algorithm using \\
$\left(\mathbb{E}_{it}[y_{it+1}|I_{it}] - \int f(k_{it+1},l_{it+1}; \beta)dF_{it}(l_{it+1})-\Lambda(X_{it-1};\hat{\lambda}_{r-1})\right)$ as the argument of the smooth function $\Psi$, to obtain coefficient estimates $\hat{\beta}_r$.
\item Use $\hat{\beta}_r$ to calculate $\textrm{a}_{it} = \left(\mathbb{E}_{it-1}[y_{it}|I_{it-1}] - \int f(k_{it},l_{it}; \hat{\beta}_r)dF_{it-1}(l_{it})\right)$ and $\textrm{b}_{it} = y_{it}-f(k_{it},l_{it}; \hat{\beta}_r)$ and their difference for all periods for which both expectations (needed to calculate a), and outcomes (needed to calculate b), are observed.
\item Obtain an updated estimate of the bias function parameters $\hat{\lambda}_r$ via estimation of 
\begin{equation*}
    (\textrm{a}_{it}-\textrm{b}_{it})=\Lambda\left(X_{it-1};\lambda\right)
\end{equation*}
\item Compare $\hat{\lambda}_r$ to $\hat{\lambda}_{r-1}$. If the difference is sufficiently small the stop, if not then repeat from step \ref{item:starthere2}.
\end{enumerate}
In this case, the productivity forecast errors are given as 
\begin{align*}
(\textrm{a}_{it}-\textrm{b}_{it}) &= \left(g(\omega_{it-1})+\iota_{it}\right) - \left(g(\omega_{it-1})+\xi_{it}+\epsilon_{it}\right) \\
&= \iota_{it}-\xi_{it}-\epsilon_{it} \\
&= \Lambda\left(X_{it-1};\lambda\right)-\xi_{it}-\epsilon_{it}
\end{align*}
and since $\xi$ and $\epsilon$ are each assumed i.i.d. and mean-zero, one can use the $(\textrm{a}_{it}-\textrm{b}_{it})$ terms to estimate the parameters of the productivity bias function $\Lambda$.

An alternative possibility is that firms' productivity bias is a function of their previous productivity. This is analogous to the ``excess sensitivity'' or ``over-extrapolation'' of output expectations discussed in literature on managerial and financial expectations (e.g., \cite{Barrero}). In our context, this type of bias means firms expect their productivity to increase/decrease between the current and next period if it has increased/decreased in recent periods. This type of bias is hard to accommodate if it specifically relates to firms' persistent productivity (i.e.,$\omega_{it}$). If, however, firms' biased beliefs over next-period productivity bias are determined by their current and recent \emph{overall} productivity (i.e., $\omega_{it}+\epsilon_{it}$), one could modify the algorithm above accordingly.
\begin{enumerate}
\item Pick an initial vector of bias function parameters, $\hat{\lambda}_0$, and an initial vector of productivity terms (of length $N \times T$), $\hat{\varepsilon}_0$.
\item For iteration $r$, implement the NPR estimation algorithm using \\
$\left(\mathbb{E}_{it}[y_{it+1}|I_{it}] - \int f(k_{it+1},l_{it+1}; \beta)dF_{it}(l_{it+1})-\Lambda(X_{it-1};\hat{\lambda}_{r-1})\right)$ as the argument of the smooth function $\Psi$, to obtain coefficient estimates $\hat{\beta}_r$, where $X_{it}$ includes $\hat{\varepsilon}_{it,r}$, $\hat{\varepsilon}_{it-1,r}$ and any other firm attributes that are believed to determine bias in firms' expectations of their next-period productivity.
\item \label{item:calcresids} Use $\hat{\beta}_r$ to calculate $\textrm{a}_{it} = \left(\mathbb{E}_{it-1}[y_{it}|I_{it-1}] - \int f(k_{it},l_{it}; \hat{\beta}_r)dF_{it-1}(l_{it})\right)$ and $\textrm{b}_{it} = y_{it}-f(k_{it},l_{it}; \hat{\beta}_r)$. Calculate b) for all periods in which outcomes are observed and the difference (a-b)  for all periods that both expectations (needed to calculate a), and outcomes (needed to calculate b), are observed. 
\item Obtain an updated estimate of the bias function parameters $\hat{\lambda}_r$ via estimation of 
\begin{equation*}
    (\textrm{a}_{it}-\textrm{b}_{it})=\Lambda\left(X_{it-1};\lambda\right)
\end{equation*}
Obtain an updated estimate of the productivity terms as
 \begin{equation*}
    \hat{\varepsilon}_{it,r} = y_{it}-f(k_{it},l_{it}; \hat{\beta}_r)
\end{equation*}
\item Compare $\hat{\lambda}_r$ to $\hat{\lambda}_{r-1}$ and $\hat{\varepsilon}_r$ to $\hat{\varepsilon}_{r-1}$, where $\hat{\varepsilon}_r$ is a vector containing the observation-specific $\hat{\varepsilon}_{it,r}$. If the difference across both vectors is sufficiently small the stop, if not then repeat from step \ref{item:starthere2}.
\end{enumerate}

\subsection*{Case 4: Imperfect Knowledge of the Production Technology}
If firms have imperfect knowledge of the production function parameters then, in the case of Cobb-Douglas technology,
\begin{equation}\label{eq:prodfunct_incorrectparams}
\mathbb{E}_{it}[y_{it+1}|\Omega_{it}]=\tilde{\beta}_{i0}+\tilde{\beta}_{ik}k_{it+1}+\tilde{\beta}_{il}\mathbb{E}_{it}[l_{it+1}|\Omega_{it}]+\mathbb{E}_{it}[\omega_{it+1}|\Omega_{it}],
\end{equation}
with 
\begin{equation*}
\tilde{\beta}_{ix}=\beta_{x}+\upsilon_{ix} \qquad \mathbb{E}[\upsilon_{ix}]=0, \qquad \mathbb{E}[\upsilon_{ix}^2]=\sigma^2_{\upsilon_x},
\end{equation*}
for $x \in \left\{0,k,l\right\}$. Here the $\upsilon$ terms capture firms' imperfect knowledge of the production technology. In this context, rearranging equation (\ref{eq:prodfunct_incorrectparams}) to isolate $\mathbb{E}_{it}[\omega_{it+1}|\Omega_{it}]$ gives
\begin{equation}\label{eq:eomega_incorrectparams}
\begin{split}
\mathbb{E}_{it}[\omega_{it+1}|\Omega_{it}]=g(\omega_{it}) &= \mathbb{E}_{it}[y_{it+1}|\Omega_{it}]-\left(\tilde{\beta}_{i0}+\tilde{\beta}_{ik}k_{it+1}+\tilde{\beta}_{il}\mathbb{E}_{it}[l_{it+1}|\Omega_{it}]\right) \\
&= \mathbb{E}_{it}[y_{it+1}|\Omega_{it}]-\left(\beta_{0}+\beta_{k}k_{it+1}+\beta_{l}\mathbb{E}_{it}[l_{it+1}|\Omega_{it}]\right) \\
& \qquad\qquad\qquad\quad -\left(\upsilon_{i0}+\upsilon_{ik}k_{it+1}+\upsilon_{il}\mathbb{E}_{it}[l_{it+1}|\Omega_{it}]\right) \\
&= \mathbb{E}_{it}[y_{it+1}|\Omega_{it}]-\left(\beta_{0}+\beta_{k}k_{it+1}+\beta_{l}\mathbb{E}_{it}[l_{it+1}|\Omega_{it}]\right) \\
& \qquad\qquad\qquad\quad -\Lambda\left(k_{it+1},\mathbb{E}_{it}[l_{it+1}|\Omega_{it}];\Upsilon_{i}\right),
\end{split}
\end{equation}
where $\Upsilon_{i}=\left\{\upsilon_{i0},\upsilon_{ik},\upsilon_{il}\right\}$ are the imperfect information terms that are unobserved by the econometrician.

The last equality of equation \eqref{eq:eomega_incorrectparams} highlights that this is very similar to case 3 above, where bias in firms' expected productivity is a function of observables. As in this case, the baseline NPR algorithm can be embedded within an outer iteration loop to recover estimates of the $\Upsilon_{i}$ imperfect information terms. A nuance is that in this case, and assuming imperfect knowledge of the production function parameters varies across firms, the function $\Lambda\left(\cdot\right)$ should be estimated as a random coefficients model. 

The exposition here assumes Cobb-Douglas technology but extends to any (linear-in-parameters) production function where imperfect knowledge of the functions' parameters can be rewritten as the function evaluated at the true parameter values and an additive term capturing firms' imperfect knowledge. However, we note that this is a particular case of imperfect information that confines firms' imperfect knowledge to the parameters of the production function rather than its functional form. Imperfect information of the latter type is generally insurmountable. 

\section{Monte Carlo Setup}\label{adix:mc_setup} 

Our Monte Carlo setup closely follows that of \citet{ackerberg_identification_2015}, the details of which are described in their Appendix. We repeat the key details here to ease readers' understanding of our results.

\subsection{Production Function and Productivity Process}

We consider a gross output production function that is Leontief in the material input:
\begin{equation*}
    Y_{it}=\min \left\{ \beta_0K_{it}^{\beta_k}L_{it}^{\beta_l}e^{\omega_{it}},\beta_mM_{it}\right\} e^{\epsilon_{it}},
\end{equation*}

\noindent where $\beta_0=1$, $\beta_k=0.4$, $\beta_l=0.6$ and  $\beta_m=1$. $\epsilon_{it}$ is a mean-zero measurement error distributed i.i.d. over firms and time with standard deviation 0.1. The productivity shock $\omega_{it}$ follows an AR(1) process:

\begin{equation*}
    \omega_{it}=\rho\omega_{it-1}+\xi_{it},
\end{equation*}
with $\rho=0.7$. The standard deviation of $\omega_{it}$ is constant over time and equal to 0.3, which is achieved via the parameterization of the normally distributed innovation $\xi_{it}$ and the initial values $\omega_{i0}$.

\subsection{Firm Choices}

Firms' labor and material input choices are static in the sense that a firm's choice in period $t$ has no implications for their choices in subsequent periods. This implies there are no adjustment costs in labor or materials which, combined with the other assumptions of our Monte Carlo setup, yield analytic solutions to the firms' decision problem.\footnote{Alternatively one could incorporate adjustment costs and solve the firms' problem numerically as in \citet{bond_adjustment_2005}.}

ACF consider several DGPs that differ in the timing of firms' labor choice. Specifically, ACF consider DGPs where firms' labor input in period $t$ is chosen in an intermediate period $(t-1+b)$ with $0<b<1$, without full knowledge of the productivity shock $\omega_{it}$. ACF include this dimension as their estimator is robust to such input timing, whereas the LP estimator is not. While our Monte Carlo algorithm can accommodate such timing, and we indeed replicate ACF's analysis below, we do not focus on this DGP in the results reported in the main text. Our baseline DGP therefore features firms that choose labor and materials concurrently in period $t$, with full knowledge of current-period productivity $\omega_{it}$. Similarly, while ACF consider DGPs featuring firm-specific wages, we do not focus on them in our simulations with firms instead facing a common time-invariant wage.

Under these assumptions, the labor choice that maximizes firms' expected profits is
\begin{equation} \label{eq:opt_labor} L^{\ast}_{it}=\left(\beta_0\beta_lK_{it}^{\beta_k}e^{\omega_{it}}\right)^{1/(1-\beta_l)},
\end{equation}
and the optimal level of materials is
\begin{equation} \label{eq:opt_int}
M^{\ast}_{it}=\beta_0K_{it}^{\beta_k}L_{it}^{\ast\beta_l}e^{\omega_{it}}.
\end{equation}
Capital is a dynamic input that evolves according to 
\begin{equation*}
    K_{it}=(1-\delta)K_{it-1}+I_{it-1},
\end{equation*}
where $\delta=0.2$ represents depreciation. Investment is subject to convex, firm-specific adjustment costs given by 
\begin{equation*}
    c_{i}(I_{it})=\frac{\phi_i}{2}I^2_{it},
\end{equation*}
where $1/ \phi_i$ is distributed log-normally across firms with a standard deviation ($\sigma$) of 0.6. While such firm-specific heterogeneity in capital adjustment costs renders the OP estimator inconsistent in all DGPs, it is included to generate variation in capital across firms comparable to that observed in both ACF's and our data.

Under the assumptions described above, ACF extend the work of \citet{syverson_market_2001} and \citet{van_biesebroeck_robustness_2007} to obtain an analytical solution for optimal investment using Euler equation techniques:
\begin{equation}\label{eq:mc_optinv}
    \begin{split}
    I^{\ast}_{it}= &\frac{\beta}{\phi_i}  \sum_{\tau=1}^{\infty} \left\{(\beta(1-\delta))^\tau \left(\frac{\beta_k}{1-\beta_l}\right)\beta_0^{1/(1-\beta_l)} \right. \\
    & \left. \times \exp \left[\left(\frac{1}{1-\beta_l}\right)\rho^{\tau+1}\omega_{it}+\frac{1}{2}\left(\frac{1}{1-\beta_l}\right)^2\sum_{s=0}^{\tau}\rho^2(\tau-s)\sigma^2_{\xi} \right] \right\} ,
    \end{split}
\end{equation}
where $\beta=0.95$ is the discount factor.\footnote{The assumption of constant returns to scale means that optimal investment does not depend on the current capital stock.}

As emphasized in the main text, the NPR estimator is robust to deviations from invertibility in either labor, materials or investment. ACF highlight the consistency of the LP estimator requires optimization error in labor but is undermined by optimization error in any other firm choices. The consistency of the ACF estimator under multiple types of optimization error is unclear. The DGPs examined in our Monte Carlo simulations therefore include optimization error in one or more inputs, which we model as
\begin{equation*}
    X_{it}=X^{\ast}_{it}e^{\xi^x_{it}},
\end{equation*}
for $X\in[L,I,M]$ where $\xi^x_{it}$ is normally-distributed i.i.d. optimization error with mean zero and standard deviation $\sigma_{\xi^L}=\sigma_{\xi^I}=\sigma_{\xi^M}=0.37$ for labor, investment and intermediates (see footnote \ref{fn:optm}).

A detail to note is that optimal investment will have a different analytical solution from equation (\ref{eq:mc_optinv}) when labor deviates from optimality. Specifically, when $\sigma_{\xi^L}^2>0$, ACF derive that optimal investment becomes 
\begin{equation}\label{eq:mc_optinv_shocks}
    \begin{split}
    I^{\ast}_{it}= &\frac{\beta}{\phi_i}  \sum_{\tau=1}^{\infty} \left\{ (\beta(1-\delta))^\tau \left(\frac{\beta_k}{1-\beta_l}\right)\beta_0^{1/(1-\beta_l)} \right. \\
    & \left. \times \exp \left[\left(\frac{1}{1-\beta_l}\right)\rho^{\tau+1}\omega_{it}+\frac{1}{2}\left(\frac{1}{1-\beta_l}\right)^2\sum_{s=0}^{\tau}\rho^2(\tau-s)\sigma^2_{\xi} \right] \right\} \\
    & \times \left[\beta_l^{\beta_l/(1-\beta_l)}e^{(1/2)\beta^2_l\sigma_{\xi^L}^2}-\beta_l^{1/(1-\beta_l)}e^{(1/2)\sigma_{\xi^L}^2}\right].
    \end{split}
\end{equation}
This is identical to equation~\eqref{eq:mc_optinv} except for the square-bracketed term, which is a constant that multiplies the entire sum. It captures the net effect of labor optimization error on expected profits, with the first component reflecting the effect on revenue and the second the effect on costs. Since $\xi^L$ is independent of $\omega$, the forward-iteration and expectations calculations that generate the horizon-varying terms inside the sum are unaffected. 

In the majority of DGPs we consider, firm wages are constant and uniform across firms. In Table \ref{tab:mcrun_summtab_acftab1}, where we replicate the DGPs of ACF, firm-varying wages introduce additional terms into conditions \eqref{eq:opt_labor} and \eqref{eq:mc_optinv_shocks}, as detailed in their Appendix A.

\subsection{Expectations}\label{app:mc_expectations}

The NPR estimator applied to a Cobb-Douglas production function requires data on one-period-ahead expectations of log labor and log output. To simulate these expectations, we rely on the expressions for firm choices described above.\footnote{The derivations here assume that firms take optimization error into account when forming expectations of next-period labor. However, since the optimization error becomes additive in the equation for log labor and is mean-zero, the expression is the same if one instead were to suppose that firms are ``naive'' in the sense that they continually expected optimal choices to be realized despite repeated experience to the contrary.} Specifically, we have
\begin{equation*}
    \mathbb{E}_{it}[\omega_{it+1}|\Omega_{it}]=\rho\omega_{it},
\end{equation*}
and hence 
\begin{equation*}
    \begin{split}
       \mathbb{E}_{it}\left[\textrm{ln}\left(L_{it+1}\right)|\Omega_{it}\right] &= \mathbb{E}_{it}\left[\textrm{ln}\left(L^{\ast}_{it+1}e^{\xi^{L}_{it+1}}\right)\right] \\ &=\mathbb{E}_{it}\left[\textrm{ln}\left(L^{\ast}_{it+1}\right)\right]+\mathbb{E}_{it}\left[\xi^{L}_{it+1}\right] \\
       &=\mathbb{E}_{it}\left[\textrm{ln}\left(\left(\beta_0\beta_l K_{it+1}^{\beta_k}e^{\omega_{it+1}}\right)^{1/(1-\beta_l)}\right)\right]+0 \\
        &=\tfrac{1}{(1-\beta_l)}\left(\textrm{ln}\left(\beta_0\beta_l\right)+\beta_kk_{it+1}+\mathbb{E}_{it}\left[\omega_{it+1}\right]\right) \\
         &=\tfrac{1}{(1-\beta_l)}\left(\textrm{ln}\left(\beta_0\beta_l\right)+\beta_kk_{it+1}+\rho\omega_{it}\right).       
   \end{split}
\end{equation*}
This in turn implies 
\begin{equation*}
    \begin{split}
   \mathbb{E}_{it}\left[\textrm{ln}\left(Y_{it+1}\right)|\Omega_{it}\right]&=  \mathbb{E}_{it}\left[\textrm{ln}\left(\min \left\{ \beta_0K_{it+1}^{\beta_k}L_{it+1}^{\beta_l}e^{\omega_{it+1}},\beta_mM_{it+1}\right\} e^{\epsilon_{it+1}}\right)|\Omega_{it}\right] \\
   &=  \mathbb{E}_{it}\left[\min \left\{ \textrm{ln}\left(\beta_0K_{it+1}^{\beta_k}L_{it+1}^{\beta_l}e^{\omega_{it+1}}\right),\textrm{ln}\left(\beta_mM_{it+1}\right)\right\}|\Omega_{it}\right]+ \mathbb{E}_{it}\left[\epsilon_{it+1}|\Omega_{it}\right] \\ 
   &=  \mathbb{E}_{it}\left[\min \left\{ \textrm{ln}\left(\beta_0K_{it+1}^{\beta_k}L_{it+1}^{\beta_l}e^{\omega_{it+1}}\right),\textrm{ln}\left(\beta_m\beta_0K_{it+1}^{\beta_k}L_{it+1}^{\beta_l}e^{\omega_{it+1}}e^{\xi^{M}_{it+1}}\right)\right\}|\Omega_{it}\right],
   \end{split}
\end{equation*}
where the replacement of $\textrm{ln}\left(\min \left\{a,b\right\}\right)$ with $\min \left\{\textrm{ln}(a),\textrm{ln}(b)\right\}$ that occurs in the transition from the first to the second equality implicitly assumes the arguments of the min function are always strictly positive. To evaluate the right hand side expectation we leverage results in \citet{nadarajah_exact_2008}, specifically their equation (11). This states that for $Y=\min(X_1,X_2)$, where $(X_1,X_2)$ is a bivariate Gaussian random vector with means $(\mu_1,\mu_2)$, variances $(\sigma^2_1,\sigma^2_2)$ and correlation coefficient $\rho$, we have:

\begin{equation}\label{eq:emean_min}
    \mathbb{E}\left[Y\right]=\mu_1\Phi\left(\frac{\mu_2-\mu_1}{\theta}\right)+\mu_2\Phi\left(\frac{\mu_1-\mu_2}{\theta}\right)-\theta\phi\left(\frac{\mu_2-\mu_1}{\theta}\right),
\end{equation}
where $\theta=\sqrt{\sigma^2_1+\sigma^2_2-2\rho\sigma_1\sigma_2}$.
Take then $X_1 \equiv \ln\left(\beta_0\right)+\beta_kk_{it+1}+\beta_ll_{it+1}+\omega_{it+1}$ and $X_2 \equiv \ln\left(\beta_m\right)+m_{it+1}$.  Conditional on $\Omega_{it}$, $(X_1,X_2)$ is bivariate normal with $\mu_1=\mu_2\equiv\mu=\mathbb{E}\left[\textrm{ln}\left(\beta_0K_{it+1}^{\beta_k}L_{it+1}^{\beta_l}e^{\omega_{it+1}}\right)|\Omega_{it}\right]=\mathbb{E}[\beta_m+m_{it+1}|\Omega_{it}]$ since optimality for the Leontief production function implies that $\textrm{ln}\left(\beta_0K_{it+1}^{\beta_k}L_{it+1}^{\beta_l}e^{\omega_{it+1}}\right)=\textrm{ln}\left(\beta_mm_{it+1}\right)$ and the (log-) optimization error in intermediates has mean zero. Using equations \eqref{eq:opt_labor} and \eqref{eq:opt_int} we can also obtain that $\sigma_1^2 = [\beta_l^2 /(1-\beta_l)^2+1]\sigma^2_\xi$, $\sigma^2_2 = \sigma_1^2 + \sigma^2_{\xi^M}$ and $\rho \sigma_1 \sigma_2 = \sigma_1^2$. This means that $\theta = \sqrt{ \sigma^2_1 + \sigma_1^2 + \sigma^2_{\xi^M} - 2 \sigma^2_1 } = \sigma_{\xi^M}$. Applying equation \eqref{eq:emean_min} to our context yields
\begin{equation*}\label{eq:e_y}
    \begin{split}        \mathbb{E}_{it}\left[\textrm{ln}\left(Y_{it+1}\right)|\Omega_{it}\right]&=\mathbb{E}_{it}\left[\min\left(\beta_0+\beta_kk_{it+1}+\beta_ll_{it+1}+\omega_{it+1},\beta_m+m_{it+1}\right)|\Omega_{it}\right] \\
        &=\mu\Phi\left(0\right)+\mu\Phi\left(0\right)-\sigma_{\xi^M}\phi\left(0\right) \\
        &\approx0.5\mu+0.5\mu-0.3989\sigma_{\xi^M} \\
        &=\mu-0.3989\sigma_{\xi^M}.
    \end{split}
\end{equation*}

\subsubsection*{Biased Expectations}

Section~\ref{sect:mc_extensions} shows that biased expectations can undermine the NPR estimator by violating the scalar unobservable condition it relies on. To evaluate the quantitative severity of this problem, we simulate data under several DGPs in which firms' expectations deviate from the rational benchmark. We consider three distinct sources of expectational biases, which differ in terms of the variable affected. Unlike measurement error in expectations, which solely affects the data observed by the econometrician, biased expectations propagate into firm decisions, demanding modification of the closed form solutions of \citet{van_biesebroeck_robustness_2007}. 

\paragraph{Biased productivity expectations.} Bias in firms' expected future productivity causes the rational conditional expectation $\mathbb{E}_{it}[\omega_{it+1}|\Omega_{it}]=\rho\omega_{it}$ to be replaced with
\begin{equation}\label{eq:biased_omega}
    \tilde{\mathbb{E}}_{it}[\omega_{it+1}|\Omega_{it}]=\rho\omega_{it}+\iota_{it},
\end{equation}
where $\iota_{it}$ is a firm-specific bias term. In the simulations, $\iota_{it}\sim N(0,\sigma^2_\iota)$ is drawn independently across firms and, in most DGPs, across periods.\footnote{The exception is the DGP in which we consider productivity expectations as a function of a constant-firm-specific variable. This corresponds to the persistent firm-level bias discussed in Appendix~\ref{app:npr_biased}, Case~2.} The variance of bias shocks is calibrated as
\begin{equation*}
    \sigma^2_\iota = c_\omega \cdot \textrm{Var}_W\!\left(\rho\omega_{it}\right),
\end{equation*}
where $\textrm{Var}_W(\cdot)$ denotes the within-firm variance over the sample period and $c_\omega$ is the parameter reported in the tables and figure axes.

The biased productivity expectation enters the firm's dynamic optimization problem. In the analytic solution for optimal investment (equation~\eqref{eq:mc_optinv_shocks}), the productivity forecast enters via $\rho^{\tau+1}\omega_{it}$; the biased belief effectively replaces $\omega_{it}$ with $\omega_{it}+\iota_{it}/\rho$. The resulting capital stock therefore differs from the rational-expectations benchmark, and this in turn affects realized labor (which depends on capital via the static optimality condition), materials and output.

The biased productivity belief also propagates into the firm's reported expectations of labor and output. The expressions for $\mathbb{E}_{it}[\textrm{ln}(L_{it+1})|\Omega_{it}]$ and $\mathbb{E}_{it}[\textrm{ln}(Y_{it+1})|\Omega_{it}]$ derived in the preceding subsection are recomputed with $\tilde{\mathbb{E}}_{it}[\omega_{it+1}]$ in place of $\mathbb{E}_{it}[\omega_{it+1}]$ and with the biased capital stock. 

\paragraph{Biased labor expectations.} The second type of expectational 
bias we consider adds a mean-zero idiosyncratic shock to the firm's 
expectation of next-period log labor:
\begin{equation}\label{eq:biased_labor}
    \tilde{\mathbb{E}}_{it}[\textrm{ln}(L_{it+1})|\Omega_{it}]
    =\mathbb{E}_{it}[\textrm{ln}(L_{it+1})|\Omega_{it}]+\eta^l_{it},
\end{equation}
where $\eta^l_{it}\sim N(0,\sigma^2_{\eta^l})$ and 
$\sigma^2_{\eta^l}=c_l\cdot\textrm{Var}_W\!\left(
\mathbb{E}_{it}[\textrm{ln}(L_{it+1})]\right)$. Since $\eta^l_{it}$ 
is additive in log labor, it corresponds to a multiplicative bias 
$e^{\eta^l_{it}}$ in the exponential of expected log labor.

This bias propagates into the firm's investment decision by the same 
mechanism as the labor optimization error described above. The firm 
perceives its future labor as 
$\tilde{L}_{it+\tau}=L^{\ast}_{it+\tau}\,e^{\eta^l_{it}}\,
e^{\xi^L_{it+\tau}}$, where $\eta^l_{it}$ is the current-period bias 
that the firm extrapolates to all future horizons. Substituting into 
the profit function, both revenue and cost terms remain proportional 
to $K^{\beta_k/(1-\beta_l)}=K$ under constant returns to scale (CRS), so the capital stock drops 
out of the Euler equation. Since $\eta^l_{it}$ is ``known'' 
to the firm, it factors out of the expectations over $\xi^L$, and the square-bracketed 
term in equation~\eqref{eq:mc_optinv_shocks} is therefore modified to
\begin{equation*}
    \left[\beta_l^{\beta_l/(1-\beta_l)}\,e^{\beta_l\eta^l_{it}}\,
    e^{(1/2)\beta_l^2\sigma^2_{\xi^L}} 
    - \beta_l^{1/(1-\beta_l)}\,e^{\eta^l_{it}}\,
    e^{(1/2)\sigma^2_{\xi^L}}\right],
\end{equation*}
with all other components of the investment expression unchanged.

The biased capital path feeds into realized labor (chosen optimally 
given the true productivity and biased capital stock), materials and 
output. It also enters the structural component of reported expected 
labor, to which the additive bias $\eta^l_{it}$ is then added. Finally, the 
biased labor expectation propagates into the reported output 
expectation through the production technology
\begin{equation*}
    \tilde{\mathbb{E}}_{it}[\textrm{ln}(Y_{it+1})|\Omega_{it}] 
    = \beta_0 + \beta_l\tilde{\mathbb{E}}_{it}[\textrm{ln}(L_{it+1})] 
    + \beta_k k_{it+1} + \mathbb{E}_{it}[\omega_{it+1}] 
    - \phi(0)\sigma_{\xi^M}.
\end{equation*}

\paragraph{Biased output expectations.} When bias affects output expectations directly (as opposed to indirectly via the biased labor expectations case described above), output expectations are constructed by adding a mean-zero shock to the rational expectation of next-period log output:
\begin{equation}\label{eq:biased_output}
    \tilde{\mathbb{E}}_{it}[\textrm{ln}(Y_{it+1})|\Omega_{it}]=\mathbb{E}_{it}[\textrm{ln}(Y_{it+1})|\Omega_{it}]+\eta^y_{it},
\end{equation}
where $\eta^y_{it}\sim N(0,\sigma^2_{\eta^y})$ and $\sigma^2_{\eta^y}=c_y\cdot\textrm{Var}_W\!\left(\mathbb{E}_{it}[\textrm{ln}(Y_{it+1})]\right)$. In contrast to bias in productivity or in labor, bias in expected output does not propagate into firm's investment or input decisions. It can therefore be interpreted as capturing idiosyncratic noise in a firm's reported output forecast that is unrelated to the structural beliefs it uses when making input decisions.

\paragraph{Calibration and implementation.} In all three channels, the variance of the bias term is calibrated as a fraction of the within-firm variance of the corresponding true expectation over the sample window. The within-firm variance is computed once per Monte Carlo replication from unbiased data generated in a first pass, in which optimal investment, capital, labor and expectations are simulated under the rational benchmark. The bias draws are then used to re-simulate investment, capital, labor and expectations in a second pass. This two-pass procedure ensures that the bias magnitudes are expressed in units comparable across channels and relative to the variation that is informative for identification. 

\subsection{Simulation Details}

We consider a panel of 1,000 firms observed over 10 time periods. For comparability with ACF's setup, our baseline specification follows their parameterization, which is intended to match key aspects of their data: 95\% of the variation in capital being across-firm (versus within-firm), and the $R^2$ of a regression of capital on labor approximately equalling 0.5. These moments are similar in our data at 93\% and 0.4 respectively.

To avoid results depending on the arbitrary initial distribution of capital across firms, we initialize all firms with $K_{i0}=e^{-10}\approx 0$. We simulate firms for 100 periods and select the 10 periods in our estimation panel as the last of these, by which time the impact of the initial values appeared minimal.

\subsection{Replication of ACF Simulations} \label{subsec:Replication_ACF}

To confirm correct implementation of the MC routine, Table \ref{tab:mcrun_summtab_acftab1} repeats key DGPs of ACF's Table 1. The table shows moments of parameter estimates obtained by applying the OP, LP, ACF and NPR estimators to data simulated under the three distinct DGPs considered in \citet{ackerberg_identification_2015}. These are: DGP 1 which has serial correlation in wages and labor chosen at $t-b$, DGP 2 which has optimization error in labor and DGP 3 which combines both DGP 1 and DGP 2. The coefficients obtained from an OLS levels regression of $y_{it}$ on a constant, $l_{it}$ and $k_{it}$ are also provided for comparison. Each panel of the table contains results pertaining to the DGP indicated in the panel heading, while the results for the various estimators are given in rows. Within each DGP we show results when materials has no measurement error (Panels A, C and E) and when materials is subject to measurement error (B, D and F). 

As expected, the OLS estimates and OP results are biased throughout, with bias in the latter due to firm-specific capital adjustment costs violating OP's monotonicity condition. As shown by ACF, the LP estimator is sensitive to assumptions regarding the timing of firms' labor decision and the presence of firm-specific wages: whereas LP out-performs ACF when labor is chosen at time $t$, wages are homogeneous across firms and there is optimization error in labor (`DGP 2' in the table), it fails to recover consistent estimates when labor is chosen at some intermediate `$t-b$' period and/or wages are heterogeneous across firms. The first sub-panel within each DGP panel shows the ACF estimator is robust to DGPs 1 and 2, but becomes biased when the assumptions of the first two DGPs occur in combination (DGP 3). The second sub-panels within each DGP panel show the performance of both the LP and ACF estimators deteriorate as measurement error is added to firms' material input. The NPR estimator, by contrast, achieves highly precise and consistent estimates across all DGPs and is unaffected by measurement error in materials. 

\begin{table}[H]
\caption{ACF Monte Carlo Results}
\label{tab:mcrun_summtab_acftab1}
\centering
\begin{adjustbox}{max width=\textwidth}
    \begin{tabular}{l|cccc|cccc}
\hline\hline
& \multicolumn{4}{c|}{$\beta_{l}=0.6$} & \multicolumn{4}{c}{$\beta_{k}=0.4$} \\
& Mean & Median & S.D. & MSE & Mean & Median & S.D. & MSE \\
\hline
& \multicolumn{8}{c}{\textbf{DGP 1 - Serially Correlated Wages}} \\
& \multicolumn{8}{c}{\textbf{and Labor Set at Time} $ t-b $} \\ 
\hline
& \multicolumn{8}{c}{\textbf{A. No materials measurement error}} \\
\hline
NPR &0.600 &0.600 &0.003 &0.000 &0.395 &0.398 &0.016 &0.000 \\
OLS &0.944 &0.944 &0.004 &0.118 &0.065 &0.065 &0.006 &0.112 \\
OP &0.824 &0.824 &0.005 &0.050 &0.332 &0.332 &0.003 &0.005 \\
LP &-0.000 &-0.000 &0.005 &0.360 &1.121 &1.123 &0.028 &0.521 \\
ACF &0.599 &0.600 &0.010 &0.000 &0.401 &0.401 &0.015 &0.000 \\
\hline
& \multicolumn{8}{c}{\textbf{B. Materials measurement error} $\sim N(0,0.5) $} \\
\hline
NPR &0.600 &0.600 &0.003 &0.000 &0.395 &0.398 &0.016 &0.000 \\
OLS &0.944 &0.944 &0.004 &0.118 &0.065 &0.065 &0.006 &0.112 \\
OP &0.824 &0.824 &0.005 &0.050 &0.332 &0.332 &0.003 &0.005 \\
LP &0.754 &0.754 &0.007 &0.024 &0.291 &0.291 &0.011 &0.012 \\
ACF &0.628 &0.628 &0.010 &0.001 &0.405 &0.404 &0.015 &0.000 \\
\hline
& \multicolumn{8}{c}{\textbf{DGP 2 - Optimization Error in Labor}} \\
\hline
& \multicolumn{8}{c}{\textbf{C. No materials measurement error}} \\
\hline
NPR &0.600 &0.600 &0.003 &0.000 &0.400 &0.400 &0.005 &0.000 \\
OLS &0.920 &0.920 &0.002 &0.103 &0.092 &0.093 &0.004 &0.095 \\
OP &0.842 &0.843 &0.004 &0.059 &0.002 &0.000 &0.025 &0.159 \\
LP &0.600 &0.600 &0.003 &0.000 &0.400 &0.400 &0.013 &0.000 \\
ACF &0.600 &0.599 &0.009 &0.000 &0.400 &0.400 &0.015 &0.000 \\
\hline
& \multicolumn{8}{c}{\textbf{D. Materials measurement error} $\sim N(0,0.5) $} \\
\hline
NPR &0.600 &0.600 &0.003 &0.000 &0.400 &0.400 &0.005 &0.000 \\
OLS &0.920 &0.920 &0.002 &0.103 &0.092 &0.093 &0.004 &0.095 \\
OP &0.842 &0.843 &0.004 &0.059 &0.002 &0.000 &0.025 &0.159 \\
LP &0.797 &0.797 &0.003 &0.039 &0.220 &0.220 &0.010 &0.033 \\
ACF &0.620 &0.620 &0.013 &0.001 &0.405 &0.404 &0.018 &0.000 \\
\hline
& \multicolumn{8}{c}{\textbf{DGP 1 plus DGP 2}} \\
\hline
& \multicolumn{8}{c}{\textbf{E. No materials measurement error}} \\
\hline
NPR &0.600 &0.600 &0.002 &0.000 &0.400 &0.401 &0.009 &0.000 \\
OLS &0.863 &0.863 &0.004 &0.069 &0.159 &0.159 &0.007 &0.058 \\
OP &0.758 &0.759 &0.004 &0.025 &0.356 &0.356 &0.003 &0.002 \\
LP &0.473 &0.473 &0.003 &0.016 &0.588 &0.588 &0.016 &0.036 \\
ACF &0.596 &0.595 &0.019 &0.000 &0.406 &0.407 &0.023 &0.001 \\
\hline
& \multicolumn{8}{c}{\textbf{F. Materials measurement error} $\sim N(0,0.5) $} \\
\hline
NPR &0.600 &0.600 &0.002 &0.000 &0.400 &0.401 &0.009 &0.000 \\
OLS &0.863 &0.863 &0.004 &0.069 &0.159 &0.159 &0.007 &0.058 \\
OP &0.758 &0.759 &0.004 &0.025 &0.356 &0.356 &0.003 &0.002 \\
LP &0.677 &0.677 &0.005 &0.006 &0.386 &0.386 &0.012 &0.000 \\
ACF &0.608 &0.608 &0.007 &0.000 &0.431 &0.432 &0.013 &0.001 \\
\hline
\end{tabular}
\end{adjustbox}
\medskip
    \caption*{{\scriptsize Note: The table contains summary statistics calculated over 500 Monte Carlo simulations. The true values of $\beta_l$ and $\beta_k$ are 0.6 and 0.4 respectively.}}
\end{table}

\section{Supplementary Data and Results}\label{adix:supp_results}
\subsection{Expectations Survey Data}
As noted in the main text the MES has been extensively used by researchers, mainly to examine the questions on management practices. The description and documentation of the full surveys are available from the ONS UK Data Service: \url{https://datacatalogue.ukdataservice.ac.uk/studies/study/8557#documentation}. For example, the specific 2017 survey is available for download at: \url{https://doc.ukdataservice.ac.uk/doc/8557/mrdoc/pdf/8557_mes_2016_questionnaire_production.pdf}.

We reproduce the key expectations questions for ``turnover'', which is the British term for sales revenue, in Figure \ref{fig:MES_expectations_q} below. The format is identical for other inputs such as employment, investment and intermediate inputs. As noted in the main text, the 2017 MES was conducted by paper, but the 2020 wave was wholly online.

The MES and several other surveys also collect firms' expectations about aggregate macroeconomic variables. Examples include the Survey on Inflation and Growth Expectations by the Bank of Italy and the Business Tendency Survey in \citet{dovernetal2023}. While we do not leverage such aggregate expectations data here, we conjecture that they offer additional information that may aid production function estimation, for example by accommodating specific processes for expected productivity (as discussed in Appendix \ref{adix:expectations}), which must be ruled out for consistency of our baseline estimator.  

\begin{figure}[H]
    \centering
    \caption{MES 2017 sales Questions}\label{fig:MES_expectations_q}
    \includegraphics[width=\textwidth]{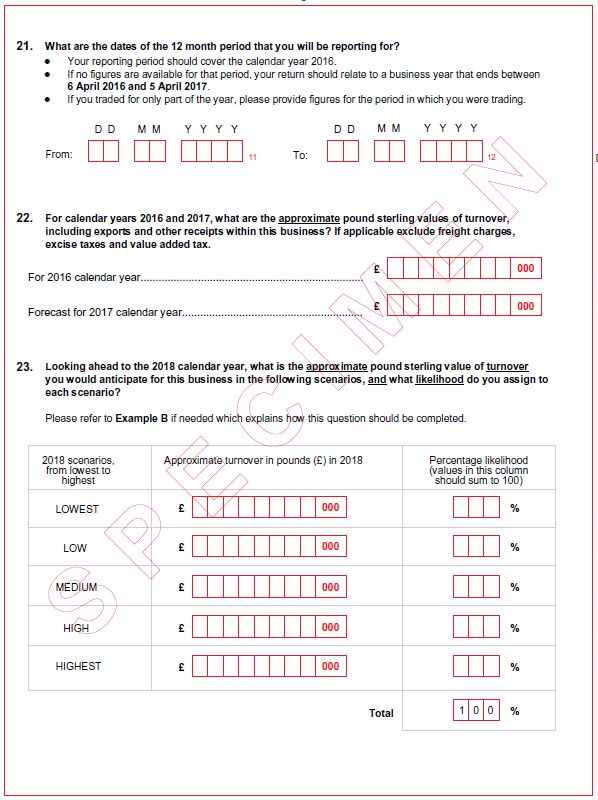}
\end{figure}

\subsection{Other Robustness Tests Relating to NPR Estimation}\label{adix:expectation_robustness}

Table \ref{tab:skewness} shows the skewness moments of subjective expectations for log employment and log output. Skewness for the normal distribution is zero, and indeed we find that the measured skewness in the raw data is concentrated around zero.

\begin{table}[H]
      \caption{Subjective Distribution Skewness Moments}\label{tab:skewness}
       \centering
       \begin{adjustbox}{max width=\textwidth}
           \begin{tabular}{l|ccccccc|c}
\hline \hline
& p5 & p10 & p25 & p50 & p75 & p90 & p95 & N \\
\hline
& \multicolumn{7}{c|}{ \textbf{(A) Electronics}} & \\\hline
\hspace{.4cm} Sales &  -0.001 &  -0.000 &  -0.000 &  -0.000 &   0.000 &   0.000 &   0.000 &     424 \\ 
\hspace{.4cm} Employment &  -0.000 &  -0.000 &  -0.000 &  -0.000 &   0.000 &   0.000 &   0.000 &     424 \\ 
\hline
& \multicolumn{7}{c|}{\textbf{(B) Retail}} & \\\hline
\hspace{.4cm} Sales &  -0.001 &  -0.000 &  -0.000 &  -0.000 &   0.000 &   0.000 &   0.000 &    1834 \\ 
\hspace{.4cm} Employment &  -0.001 &  -0.000 &  -0.000 &  -0.000 &   0.000 &   0.000 &   0.000 &    1834 \\ 
\hline
& \multicolumn{7}{c|}{\textbf{(C) Restaurants}} & \\\hline
\hspace{.4cm} Sales &  -0.004 &  -0.001 &  -0.000 &  -0.000 &   0.000 &   0.000 &   0.000 &     363 \\ 
\hspace{.4cm} Employment &  -0.003 &  -0.001 &  -0.000 &  -0.000 &   0.000 &   0.000 &   0.000 &     363 \\ 
\hline \hline
\end{tabular}

        \end{adjustbox}
        \caption*{\footnotesize{Note: The table shows moments of skewness calculated as $(\mathbb{E}[X^3]-3\mathbb{E}[X]\sigma^2-(\mathbb{E}[X])^3)/\sigma^3$}, where moments $\mathbb{E}[X^3]$, $\mathbb{E}[X]$ and $\sigma$ are firm-specific and calculated from the five-point subjective distributions reported in the MES. $X$ is either log sales or log employment, as indicated by row titles. The columns contain fuzzy percentiles of the skewness value calculated across all firms in the industry-specific sample denoted by panel titles.}
\end{table}

Table \ref{tab:backfit_results_ecomp} presents NPR production function estimates obtained using alternative constructions of the subjective expectation of the log of output and of the log of employment. Column (1) contains our baseline results and the other four are alternative methods, generating reasonably similar results. This demonstrates that the results presented in the main text are relatively robust to the specific method of constructing firms' expectations.

\begin{table}[H]
      \caption{NPR Production Function Coefficient Estimates Using Alternative Expectations Variables}\label{tab:backfit_results_ecomp}
       \centering
       \begin{adjustbox}{max width=\textwidth}
           \begin{tabular}{l|ccccc}
\hline\hline
 &  &  &  &  &  Weighted \\
 &  Baseline  &  CDF  &  SF  &  Beta  & Average \\ \hline
& \multicolumn{5}{c}{\textbf{(A) Electronics}} \\ \hline
 $ \beta_l $   &  0.83$ ^{\ast\ast\ast} $  &  0.84$ ^{\ast\ast\ast} $  &  0.77$ ^{\ast\ast\ast} $  &  0.86$ ^{\ast\ast\ast} $  &  0.86$ ^{\ast\ast\ast} $ \\
 &(0.07)&(0.07)&(0.12)&(0.11)&  (0.12)  \\
 $ \beta_k $   &  0.24$ ^{\ast\ast\ast} $  &  0.24$ ^{\ast\ast\ast} $  &  0.36$ ^{\ast} $  &  0.27$ ^{\ast} $  &  0.28$ ^{\ast} $ \\
 &(0.06)&(0.05)&(0.09)&(0.09)&  (0.09)  \\
 \hline
 $ \beta_l+\beta_k $  & 1.08 & 1.08 & 1.14 & 1.13 &  1.14 \\
 CRS   & 0.10 & 0.08 & 0.23 & 0.31 &  0.43 \\
 \hline
 N obs.   & 424 & 424 & 424 & 424 &    422 \\
 N firms   & 374 & 374 & 374 & 374 &    372 \\ \hline
& \multicolumn{5}{c}{\textbf{(B) Retail}} \\ \hline
 $ \beta_l $   &  0.75$ ^{\ast\ast\ast} $  &  0.76$ ^{\ast\ast\ast} $  &  0.76$ ^{\ast\ast\ast} $  &  0.77$ ^{\ast\ast\ast} $  &  0.78$ ^{\ast\ast\ast} $ \\
 &(0.07)&(0.08)&(0.07)&(0.12)&  (0.12)  \\
 $ \beta_k $   &  0.20$ ^{\ast\ast\ast} $  &  0.20$ ^{\ast\ast\ast} $  &  0.19$ ^{\ast\ast\ast} $  &  0.15$ ^{\ast\ast} $  &  0.15$ ^{\ast\ast\ast} $ \\
 &(0.04)&(0.04)&(0.04)&(0.06)&  (0.05)  \\
 \hline
 $ \beta_l+\beta_k $  & 0.95 & 0.96 & 0.94 & 0.92 &  0.93 \\
 CRS   & 0.29 & 0.30 & 0.30 & 0.53 &  0.47 \\
 \hline
 N obs.   & 1834 & 1834 & 1834 & 1834 &   1809 \\
 N firms   & 1633 & 1633 & 1633 & 1633 &   1612 \\ \hline
& \multicolumn{5}{c}{\textbf{(C) Restaurants}} \\ \hline
 $ \beta_l $   &  0.90$ ^{\ast\ast\ast} $  &  0.90$ ^{\ast\ast\ast} $  &  0.91$ ^{\ast\ast\ast} $  &  0.92$ ^{\ast\ast\ast} $  &  0.91$ ^{\ast\ast\ast} $ \\
 &(0.07)&(0.06)&(0.09)&(0.09)&  (0.07)  \\
 $ \beta_k $   & 0.11 & 0.12 & 0.09 & 0.09 &  0.12 \\
 &(0.07)&(0.07)&(0.08)&(0.08)&  (0.08)  \\
 \hline
 $ \beta_l+\beta_k $  & 1.01 & 1.02 & 1.00 & 1.01 &  1.03 \\
 CRS   & 0.84 & 0.84 & 0.76 & 0.83 &  0.80 \\
 \hline
 N obs.   & 363 & 363 & 363 & 363 &    348 \\
 N firms   & 338 & 338 & 338 & 338 &    323 \\
\hline\hline\end{tabular}
        \end{adjustbox}
        \caption*{\footnotesize{Note: dependent variable is log sales. Parentheses contain standard errors calculated from 100 bootstrap replications. Column titles indicate method used to construct subjective expectations variables. Baseline denotes the baseline construction described in Section \ref{subsect:data_expectations}. `CDF/SF/Beta' denotes fitted expectations using the lognormal CDF/lognormal survival function/beta CDF-survival function mean approach respectively. `Weighted Average' denotes expectations calculated as a weighted average across scenario values. Row `CRS' contains the p-value from a test that the labor and capital coefficient sum to 1. All specifications include survey year dummies. $^{\ast}$/$^{\ast\ast}$/$^{\ast\ast\ast}$ denote significance at the 10/5/1 percent level respectively.}}
\end{table}

\subsection{Translog Production Function}\label{adix:prodfunct_results}

The parameters of a translog production function can be estimated via the basic NPR algorithm outlined in Section \ref{sect:methodology} by setting 

\begin{equation*}
\begin{split}
y_{it}&=f(k_{it},l_{it}; \beta)+\omega_{it}+\epsilon_{it} \\
&=\beta_0+\beta_kk_{it}+\beta_ll_{it}+\beta_{k2}k^2_{it}+\beta_{l2}l^2_{it}+\beta_{kl}k_{it}l_{it}+\omega_{it}+\epsilon_{it}.
\end{split}
\end{equation*}

In this case, construction of the argument of the non-linear function NPR uses to control for $\omega$ requires data on $\mathbb{E}[l^2_{it+1}|\Omega_{it}]$. We calculate this as
\begin{equation*}
\begin{split}
    \mathbb{E}[l^2_{it+1}|\Omega_{it}]&=\textrm{Var}[l_{it+1}|\Omega_{it}]+\mathbb{E}[l_{it+1}|\Omega_{it}]^2 \\
    &=\sigma_l^2+\mu_l^2,
\end{split}
\end{equation*}
where $\mu_l$ and $\sigma_l^2$ are the mean and variance of the lognormal distributions we fit to firms' subjective expectations of year-ahead labor. Our assumptions on capital mean $k_{it+1}\in \Omega_{it}$, and hence the other additional terms in the control function are straightforward to obtain as $k^2_{it+1}$ and $k_{it+1}\mathbb{E}[l_{it+1}|\Omega_{it}]$. 

Tables \ref{tab:backfit_results_tl_electronics}-\ref{tab:backfit_results_tl_foodbev_serv} contain translog parameter estimates obtained via NPR, ACF and various OLS estimators. The $\bar{l}$ and $\bar{k}$ rows contain the estimation sample means of $l$ and $k$ respectively, which are used to calculate the mean partial derivatives given in the $ \frac{\partial y}{\partial l} $ and $ \frac{\partial y}{\partial k} $ rows. The `Cobb-Douglas Wald' row gives the Wald statistic of a test that the non-Cobb-Douglas parameters $(\beta_{l2},\beta_{k2},\beta_{lk})$ are all equal to zero and Row `Cobb-Douglas P-Value' gives the corresponding p-value. These rows show the NPR estimates fail to reject Cobb-Douglas technology in the retail and restaurants industries, whereas Cobb-Douglas technology is rejected by NPR in the electronics industry. 

\begin{table}[H]
       \caption{Electronics: Translog Production Function Coefficient Estimates}
       \label{tab:backfit_results_tl_electronics}
        \centering
        \begin{adjustbox}{max width=\textwidth}
            \begin{tabular}{l|ccccc}
\hline\hline
& NPR & ACF & OLS & OLS FD & OLS FE \\
\hline
 $ \beta_{l1} $  & 1.10$ ^{\ast\ast\ast} $ & 0.99$ ^{\ast\ast\ast} $ & 0.95$ ^{\ast\ast\ast} $ & 0.68$ ^{\ast\ast} $ & 0.84$ ^{\ast\ast\ast} $ \\
 &(0.25)&(0.20)&(0.18)&(0.27)&(0.28) \\
 $ \beta_{k1} $  & 0.54$ ^{\ast\ast} $ &0.16&0.18&0.11& 0.31$ ^{\ast\ast} $ \\
 &(0.26)&(0.29)&(0.12)&(0.07)&(0.13) \\
 $ \beta_{l2} $  & 0.12$ ^{\ast\ast\ast} $ &0.06& 0.08$ ^{\ast} $ &-0.01&0.02\\
 &(0.04)&(0.14)&(0.04)&(0.04)&(0.04) \\
 $ \beta_{k2} $  &0.05&0.02& 0.05$ ^{\ast\ast\ast} $ &0.01&0.00\\
 &(0.03)&(0.03)&(0.01)&(0.00)&(0.01) \\
 $ \beta_{lk} $  & -0.19$ ^{\ast\ast\ast} $ &-0.07& -0.13$ ^{\ast\ast\ast} $ & -0.03$ ^{\ast} $ & -0.07$ ^{\ast\ast} $ \\
 &(0.06)&(0.09)&(0.04)&(0.02)&(0.03) \\
 \hline
 $ \bar{l} $  &4.01&4.01&4.01&4.01&4.00\\
 $ \bar{k} $  &7.66&7.66&7.66&7.66&7.64\\
 \hline
 $ \frac{\partial y}{\partial l} $  & 0.64$ ^{\ast\ast\ast} $ &0.96& 0.62$ ^{\ast\ast\ast} $ & 0.39$ ^{\ast\ast\ast} $ & 0.52$ ^{\ast\ast\ast} $ \\
 &(0.06)&(0.80)&(0.04)&(0.11)&(0.09) \\
 $ \frac{\partial y}{\partial k} $  & 0.50$ ^{\ast\ast\ast} $ &0.17& 0.43$ ^{\ast\ast\ast} $ & 0.08$ ^{\ast} $ &0.08\\
 &(0.06)&(0.18)&(0.03)&(0.04)&(0.06) \\
 \hline
 Cobb-Douglas $ \chi^2 $ &16.78&0.56&40.74&4.01&5.61\\
 Cobb-Douglas P-Value &0.00&0.90&0.00&0.26&0.13\\
 \hline
 N obs.  &424&424&424&424&848\\
 N firms  &374&374&374&374&374\\
\hline\hline\end{tabular}
        \end{adjustbox}
        \caption*{\footnotesize{Note: dependent variable is log sales. Parentheses contain standard errors. NPR standard errors calculated from 100 bootstrap replications. Column titles indicate estimation methods. $^{\ast}$/$^{\ast\ast}$/$^{\ast\ast\ast}$ denote significance at the 10/5/1 percent level respectively.}}
\end{table}

\begin{table}[H]
       \caption{Retail: Translog Production Function Coefficient Estimates}
       \label{tab:backfit_results_tl_wsale_retail}
        \centering
        \begin{adjustbox}{max width=\textwidth}
            \begin{tabular}{l|ccccc}
\hline\hline
& NPR & ACF & OLS & OLS FD & OLS FE \\
\hline
 $ \beta_{l1} $  & 1.01$ ^{\ast\ast\ast} $ & 1.07$ ^{\ast\ast} $ & 1.09$ ^{\ast\ast\ast} $ & 0.58$ ^{\ast\ast\ast} $ &0.27\\
 &(0.27)&(0.50)&(0.11)&(0.14)&(0.37) \\
 $ \beta_{k1} $  &0.15&-0.13& 0.37$ ^{\ast\ast\ast} $ &-0.01&0.02\\
 &(0.30)&(0.36)&(0.12)&(0.02)&(0.03) \\
 $ \beta_{l2} $  &0.01& -0.56$ ^{\ast} $ &-0.02&-0.01&0.03\\
 &(0.03)&(0.33)&(0.03)&(0.02)&(0.04) \\
 $ \beta_{k2} $  &0.02&0.02&0.00&0.00&0.00\\
 &(0.03)&(0.02)&(0.01)&(0.00)&(0.00) \\
 $ \beta_{lk} $  &-0.05&-0.04&-0.03&0.00&-0.01\\
 &(0.06)&(0.08)&(0.03)&(0.00)&(0.01) \\
 \hline
 $ \bar{l} $  &3.95&3.95&3.95&3.95&3.94\\
 $ \bar{k} $  &6.76&6.76&6.76&6.76&6.61\\
 \hline
 $ \frac{\partial y}{\partial l} $  & 0.75$ ^{\ast\ast\ast} $ &-3.64& 0.76$ ^{\ast\ast\ast} $ & 0.49$ ^{\ast\ast\ast} $ & 0.49$ ^{\ast\ast\ast} $ \\
 &(0.07)&(2.59)&(0.04)&(0.05)&(0.09) \\
 $ \frac{\partial y}{\partial k} $  & 0.20$ ^{\ast\ast\ast} $ &-0.01& 0.29$ ^{\ast\ast\ast} $ &0.00&0.02\\
 &(0.07)&(0.05)&(0.03)&(0.01)&(0.02) \\
 \hline
 Cobb-Douglas $ \chi^2 $ &2.98&3.43&31.93&0.77&1.7\\
 Cobb-Douglas P-Value &0.40&0.33&0.00&0.86&0.64\\
 \hline
 N obs.  &1834&1834&1834&1834&3668\\
 N firms  &1633&1633&1633&1633&1633\\
\hline\hline\end{tabular}
        \end{adjustbox}
        \caption*{\footnotesize{Note: dependent variable is log sales. Parentheses contain standard errors. NPR standard errors calculated from 100 bootstrap replications. Column titles indicate estimation methods. $^{\ast}$/$^{\ast\ast}$/$^{\ast\ast\ast}$ denote significance at the 10/5/1 percent level respectively.}}
\end{table}

\begin{table}[H]
       \caption{Restaurants: Translog Production Function Coefficient Estimates}
       \label{tab:backfit_results_tl_foodbev_serv}
        \centering
        \begin{adjustbox}{max width=\textwidth}
            \begin{tabular}{l|ccccc}
\hline\hline
& NPR & ACF & OLS & OLS FD & OLS FE \\
\hline
 $ \beta_{l1} $  & 1.03$ ^{\ast\ast\ast} $ & 4.55$ ^{\ast\ast\ast} $ & 1.01$ ^{\ast\ast\ast} $ & 0.66$ ^{\ast\ast\ast} $ & 0.58$ ^{\ast\ast\ast} $ \\
 &(0.19)&(1.35)&(0.10)&(0.18)&(0.19) \\
 $ \beta_{k1} $  &0.08&-0.46&-0.01&-0.23&-0.15\\
 &(0.24)&(0.70)&(0.07)&(0.17)&(0.13) \\
 $ \beta_{l2} $  &0.00& -0.45$ ^{\ast\ast\ast} $ &-0.03& -0.04$ ^{\ast} $ &-0.02\\
 &(0.03)&(0.15)&(0.02)&(0.02)&(0.02) \\
 $ \beta_{k2} $  &0.01&0.01&0.01&0.01&0.00\\
 &(0.03)&(0.05)&(0.01)&(0.01)&(0.01) \\
 $ \beta_{lk} $  &-0.03& 0.14$ ^{\ast} $ &0.02& 0.05$ ^{\ast\ast} $ & 0.04$ ^{\ast\ast} $ \\
 &(0.05)&(0.08)&(0.03)&(0.02)&(0.02) \\
 \hline
 $ \bar{l} $  &4.57&4.57&4.57&4.57&4.57\\
 $ \bar{k} $  &7.72&7.72&7.72&7.72&7.69\\
 \hline
 $ \frac{\partial y}{\partial l} $  & 0.87$ ^{\ast\ast\ast} $ & 1.52$ ^{\ast\ast} $ & 0.82$ ^{\ast\ast\ast} $ & 0.68$ ^{\ast\ast\ast} $ & 0.75$ ^{\ast\ast\ast} $ \\
 &(0.04)&(0.72)&(0.04)&(0.06)&(0.06) \\
 $ \frac{\partial y}{\partial k} $  & 0.16$ ^{\ast\ast\ast} $ &0.37& 0.20$ ^{\ast\ast\ast} $ & 0.09$ ^{\ast} $ & 0.07$ ^{\ast} $ \\
 &(0.04)&(0.23)&(0.04)&(0.05)&(0.04) \\
 \hline
 Cobb-Douglas $ \chi^2 $ &0.69&8.77&9.63&5.8&5.56\\
 Cobb-Douglas P-Value &0.88&0.03&0.02&0.12&0.13\\
 \hline
 N obs.  &363&363&363&363&726\\
 N firms  &338&338&338&338&338\\
\hline\hline\end{tabular}
        \end{adjustbox}
        \caption*{\footnotesize{Note: dependent variable is log sales. Parentheses contain standard errors. NPR standard errors calculated from 100 bootstrap replications. Column titles indicate estimation methods. $^{\ast}$/$^{\ast\ast}$/$^{\ast\ast\ast}$ denote significance at the 10/5/1 percent level respectively.}}
\end{table}
    
\end{document}